\documentclass{aastex631}

\usepackage{amsthm}
\usepackage{amsmath}
\usepackage{multirow}
\usepackage{hyperref}
\usepackage{float}
\usepackage{subfigure}

\newcommand{\be}{\begin{equation}}
\newcommand{\ee}{\end{equation}}
\newcommand{\bdm}{\begin{displaymath}}
\newcommand{\edm}{\end{displaymath}}
\newcommand{\bea}{\begin{eqnarray}}
\newcommand{\eea}{\end{eqnarray}}

\newcommand{\halb}{\frac{1}{2}}

\renewcommand{\u}{\mathbf{u}}
\newcommand{\U}{\mathbf{U}}
\newcommand{\q}{\mathbf{q}}
\newcommand{\Q}{\mathbf{Q}}
\newcommand{\w}{\mathbf{w}}
\newcommand{\m}{\mathbf{m}}
\newcommand{\F}{\mathbf{F}}
\newcommand{\f}{\mathbf{f}}
\newcommand{\g}{\mathbf{g}}
\newcommand{\h}{\mathbf{h}}
\newcommand{\B}{\mathbf{B}}
\newcommand{\A}{\mathbf{A}}

\newcommand{\ie}{\textit{i.e.}~}

\begin{document}

\title{Well-balanced high order finite difference WENO schemes for a first-order Z4 formulation of the Einstein field equations} 

\correspondingauthor{Dinshaw Balsara}
\email{dbalsara@nd.edu}

\author[0000-0003-3309-1052]{Dinshaw Balsara}
\affiliation{Department of Physics and Astronomy, University of Notre Dame, Notre Dame, IN, United States}
\affiliation{ACMS Department, University of Notre Dame, Notre Dame, USA}

\author[0000-0003-2849-9045]{Deepak Bhoriya}
\affiliation{Department of Physics and Astronomy, University of Notre Dame, Notre Dame, IN, United States}

\author[0000-0003-4307-6809]{Olindo Zanotti}
\affiliation{Laboratory of Applied Mathematics, Department of Civil, Environmental and Mechanical Engineering, University of Trento, Via Mesiano 77, I-38123 Trento, Italy}

\author[0000-0002-8201-8372]{Michael Dumbser}
\affiliation{Laboratory of Applied Mathematics, Department of Civil, Environmental and Mechanical Engineering, University of Trento, Via Mesiano 77, I-38123 Trento, Italy}

\begin{abstract}

   We develop a new class of high--order accurate well-balanced 
   finite difference (FD) Weighted Essentially Non-Oscillatory (WENO) methods for numerical general relativity, 
   which can be applied to any first--order reduction of the Einstein field equations, even if non--conservative terms are present.
   We choose the first--order non--conservative Z4 formulation of the Einstein equations, which
   has a built--in cleaning procedure that accounts for the Einstein constraints and that
   has already shown its ability in keeping stationary solutions stable over long timescales.
   By introducing auxiliary variables, the vacuum Einstein equations in first--order form constitute a PDE system of 54 equations that is naturally non-conservative. We show how to design FD--WENO schemes that can handle non-conservative products.
   Different variants of FD--WENO are discussed, with an eye to their suitability for higher order accurate formulations for numerical general relativity.
   We successfully solve a set of fundamental tests of numerical general relativity with up to ninth order spatial accuracy. 
   Due to their intrinsic robustness, flexibility and ease of implementation, finite difference WENO schemes can effectively replace traditional central finite differencing in any first--order
   formulation of the Einstein field equations, without any artificial viscosity. 
   When used in combination with \emph{well--balancing}, the new numerical schemes preserve stationary equilibrium solutions of the Einstein equations exactly. This is particularly relevant in view of the numerical study of the quasi-normal modes of oscillations of relevant astrophysical sources.
   In conclusion, general relativistic high--energy astrophysics could benefit from this new class of numerical schemes and the ecosystem of desirable capabilities built around them.

\end{abstract}



\section{Introduction}
\label{introduction}
 In recent years, many of the equations that are needed in computational astrophysics have seen the development of higher order methods for their solution. This harkens to the fact that the corresponding high accuracy is needed in response to more accurate observations; General Relativity (GR) is an emblematic example of this fact.
 Indeed, the third generation of 
 gravitational wave detectors~\citep{Luck2022}, such as the Einstein Telescope in Europe and the Cosmic Explorer in US, with a planned sensitivity $h$
 in the range $10^{-25}\div10^{-24}\,\rm{Hz}^{-1/2}$, will soon call for more accurate numerical solution methods in Numerical Relativity (NR). 
 
As is well known, the Einstein equations present inbuilt challenges that do not show up in many of the other governing equations commonly considered in computational astrophysics. 
For one thing, the system naturally arises as a second order PDE system, whereas our best numerical tools have been developed for first order PDE systems. 
{
In this respect, it is interesting to note that, while the generalized harmonic formulation
exists as a first order system since quite a while~\citep{Lindblom2006}, having already allowed for 
highly relevant scientific results (see, among others, \cite{Boyle2007,Duez2008,Scheel2009,Szilagyi2014,Tichy2023,Deppe2024} and references therein),
the whole family represented by the BSSNOK/Z4/CCZ4 formulations~\citep{Nakamura87,Shibata95,Baumgarte99,Alcubierre:2008,Baumgarte2010,Bona-and-Palenzuela-Luque-2005:numrel-book, Alic:2009} 
seems to have shown a stronger inertia in performing the transition from second--order to first--order 
implementations, with notable but rare progresses reported by 
\cite{Bona:2004}, \cite{Brown2012}, \cite{Dumbser2017strongly}.}
The latter, in particular, applied Discontinuous Galerkin (DG) schemes to the conformal and covariant Z4 formulation of the Einstein equations (FO-CCZ4) of \cite{Alic:2011a}.
More recently,  \cite{DumbserZanottiGaburroPeshkov2023} have obtained promising results after using DG methods in a well--balanced first--order implementation of the damped version of the Z4 formulation proposed by
\cite{Gundlach2005:constraint-damping}, henceforth referred to as the FO-Z4 formulation.
We recall that the Z4 formulation of the Einstein equations was originally proposed by \cite{Bona:2003fj,Bona:2004} to automatically account for
a proper treatment of the
Einstein constraints through the addition of a four vector $z^\mu$, in a rather similar way to what is  done in the divergence cleaning approach by~\cite{MunzCleaning} and \cite{Dedner:2002} for the Maxwell and magnetohydrodynamics (MHD) equations. 
Later on, this approach has been combined with the conformal decomposition of the metric, which is missing in the original Z4 formulation, to obtain the so--called CCZ4 and Z4c formulations~\citep{Alic:2011a,Hilditch2012a,Hilditch2012b, Alic2013,Dumbser2017strongly,Peterson_2023}.

On the computational side, one can expect that,
for various reasons,
finite difference (FD) schemes will remain the preferred
choice by the NR community still for many years to come.
When a second--order formulation of the  Einstein equations 
is adopted, central finite difference schemes are 
the most natural choice, which is  indeed the case for
such popular codes as
Einstein--Toolkit~\citep{Loffler2012}, LazEv~\citep{Zlochower2005,Lousto2023}, BAM~\citep{Bruegmann2008,Thierfelder2011},
McLachlan~\citep{McLachlan:web}, GRChombo~\citep{Clough_2015},
AMReX~\citep{Peterson_2023}, SACRA~\citep{Yamamoto2009,Kiuchi2017}.
On the other hand, when a first--order 
formulation of the  Einstein equations is available, 
a natural temptation arises: namely to migrate from central finite difference schemes
to finite difference  Weighted Essentially Non-Oscillatory 
(WENO) methods, due to their superior robustness in the presence of shock waves and singularities, maintaining at the same time an excellent performance in terms of accuracy.
However,  original WENO methods were specifically devised for first--order
hyperbolic systems {\em in conservation form}, and they do therefore not fit straightforwardly into
the non--conservative form of the
first--order Einstein equations.
In fact, the usage of FD--WENO methods
in the relativistic context has been so far limited to the solution of 
the (conservative) 
term $\nabla_\mu T^{\mu\nu}=0$, either in special relativity or in general 
relativistic but stationary spacetimes~\citep{DelZanna2007,Kailiang2015,Inghirami2016}.
It is interesting to note that 
the Z4 formulation of the Einstein equations
was actually originally proposed in
first--order conservative form~\citep{Bona:2004,Bona-and-Palenzuela-Luque-2005:numrel-book}, though it does not seem that it has ever been implemented by using WENO finite difference methods.

As we have just discussed, the very large size of the FO-Z4 system, as well as its non-conservative aspect, restricts our choice of numerical methods that can be applied to it. As realized in \cite{Dumbser2017strongly,Dumbser2020GLM} and \cite{DumbserZanottiGaburroPeshkov2023}, path-conservative Discontinuous Galerkin (DG) schemes may be one route to higher order for the Einstein equations, but DG schemes tend to be prohibitively expensive, both in terms of their memory usage as well as in terms of computational complexity. This is because a DG scheme retains all the higher order modes of a hyperbolic system and for a gigantic hyperbolic system, like the one being considered, that exacerbates the memory consumption. All these modes have also to be evolved in time with the result that the computational complexity of evaluating so many evolutionary equations for each of the modes of a DG scheme can again become prohibitive. Finite volume Weighted Essentially Non-Oscillatory (WENO) schemes are another alternative route to higher order and recently some efficient options have been offered, see \cite{balsara2023efficient}, and the supplement of that paper, as well as \cite{AMR3DCL}. However, the finite volume WENO schemes require the reconstruction of the same number of modes as a DG scheme of the same order. Consequently, finite volume WENO schemes also have a prohibitive memory usage, though their computational complexity and timestep restriction are much better than DG schemes of the same order. The only remaining path to high order accuracy at low computational cost is therefore through finite difference WENO schemes \citep{shu1988efficient,shu1989efficient,jiang1996efficient,balsara2000monotonicity,balsara2016efficient,balsara2023FDWENO,balsara2024a,balsara2024b}. Compared to their finite volume and DG cousins, finite difference schemes have a memory usage and computational complexity that can be described as downright Spartan. Unlike DG schemes, they offer a robust CFL even at higher orders. As shown in \cite{balsara2024b}, the ninth order finite difference WENO schemes are not much more expensive compared to their third order variants, with the result that the path to progressively higher order does not incur much additional expense in terms of computational complexity. This owes to the dimension-by-dimension approach that is used in finite difference WENO schemes.  

It is worth noting that there are two variants of finite difference WENO schemes. The well-known one, which we can refer to as the classical finite difference WENO scheme (here referred to as FD--WENO) stems from \cite{shu1989efficient}. For a long time, it was only available in a form that was suitable for conservation laws. In that form, it is quite useless for first--order non--conservative formulations of 
the Einstein equations, such as those considered by
\cite{Dumbser2017strongly}, \cite{DumbserZanottiGaburroPeshkov2023} or \cite{Brown2012}.
Recently, the FD--WENO schemes have also been extended to hyperbolic PDE systems with non-conservative products \citep{balsara2023FDWENO}. In a paper that slightly predates \cite{shu1989efficient}, \cite{shu1988efficient} had also presented an alternative formulation of finite difference WENO schemes, but for a long time these methods had remained inaccessible to the broader community. This is partly owing to the mathematically recondite nature of the original presentation and partly owing to the fact that the original formulation in \cite{shu1988efficient} was far from broadly usable. We will refer to this alternative strain of finite difference WENO schemes as the AFD--WENO schemes. The original AFD--WENO schemes were, therefore, eclipsed by the vastly more popular FD--WENO schemes. A paper by \cite{merriman2003understanding} tried to make the underlying mathematics of the AFD--WENO schemes more accessible. These days, of course, the mathematics that underlies AFD--WENO schemes has become very accessible; please see Section 2 from \cite{balsara2024a} as well as its Appendix A, which provides a computer algebra system-based derivation of the scheme. The impetus for AFD--WENO schemes came when \cite{cai2008performance} and \cite{nonomura2010freestream} showed that FD--WENO schemes could not preserve a free stream condition on geometrically complex meshes.  In \cite{jiang2013,jiang2014free} it was realized that FD--WENO suffered from an inability to preserve a free stream condition because it was based on flux reconstruction. AFD--WENO schemes were found to be free of such limitations. However, even in \cite{zheng2021high}, controlling the Gibbs phenomenon that arises from the higher order flux derivatives in AFD--WENO was found to be an elusive enterprise. A full resolution of controlling the Gibbs phenomenon only emerged in \cite{balsara2024a} where a different type of WENO interpolation was invented for that purpose. The paper by \cite{balsara2024a} still presented AFD--WENO schemes for conservation laws. The extension of these AFD--WENO methods to hyperbolic systems with non-conservative products was finally accomplished in \cite{balsara2024b}. The upshot of this paragraph is that we now have two very proficient methods for treating hyperbolic PDEs with a large number of non-conservative products, such as arises in NR. We have the FD--WENO methods from \cite{balsara2023FDWENO} and we have the AFD--WENO methods from \cite{balsara2024b}.  

Modern astrophysical codes rely on capabilities that go beyond the baseline scheme. A good example would be well-balancing, which makes it possible to approach steady state on moderately resolved meshes, see \cite{kappeli2022well} for a review and \cite{Gaburro2021WBGR1D} for a first successful application of well-balancing to numerical general relativity. Because FD--WENO schemes in conservation form have been around for a while, well-balancing has been developed for these methods \citep{xing2011high}. In \cite{balsara2023FDWENO} we showed that FD--WENO schemes for non-conservative products also tend to be well-balanced if contact-discontinuity preserving Riemann solvers are used. For AFD--WENO schemes in conservation form, \cite{xu2024} have shown that they can be formulated so as to be well-balanced and can also preserve moving equilibria provided one is willing to go through the cumbersome process of identifying those equilibria. In recent work, we have seen the development of well-balanced methods that preserve moving equilibria \citep{xu2024}. Another capability that goes beyond the baseline scheme is positivity preservation; which is also referred to these days as the physical condition preserving (PCP) property. For FD--WENO schemes in conservation form, \cite{hu2013positivity} have developed positivity preserving formulations. This author and his collaborators have also developed PCP formulations for AFD--WENO schemes \citep{bhoriya2024}. Astrophysical codes also have geometrical constraints that require divergence-preservation \citep{balsara1999staggered,balsara2010multidimensional,balsara2012self,balsara2014multidimensional} and curl-preservation \citep{balsara2021curl,balsara2022neumann} and the author and his collaborators are extending these capabilities to include AFD--WENO methods. We see, therefore, that there is an ecosystem of ancillary algorithms that have been built, and are being built, around finite difference WENO methods that make them very useful for NR. 

The plan of the paper is the following. In Section~\ref{sec:Z4} we
recall the essential features of the first order Z4 formulation of the Einstein field equations;
Section~\ref{sec:scheme} contains the core novelties of this paper, by presenting a 
new class of finite difference WENO schemes, whose validation is reported in 
Section~\ref{sec:tests}. Finally, Section~\ref{sec:conclusions} contains the conclusions of our work.

We assume a signature $\{-,+,+,+\}$
for the space-time metric and we use Greek letters
$\mu,\nu,\lambda,\ldots$ (running from 0 to 3) for
four-dimensional space-time tensor components, while
Latin letters $i,j,k,\ldots$ (running from 1 to 3) 
are employed for three-dimensional spatial tensor
components. Moreover, we set $G=c=1$.

\section{The damped first--order Z4 system of the Einstein equations}
\label{sec:Z4}
As usual for 3+1 formulations of the Einstein equations, the  spacetime is foliated 
through $\Sigma_t=const$ hypersurfaces as~\citep{Alcubierre:2008,Baumgarte2010,Gourgoulhon2012,Rezzolla_book:2013}
\begin{equation}
\label{eq:ds2_3p1}
ds^{2} = g_{\mu\nu}dx^{\mu}dx^{\nu}=  -(\alpha^{2}-\beta_{i}\beta^{i}) dt^{2}+ 
2 \beta_{i} dx^{i} dt + \gamma_{ij} dx^{i}dx^{j} \,,
\end{equation}
where $g_{\mu\nu}$ is the spacetime metric tensor,
$\alpha$ is the lapse, $\beta^i$ is the shift and $\gamma_{ij}$ is the metric of the three dimensional space. Within this framework,
we adopt the
damped version of the Z4 formulation of the vacuum Einstein equations, originally 
proposed by~\cite{Gundlach2005:constraint-damping} and recently
reformulated with minor modifications by \cite{DumbserZanottiGaburroPeshkov2023}, \ie
\begin{equation}
\label{Trento1-Z4-1}
G_{\mu\nu} + \nabla_\mu z_\nu + \nabla_\nu z_\mu  -  \nabla_\pi z^\pi g_{\mu\nu} - \kappa_1(n_\mu z_\nu + n_\nu z_\mu) - \kappa_2  n_\pi z^\pi g_{\mu\nu}
=0\,,
\end{equation}
where $G_{\mu\nu}$ is the Einstein tensor and $n^\mu$ is the four--velocity of the Eulerian observer.
The four vector $z^\mu=\Theta n^{\mu}+ Z^{\mu}$ is artificially introduced to {\emph{clean}} the violations of the Einstein constraints. {We therefore notice that $\Theta$ and $Z^\mu$ are the 3+1 representations of $z^\mu$.}
The two constant coefficients $\kappa_1$ and $\kappa_2$ can act as damping mechanisms over 
the four vector $z^\mu$. Not present in the original formulation 
by \cite{Bona:2003fj,Bona:2003qn}, such coefficients were introduced by 
\cite{Gundlach2005:constraint-damping} and slightly modified by \cite{DumbserZanottiGaburroPeshkov2023}
in such a way to avoid their mutual multiplication in the resulting PDE system.
The {\emph {extrinsic curvature}} of the hypersurface $\Sigma_t$, a crucial  quantity of differential geometry,
is one of the primary variables of the Z4 formulation and it is defined as 
\begin{equation}
K_{\mu\nu}=-\gamma^\alpha_\mu  \nabla_\alpha n_\nu=-\nabla_\mu n_\nu - n_\mu a_\nu\,,
\end{equation}
where $a_\mu = n^\nu \nabla_\nu n_\mu$ is the acceleration of the
Eulerian observer.
Upon the introduction of 30 auxiliary variables involving first derivatives of the metric terms, namely
\begin{align}
\label{eq:Auxiliary}
A_i := \partial_i\ln\alpha = \frac{\partial_i \alpha }{\alpha}\,, \qquad
B_k^{\,\,i} := \partial_k\beta^i\,,
\qquad
D_{kij} := \frac{1}{2}\partial_k\gamma_{ij}\,, \qquad
\end{align}
and after fixing the gauge conditions in a pretty standard way as~\citep{Baumgarte2010,Faber2007}
\begin{eqnarray}
\label{eqn.slicing}
\partial_t \ln\alpha - \beta^k{\partial_k\ln\alpha} &=& -g(\alpha)\alpha(K-K_0-2c\Theta)\,,\\
\label{g-driver1}
\partial_t \beta^i   &=& \frac{3}{4}b^i, \\
\label{g-driver2}
\partial_t b^i&=&\partial_t \hat\Gamma^i - \eta b^i\,,
\end{eqnarray}
the vacuum Einstein equations \eqref{Trento1-Z4-1} lead to the following first order system of hyperbolic PDEs:
\begin{align}
\label{eqn.gamma}
&\partial_t\gamma_{ij} - \beta^k\partial_k\gamma_{ij}=\gamma_{ik} B_{j}^{\,\,k}  + \gamma_{kj} B_{i}^{\,\,k} - 2\alpha K_{ij}\,,\\
\label{eqn.Kij}
&\partial_t K _{ij} - \beta^k \partial_k K_{ij}  + \alpha \partial_{(i} {A}_{j)}
- \alpha {\gamma}^{kl} \left( \partial_{(k} {D}_{i)jl}  - \partial_{(k} {D}_{l)ij} \right)
+ \alpha {\gamma}^{kl} \left( \partial_{(j} {D}_{i)kl}  - \partial_{(j} {D}_{l)ik} \right)
- 2\alpha  \partial_{(i} {Z}_{j)}
 =  K_{ki} B_j^{\,\,k} + K_{kj} B_i^{\,\,k}  \nonumber \\
&- \alpha A_i A_j + \alpha \Gamma^k_{ij} A_k +  
\alpha\bigg[  - 2 {\gamma}^{kn} {\gamma}^{pl} D_{knp}  \left( D_{ijl} + D_{jil} - D_{lij} \right)   
              + 2 {\gamma}^{kn} {\gamma}^{pl} D_{jnp}  \left( D_{ikl} + D_{kil} - D_{lik} \right) \nonumber \\
& + \Gamma^m_{lm} \Gamma^l_{ij} - \Gamma^m_{lj} \Gamma^l_{im}     \bigg] 
- 2\alpha\Gamma^k_{ij} Z_k     
-  \alpha\Theta\gamma_{ij}(\kappa_1 + \kappa_2)  
- 2 \alpha K_{il} \gamma^{lm} K_{mj} +  \alpha K_{ij}(K - 2 \Theta  )   \,, \\
\label{eqn.theta}
&\partial_t \Theta  - \beta^k\partial_k\Theta  -  \frac{1}{2}\alpha {e^2} \left[ \gamma^{ij} {\gamma}^{kl} \left( \partial_{(k} {D}_{i)jl}  - \partial_{(k} {D}_{l)ij} \right) - \gamma^{ij}{\gamma}^{kl} \left( \partial_{(j} {D}_{i)kl}  - \partial_{(j} {D}_{l)ik} \right) + 2 \gamma^{ij}\partial_{i} Z_{j}\right]
= \nonumber \\
& = \frac{\alpha}{2}\, e^2 \, 
 \bigg[- 2 \gamma^{ij}{\gamma}^{kn} {\gamma}^{pl} D_{knp}  \left( D_{ijl} + D_{jil} - D_{lij} \right)   
       + 2 \gamma^{ij}{\gamma}^{kn} {\gamma}^{pl} D_{jnp}  \left( D_{ikl} + D_{kil} - D_{lik} \right)\nonumber \\
&+\gamma^{ij}\left( \Gamma^m_{lm} \Gamma^l_{ij} - \Gamma^m_{lj} \Gamma^l_{im}   \right) + K^2 - K_{ij}\,K^{ij} \bigg] + 
\alpha\left[ -\gamma^{ij}\,\Gamma^k_{ij}Z_k 
  - Z^k A_k   \right] - \alpha \Theta K  
				- \alpha\Theta(2\kappa_1+ \kappa_2)\,,  \\
\label{eqn.zeta}
&\partial_t Z_i  - \beta^k\partial_k Z_i - \alpha \partial_i\Theta -\alpha\left[ \gamma^{jm}\partial_jK_{mi} - \gamma^{mn}\partial_i K_{mn}  \right]
  = Z_k\,B_i^{\,\,k}  +  \alpha\, \bigg[-\gamma^{jm}(\Gamma^n_{jm} K_{ni}+\Gamma^n_{ji} K_{mn}) \nonumber \\
	&+\gamma^{mn}(\Gamma^l_{im} K_{ln}+\Gamma^l_{in} K_{ml}) \bigg] + \alpha[ - 2\, {K_i}^j\, Z_j  -  \Theta\, A_i
-\kappa_1 Z_i]\,,\\
\label{eqn.A}
& \partial_t A_{i} - {\beta^k \partial_k A_i} + \alpha g(\alpha) \left( \gamma^{mn}\partial_i K_{mn} - \partial_i K_0 - 2c \partial_i \Theta \right)
=-\alpha A_i \left( K - K_0 - 2 \Theta c \right) \left( g(\alpha) + \alpha g^\prime(\alpha)  \right) 
	\\
	&+  2\alpha g(\alpha) K^{jk}D_{ijk}  +   B_i^{\,\,k} ~A_{k} \,,
\\
\label{eqn.B}
& \partial_t B_k^{\,\,i}  - s\left(  \frac{3}{4} \partial_k b^i  
- \alpha^2 \mu \, \gamma^{ij} \gamma^{nl} \left( \partial_k D_{ljn} - \partial_l D_{kjn} \right)  \right)
=		0 \,,
\\
\label{eqn.D}
&
\partial_t D_{kij} - {\beta^l \partial_l D_{kij}} 
         - \frac{1}{2} {\gamma}_{mi} \partial_{(k} {B}_{j)}^m
         - \frac{1}{2} {\gamma}_{mj} \partial_{(k} {B}_{i)}^m
				 +  \alpha \partial_k {K}_{ij} = B_k^{\,\,m} D_{mij} + B_j^{\,\,m} D_{kmi} + B_i^{\,\,m} D_{kmj} - \alpha A_k  {K}_{ij}\,.
%
\end{align}
The equations~\eqref{eqn.gamma}--\eqref{eqn.D}, augmented
by the gauge conditions
\eqref{eqn.slicing}--\eqref{g-driver2},
form a non-conservative system which can be written as
\begin{equation}
\label{eqn.pde.mat.preview}
\frac{\partial \U }{\partial t} +
\A_i(\U) \frac{\partial \U}{\partial x_i} 
 = {\bf{S}}(\U),
 \qquad 
 \textnormal{or, equivalently,}
 \qquad 
\frac{\partial \U }{\partial t}  +
\A(\U) \cdot \nabla \U  
= {\bf{S}}(\U),
\end{equation}
where $\U$ is the state vector, composed of 54 dynamical variables, \ie 10 for the lapse, the shift vector and the metric components, 6 for $K_{ij}$,  {3 for the  three vector $Z_{i}$, 1 for the scalar $\Theta$}, 3 for $A_i$, 9 for $B_i^{\,\,j}$, 18 for $D_{ijk}$, 1 for $K_0$ and 3 for $b^i$.
The source term $\bf S(\U)$ contains algebraic terms only. The hyperbolic nature
of \eqref{eqn.pde.mat.preview} has been proved by \cite{DumbserZanottiGaburroPeshkov2023}
after a careful analysis of the matrix $\A(\U)$ via mathematical software packages, and 
the interested reader is
pointed to that work, and especially the Appendix in that work, 
for further details. We just recall here that, as already noticed by \cite{Dumbser2017strongly},
hyperbolicity is helped 
by moving any spatial derivatives of the metric terms
$\alpha$, $\beta^i$ and ${\gamma}_{ij}$ to the right hand side 
via the auxiliary variables \eqref{eq:Auxiliary}.

A few additional comments regarding the  terms entering
Eq.~\eqref{eqn.gamma}--\eqref{eqn.D} are worth giving
\begin{itemize}
\item $\Gamma^i_{jk}={\gamma}^{kl} \left( D_{ijl} + D_{jil} - D_{lij} \right)$ are the Christoffel symbols of the spatial metric $\gamma_{ij}$.
\item $K$ in Eq.~\eqref{eqn.slicing} (and elsewhere) is the trace of the extrinsic curvature,
but it is not a primary variables, \ie it does not belong to the vector $\U$.
\item The function $g(\alpha)$ in Eq.~\eqref{eqn.slicing} is set to $g(\alpha)=2/\alpha$ for the \emph{1+log gauge condition}, and to $g(\alpha)=1$
for the \emph{harmonic gauge condition}.
\item The factor $c$ in Eq.~\eqref{eqn.slicing} (and elsewhere) is always zero, except for the test of the gauge wave discussed in Sect.~\ref{sec:gaugewave}.
\item The factor $s$ in Eq.~\eqref{eqn.B}, either 1 or 0, is used to switch the \emph{gamma--driver} on or off.
\item The quantities $\hat\Gamma^i$ 
in Eq.~\eqref{g-driver2} for the gamma driver are defined as 
$\hat\Gamma^i=\gamma^{jk}\,\Gamma^i_{jk} + 2 \gamma^{ij}Z_j$, but they
are not primary variables. Hence their evolution is obtained from that of the primary variables, \ie
\begin{eqnarray}
\partial_t \Gamma^i_{jk}&=&\gamma^{im}\beta^r\left[\partial_r D_{jmk}+\partial_r D_{kjm}-\partial_r D_{mjk}\right] +\partial_{(j}B_{k)}^{\,\,i}-\alpha\gamma^{im}\left(\partial_j K_{mk}+\partial_k K_{jm}-\partial_mK_{jk}\right)+\nonumber \\
&&+\gamma^{im}\big[ D_{jmn}B_{k}^{\,\,n} + D_{nmk}B_{j}^{\,\,n} + D_{knm}B_{j}^{\,\,n} + D_{njm}B_{k}^{\,\,n} - D_{mjn}B_{k}^{\,\,n} - D_{mnk}B_{j}^{\,\,n} \big] \nonumber \\
&&-\alpha \gamma^{im}\left( A_j K_{mk} + A_k K_{jm} - A_m K_{jk} \right) \nonumber \\
&&+\big[ -2\gamma^{ip}\gamma^{mq}\beta^r D_{rpq} - \gamma^{mr}B_r^{\,\,i} + 2\alpha \gamma^{ip}\gamma^{mq}K_{pq}  \big]\left( D_{jmk} + D_{kjm} - D_{mjk}  \right).
\end{eqnarray}
\item The factor $\eta$ in Eq.~\eqref{g-driver2} is a damping parameter for the gamma--driver.
\item The factor $e$ in Eq.~\eqref{eqn.theta} is the cleaning speed of the Einstein energy constraint.
\item {The factor $\mu$ in Eq.~\eqref{eqn.B} is used to insert a curl-free term that is zero on the continuum level, but that 
favours hyperbolicity of the whole PDE system.}
\end{itemize}

We are now in a position to address the discretization of the system
\eqref{eqn.pde.mat.preview} via FD--WENO schemes, to which Section~\ref{sec:scheme}
is entirely devoted.

\section{A consideration of several finite difference WENO schemes as they pertain to numerical relativity}
\label{sec:scheme}

We had briefly mentioned that finite difference WENO schemes offer the three-fold advantages of very high accuracy, very low memory usage and low computational complexity. To use them well, and also to make decisions about their strengths and weaknesses, it is very important to understand the finite difference WENO philosophy very briefly and why it offers these three-fold benefits. Consider, therefore, the finite volume schemes that are quite common in computational astrophysics. They all start with a volume-averaged representation of the solution vector in each zone. By polling the neighboring zones, a multidimensional high order finite volume scheme will use WENO reconstruction to build all the higher order moments in multiple dimensions. As the order of accuracy increases, the number of moments with cross-terms that have to be built also increases. By contrast, consider a finite difference approach to a simple one-dimensional conservation law, which we write as:
\begin{equation}
{{\partial }_{t}}\mathbf{U}+{{\partial }_{x}}\mathbf{F}\left( \mathbf{U} \right)=0\ \text{     }\Leftrightarrow \text{     }{{\partial }_{t}}\mathbf{U}=-{{\partial }_{x}}\mathbf{F}\left( \mathbf{U} \right)\,.
\label{eq:pde}
\end{equation}
A finite difference scheme will start with a mesh function, $\left\{ {{\mathbf{U}}_{i}} \right\}$ that is provided in the form of point values of the solution vector at each of the zone centers “$i$”. Here we assume a uniform mesh with zones of size $\Delta x$, taking timesteps of size $\Delta t$. Because we are starting with point values, the finite difference scheme requires us to accurately evaluate the gradient ${{\partial }_{x}}\mathbf{F}\left( \mathbf{U} \right)$ at that same zone-centered point. The transcription to multiple dimensions is, therefore, very easy because we simply want all the flux gradients in all three directions to be evaluated at the same point with sufficiently high accuracy. This shows us that finite difference schemes operate on a dimension-by-dimension basis. This is the reason why it was acceptable to only show eqn.~\eqref{eq:pde} in one dimension; because additional dimensions only contribute additively. Unlike high order finite volume approaches where a 3D finite volume scheme can be substantially more than three times costlier than a one-dimensional scheme, a 3D finite difference scheme will only be three times costlier than a one-dimensional one. We also see that the memory usage is very favorable for finite difference schemes. This is because we do not need to reconstruct all the higher order moments in all dimensions and store them in computer memory, as we would do for a finite volume scheme. A finite difference scheme only needs to retain the point values of the solution vector at the zone centers. Any higher moments that are needed will only be needed in a dimension-by-dimension fashion and can be discarded from computer memory once that dimension has been processed. By the same token, it is also possible to show that the computational complexity of a finite difference WENO scheme does not increase by much as one progresses to higher orders of spatial accuracy; please see Table V in subsection 5.4 of~\cite{balsara2024b}.

Now that the advantages of finite difference WENO have been documented, we provide very brief descriptions of these schemes here so that the reader has sufficient background, concentrated all in one place, with which to understand these schemes as they will be used for the FO-Z4 system from Sect.~\ref{sec:Z4}.

\subsection{FD-WENO Schemes for Conservation Laws} \label{subsec:fdweno}
Although our immediate goal is not to understand conservation laws, it helps to quickly document the WENO philosophy that led to FD-WENO schemes for conservation laws. (Without an understanding of conservation laws it is impossible to understand systems in non-conservative form.) The primary goal of FD-WENO for conservation laws is to try and write eqn.~\eqref{eq:pde} for any zone “$i$” on a one-dimensional mesh as
\begin{equation}
    {{\partial }_{t}}\mathbf{U}_{i}^{{}}=-\frac{{{{\mathbf{\hat{F}}}}_{i+1/2}}-{{{\mathbf{\hat{F}}}}_{i-1/2}}}{\Delta x}.
    \label{eq:fda}
\end{equation}
Here ${\mathbf{\hat{F}}}_{i+1/2}$ and ${{\mathbf{\hat{F}}}_{i-1/2}}$ are the reconstructed numerical fluxes at the zone boundaries ${{x}_{i}}+{\Delta x}/{2}\;$ and ${{x}_{i}}-{\Delta x}/{2}\;$. For this to work out as a viable high order scheme, we need the finite difference approximation (FDA) given by ${\left( {{{\mathbf{\hat{F}}}}_{i+1/2}}-{{{\mathbf{\hat{F}}}}_{i-1/2}} \right)}/{\Delta x}\;$ to approximate ${{\left( {{\partial }_{x}}\mathbf{F} \right)}_{i}}$ with very high order of accuracy. Therefore, the problem can be stated as follows. We have to start with the so--called \emph{physical fluxes} evaluated from the point values on the mesh, i.e. $\left\{ \mathbf{F}\left( {{\mathbf{U}}_{i}} \right) \right\}$, and obtain \emph{numerical fluxes} such that ${\left( {{{\mathbf{\hat{F}}}}_{i+1/2}}-{{{\mathbf{\hat{F}}}}_{i-1/2}} \right)}/{\Delta x}\;={{\left( {{\partial }_{x}}\mathbf{F} \right)}_{i}}+O\left( \Delta {{x}^{k}} \right)$ for a spatially $k^{th}$ order accurate scheme.

The denouement of the previously-posed problem comes from the fundamental theorem of integral calculus, as was first realized by~\cite{shu1989efficient}. Consider a function $\mathbf{f}\left( x \right)$ that is defined in terms of another functional $\mathcal{F}\left( \xi  \right)$ as follows:-
\begin{equation}
\mathbf{f}\left( x \right)\equiv \frac{1}{\Delta x}\int\limits_{x-\Delta x/2}^{x+\Delta x/2}{\mathcal{F}\left( \xi  \right)d\xi }. 
\label{eq:f_int}
\end{equation}
Then the fundamental theorem of integral calculus tells us that:-
\begin{equation}
\left( {{\partial }_{x}}\mathbf{f} \right){{|}_{x={{x}_{i}}}}=\frac{1}{\Delta x}\left[ \mathcal{F}\left( {{x}_{i}}+\frac{\Delta x}{2} \right)-\mathcal{F}\left( {{x}_{i}}-\frac{\Delta x}{2} \right) \right].
\label{eq:f_diff}
\end{equation}
We see, therefore, that the left hand side of eqn.~\eqref{eq:f_diff} is exactly of the same form as the right hand side of the second equation in eqn.~\eqref{eq:pde} whereas the right hand side of eqn.~\eqref{eq:f_diff} is exactly of the same form as the right hand side of eqn.~\eqref{eq:fda}. We have, therefore, found a way of connecting the gradient in eqn.~\eqref{eq:pde} to the FDA in eqn.~\eqref{eq:fda}. Consequently, we seek a function $\mathcal{F}\left( \xi  \right)$ whose value $\mathcal{F}\left( {{x}_{i+1/2}} \right)$ at the zone boundary approximates ${{\mathbf{\hat{F}}}_{i+1/2}}$ with a sufficiently high order of accuracy. Now, the very familiar problem of “reconstruction by primitive” that was introduced in~\cite{colella1984piecewise} (and extended to ENO schemes with accreditation in~\cite{harten1986some}) comes to the rescue. It says that the reconstruction polynomial for the zone boundary “$i+1/2$” that we seek should be such that it matches the condition
\begin{equation}
    \int\limits_{\xi ={{x}_{i-j}}-\Delta x/2}^{\xi ={{x}_{i-j}}+\Delta x/2}{\mathcal{F}\left( \xi  \right)d\xi }=\Delta x \ \mathbf{F}\left( {{\mathbf{U}}_{i-j}} \right)
    \label{eq:f_int_new}
\end{equation}							
for some zones “$j$” that are adjacent to the zone boundary “$i+1/2$” under consideration. If the polynomial is chosen to be of sufficient order of accuracy, the task that was identified in the previous paragraph is accomplished. We can make the polynomial sufficiently accurate by making the stencil operations alluded to in eqn.~\eqref{eq:f_int_new} wide enough. We choose $\mathcal{F}\left( \xi  \right)$ to be a consistent polynomial that is of degree “$k-1$” in order to obtain a $k^{th}$ order accurate scheme. This polynomial will be pinned down more precisely in subsequent paragraphs.

\begin{figure}[!ht]
	\begin{center}
		\includegraphics[width=0.9\textwidth,clip=]{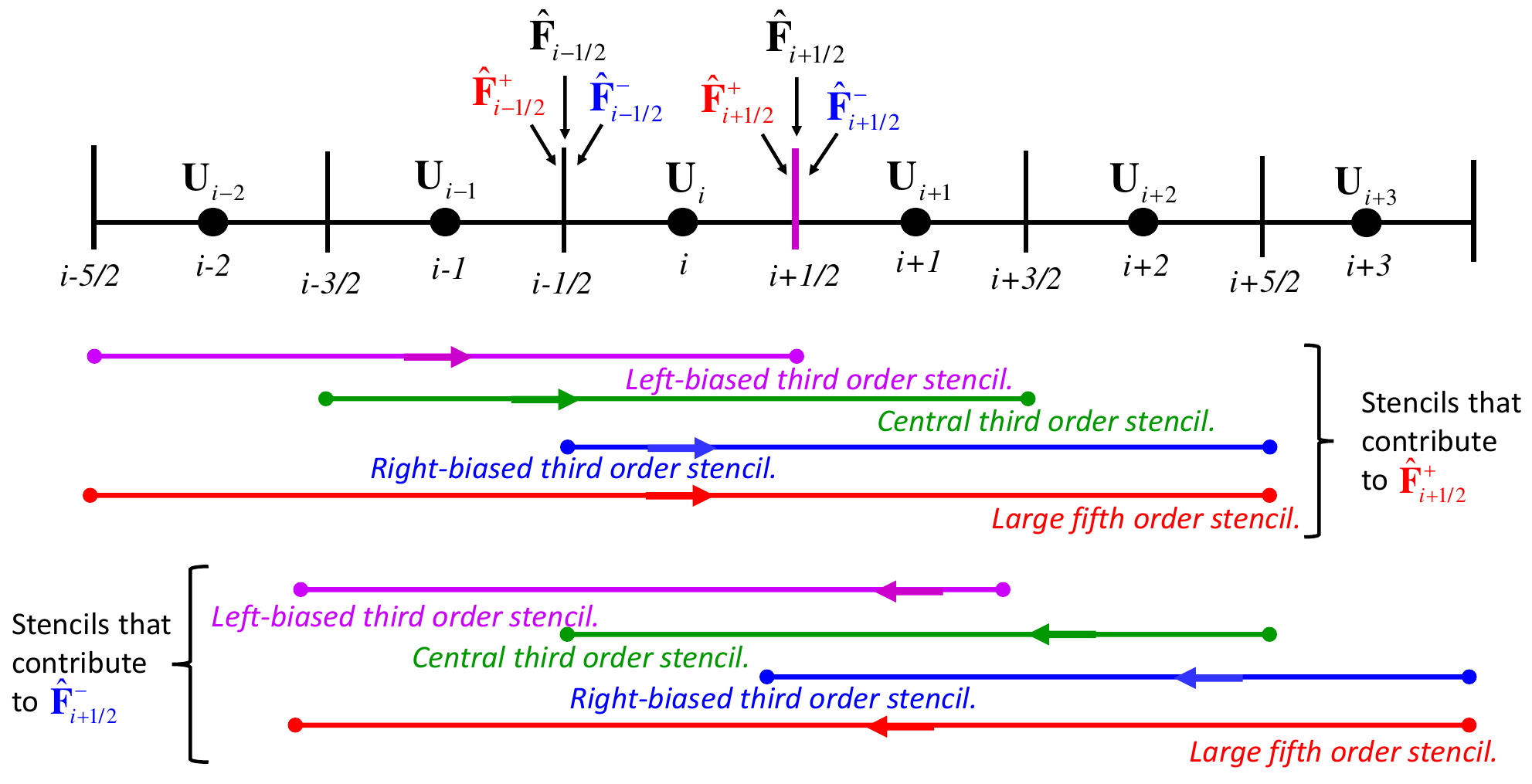}
		\caption{The schematic structure of a fifth order FD-WENO scheme in conservation form. Please focus on the purple zone boundary at $i+1/2$. The numerical flux at that zone boundary is made up of two split flux contributions $\mathbf{F}^+$ and $\mathbf{F}^-$. The stencils contributing to the reconstruction of the right-going flux $\mathbf{F}^+$ at the purple zone boundary are shown. The stencils contributing to the reconstruction of the left-going flux $\mathbf{F}^-$ at the purple zone boundary are also shown. 
.}
		\label{fig:stencil_1}
	\end{center}
\end{figure}

While the process of identifying the polynomial associated with reconstructing the fluxes was described in the previous paragraph, more is needed in order to obtain a successful scheme. Such a scheme needs to be upwinded and the reconstruction process that was briefly described in eqn.~\eqref{eq:f_int_new} needs to be based on non-linear hybridization. To appreciate the concept of upwinding, realize that any flux “$\mathbf{F}\left( \mathbf{U} \right)$” can be split as
\begin{equation}
    \mathbf{F}\left( \mathbf{U} \right)=\frac{1}{2}\left( \mathbf{F}\left( \mathbf{U} \right)+S\text{ }\mathbf{U} \right)+\frac{1}{2}\left( \mathbf{F}\left( \mathbf{U} \right)-S\text{ }\mathbf{U} \right)={{\mathbf{F}}^{\mathbf{+}}}\left( \mathbf{U} \right)+{{\mathbf{F}}^{-}}\left( \mathbf{U} \right)
    \label{eq:f_split}
\end{equation}
with the definitions
\begin{equation}
    {{\mathbf{F}}^{\mathbf{+}}}\left( \mathbf{U} \right)\equiv \frac{1}{2}\left( \mathbf{F}\left( \mathbf{U} \right)+S\text{ }\mathbf{U} \right)
    \quad \text{  and  } \quad
    {{\mathbf{F}}^{-}}\left( \mathbf{U} \right)\equiv \frac{1}{2}\left( \mathbf{F}\left( \mathbf{U} \right)-S\text{ }\mathbf{U} \right)
    \label{f_split_defs}
\end{equation}
Here the wave speed “$S$” is the maximum of the absolute values of all the signal speeds from the flux. This choice of “$S$” ensures that the Jacobian  ${\partial {{\mathbf{F}}^{\mathbf{+}}}\left( \mathbf{U} \right)}/{\partial \mathbf{U}}\;$ will always have non-negative eigenvalues and the Jacobian  ${\partial {{\mathbf{F}}^{-}}\left( \mathbf{U} \right)}/{\partial \mathbf{U}}\;$ will always have non-positive eigenvalues. (As a shorthand, we will often use ${{\mathbf{F}}^{\mathbf{+}}}\left( \mathbf{U} \right)\to {{\mathbf{F}}^{\mathbf{+}}}$ and ${{\mathbf{F}}^{-}}\left( \mathbf{U} \right)\to {{\mathbf{F}}^{-}}$.) The flux splitting in the above two equations is known as the LLF 
(locally Lax–Friedrichs)
flux splitting. Now please focus on Fig.~\ref{fig:stencil_1}. We see a mesh with zones, and zone boundaries, as well as a mesh function that is specified at zone centers. At each zone center, we can use the point value of the mesh function to also evaluate the zone-centered point value for the flux. These zone-centered flux values are to be used to obtain the numerical flux ${{\mathbf{\hat{F}}}_{i+1/2}}$ at the purple zone boundary in Fig.~\ref{fig:stencil_1}. But realize that the numerical flux ${{\mathbf{\hat{F}}}_{i+1/2}}$ is constituted by the sum of $\mathbf{\hat{F}}_{i+1/2}^{+}$ (which carries the right-going flux contributions) and $\mathbf{\hat{F}}_{i+1/2}^{-}$ (which carries the left-going flux contributions). If $\mathbf{\hat{F}}_{i+1/2}^{+}$ and $\mathbf{\hat{F}}_{i+1/2}^{-}$ are obtained via a high order, one-dimensional finite-volume-style reconstruction then the entire scheme shown in eqn.~\eqref{eq:fda} will be high order accurate in space. This paragraph has, therefore, shown how the flux splitting is to be used in FD-WENO schemes for conservation laws. Notice that in light of the discussion surrounding eqns.~\eqref{eq:f_int},~\eqref{eq:f_diff} and~\eqref{eq:f_int_new}, the reconstruction has to be a finite-volume style reconstruction. In such a reconstruction, the zone averages are known, as shown in the right hand side of eqn.~\eqref{eq:f_int_new}, and the reconstruction polynomial has to be constructed so as to match the zone averages. 

The previous paragraph addressed the issue of flux splitting. But it did not address the twin issues of upwinding and non-linear hybridization. We know that these two concepts are essential for obtaining a stable scheme. Upwinding requires that any stencils that contribute to the right-going flux $\mathbf{\hat{F}}_{i+1/2}^{+}$ must include the zone that is left of the zone boundary “$i+1/2$”. As a result, those stencils must include zone “$i$” in Fig.~\ref{fig:stencil_1}. We see that this requirement is met by the stencils shown in Fig.~\ref{fig:stencil_1}. Upwinding also requires that any stencils that contribute to the left-going flux $\mathbf{\hat{F}}_{i+1/2}^{-}$ must include the zone that is right of the zone boundary “$i+1/2$”. As a result, those stencils must include zone “$i+1$” in Fig.~\ref{fig:stencil_1}; and we see that this requirement is met by the stencils shown in Fig.~\ref{fig:stencil_1}.

Now that flux splitting and upwinding have been clarified, let us address the issue of non-linear hybridization. Over the years, there have been many interpretations of what it means to non-linearly hybridize the available stencils in a WENO scheme \citep{jiang1996efficient,balsara2000monotonicity,Levy2000,henrick2006simulations,borges2008improved,zhu2016new,balsara2016efficient,cravero2016accuracy,zhu2018new}. Fig.~\ref{fig:stencil_1} shows the WENO--AO approach of~\cite{balsara2016efficient}; where “AO” stands for adaptive order. Fig.~\ref{fig:stencil_1} focuses on a fifth order WENO-AO reconstruction, but the methods have been extended up to $11^{th}$ order. The three smaller left-biased, zone-centered and right-biased stencils cover the zones of interest; i.e., zone “$i$” and zone “$i+1$” respectively. Their non-linear hybridization would give a stable stencil that is third order accurate. For example, consider the stencils with the right arrow in Fig.~\ref{fig:stencil_1} that contribute to $\mathbf{\hat{F}}_{i+1/2}^{+}$. The smaller stencils include the left-biased stencil which includes zones $\left\{ i-2,i-1,i \right\}$, the centered stencil which includes zones $\left\{ i-1,i,i+1 \right\}$ and the right-biased stencil which includes zones $\left\{ i,i+1,i+2 \right\}$. Together, these three smaller stencils, if they are non-linearly hybridized amongst themselves, would yield a stable, spatially third order accurate reconstruction over the zone “$i$” which must be included in the upwinding of $\mathbf{\hat{F}}_{i+1/2}^{+}$. To get a fifth order accurate reconstruction over the zone “$i$”, one must make a non-linear hybridization of these smaller stencils with the larger stencil in Fig.~\ref{fig:stencil_1} which includes zones $\left\{ i-2,i-1,i,i+1,i+2 \right\}$. The smaller stencils guarantee stability, when stability becomes an issue. The larger stencil provides higher (fifth) order accuracy, when the solution is smooth enough to justify the higher accuracy. The stencils for obtaining a fifth order in space approximation for the right-going flux are shown by the stencils with a right-pointing arrow in Fig.~\ref{fig:stencil_1}. Fig.~\ref{fig:stencil_1} also displays the smaller and larger stencils used for the spatially fifth order accurate, upwinded reconstruction of the left-going flux $\mathbf{\hat{F}}_{i+1/2}^{-}$. The stencils for obtaining a fifth order in space approximation for the left-going flux are shown by the stencils with a left-pointing arrow in Fig.~\ref{fig:stencil_1}. We can see that all those stencils in Fig.~\ref{fig:stencil_1} cover the zone “$i+1$”.  The final FD-WENO scheme for conservation laws is then given by:-
\begin{equation}
    {{\partial }_{t}}\mathbf{U}_{i}^{{}}=-\frac{\left( \mathbf{\hat{F}}_{i+1/2}^{+}+\mathbf{\hat{F}}_{i+1/2}^{-} \right)-\left( \mathbf{\hat{F}}_{i-1/2}^{+}+\mathbf{\hat{F}}_{i-1/2}^{-} \right)}{\Delta x}
    \label{eq:fdweno_cons}
\end{equation}
The reader can obtain further detail associated with WENO-AO reconstruction shown in Fig.~\ref{fig:stencil_1} from Sections 2 and 3 of~\cite{balsara2016efficient}. We should also mention that the reconstruction should be done in characteristic variables in order to obtain the best quality solution. This completes our very brief description of FD-WENO for conservation laws. This FD-WENO scheme for conservation laws will prove indispensable in the next section for understanding FD-WENO schemes for systems in non-conservation form like the FO-Z4 formulation of the Einstein equations.

\subsection{FD-WENO Schemes for Hyperbolic Systems with Non-Conservative Products} \label{subsec:fdweno_noncons}

For a very long time, FD-WENO schemes were only available for conservation laws. However, first order formulations of the Einstein equations, as are nowadays presented~\citep{Brown2012,Dumbser2017strongly,DumbserZanottiGaburroPeshkov2023}, are cast in a non conservative form like \eqref{eqn.pde.mat.preview}.
Here, to simplify the description of the numerical scheme, we disregard the source terms on the right hand side and we consider
\begin{align}
    {{\partial }_{t}}\mathbf{U}+\mathbf{A}\left( \mathbf{U} \right){{\partial }_{x}}\mathbf{U}=0\ \text{     }\Leftrightarrow \text{     }{{\partial }_{t}}\mathbf{U}=-\mathbf{A}\left( \mathbf{U} \right){{\partial }_{x}}\mathbf{U}\,.
    \label{eq:pde_2}
\end{align}
Such situations are best handled in fluctuation form. The problem was that, until the advent of~\cite{balsara2023FDWENO}, fluctuation form-based FD-WENO schemes were not available for eqn.~\eqref{eq:pde_2}. Eqn. (16) in Section 4 of~\cite{balsara2023FDWENO} yields a scheme that can be applied to eqn.~\eqref{eq:pde_2}. In Section 5 of~\cite{balsara2023FDWENO} we do provide a derivation of the scheme. However, that derivation is somewhat harder to understand, so we present a much simpler derivation here. We give 
this derivation in the limit of a linear flux, $\mathbf{F}=\mathbf{AU}$, with “$\mathbf{A}$” is a constant matrix, but the final form that we will obtain will be in fluctuation form so that it can be extended to include any Riemann solver that accommodates the non-linearities at the zone boundaries. As in~\cite{balsara2023FDWENO}, we work in the limit of the LLF flux.

\begin{figure}[!ht]
	\begin{center}
		\includegraphics[width=0.9\textwidth,clip=]{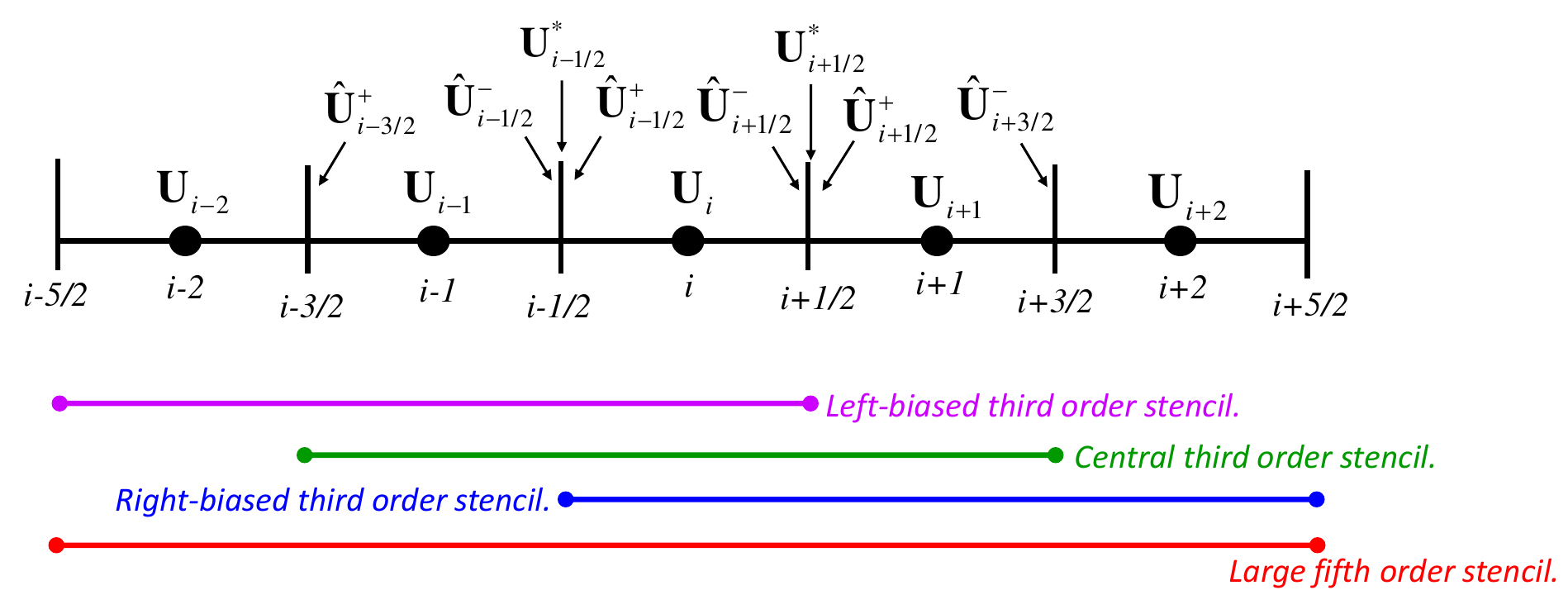}
		\caption{Panel shows part of the mesh around zone “i”. The mesh functions are collocated at the zone centers, as shown by the thick dots. The zone boundaries are shown by the vertical lines. The figure also shows the stencils associated with the zone “$i$” for the fifth order WENO--AO reconstruction/interpolation. We have three smaller third order stencils and a large fifth order stencil. The reconstructed/interpolated variables at the zone boundaries are shown with a caret. The variables with a superscript star are resolved states obtained by the pointwise application of a simple HLL or LLF Riemann solver at the zone boundaries.}
		\label{fig:stencil_2}
	\end{center}
\end{figure}

Consider Fig.~\ref{fig:stencil_2} where the solution vector has itself been reconstructed. Because we are only reconstructing the solution vector, only one reconstruction is needed within each zone. (This is different from what is shown in Fig.~\ref{fig:stencil_1}, which uses two reconstructions for each of the right-going and left-going fluxes per zone.) As a result, at each zone boundary, say the zone boundary “$i+1/2$” in Fig.~\ref{fig:stencil_2}, we have the left state $\mathbf{\hat{U}}_{i+1/2}^{-}$ and the right state $\mathbf{\hat{U}}_{i+1/2}^{+}$. Please try to understand how $\mathbf{\hat{F}}_{i+1/2}^{-}$ in Fig.~\ref{fig:stencil_1} relates to $\mathbf{\hat{U}}_{i+1/2}^{+}$ in Fig.~\ref{fig:stencil_2}. Likewise, please try to understand how $\mathbf{\hat{F}}_{i+1/2}^{+}$ in Fig.~\ref{fig:stencil_1} relates to $\mathbf{\hat{U}}_{i+1/2}^{-}$ in Fig.~\ref{fig:stencil_2}, With “$\mathbf{A}$” held constant, we first write
\begin{equation}
\begin{aligned}
  \mathbf{\hat{F}}_{i+1/2}^{-} &= \frac{1}{2}\left( \mathbf{A}-S\mathbf{I} \right)\mathbf{\hat{U}}_{i+1/2}^{+}
  \quad   ;  \quad
  \mathbf{\hat{F}}_{i+1/2}^{+}=\frac{1}{2}\left( \mathbf{A}+S\mathbf{I} \right)\mathbf{\hat{U}}_{i+1/2}^{-}
  \quad  ; \quad \\ 
 \mathbf{\hat{F}}_{i-1/2}^{-} &= \frac{1}{2}\left( \mathbf{A}-S\mathbf{I} \right)\mathbf{\hat{U}}_{i-1/2}^{+}
 \quad   ;  \quad
 \mathbf{\hat{F}}_{i-1/2}^{+}=\frac{1}{2}\left( \mathbf{A}+S\mathbf{I} \right)\mathbf{\hat{U}}_{i-1/2}^{-}
\end{aligned} \label{FDWENO_NONCONS_fluxes}
\end{equation}
It is then easy to show that:-
\begin{equation}
\begin{aligned}
  -\left( \mathbf{\hat{F}}_{i+1/2}^{-}+\mathbf{\hat{F}}_{i+1/2}^{+} \right)
  &=
  -\frac{1}{2}\left( \mathbf{A}-S\mathbf{I} \right)\mathbf{\hat{U}}_{i+1/2}^{+}-\frac{1}{2}\left( \mathbf{A}+S\mathbf{I} \right)\mathbf{\hat{U}}_{i+1/2}^{-}
  \\ 
  &= -\frac{1}{2}\left( \mathbf{A}-S\mathbf{I} \right)\mathbf{\hat{U}}_{i+1/2}^{+}+\frac{1}{2}\left( \mathbf{A}-S\mathbf{I} \right)\mathbf{\hat{U}}_{i+1/2}^{-}-\mathbf{A\hat{U}}_{i+1/2}^{-}
  \\ 
  &= -\frac{1}{2}\left( \mathbf{A}-S\mathbf{I} \right)\left( \mathbf{\hat{U}}_{i+1/2}^{+}-\mathbf{\hat{U}}_{i+1/2}^{-} \right)-\mathbf{A\hat{U}}_{i+1/2}^{-}
  \\ 
  &= -\mathbf{D}_{i+1/2}^{-}-\mathbf{A\hat{U}}_{i+1/2}^{-} 
\end{aligned} \label{FDWENO_NONCONS_flux_diff}
\end{equation}
It is also easy to show via a similar derivation that:-
\begin{equation}
\begin{aligned}
  \left( \mathbf{\hat{F}}_{i-1/2}^{-}+\mathbf{\hat{F}}_{i-1/2}^{+} \right)
  &=
  \frac{1}{2}\left( \mathbf{A}-S\mathbf{I} \right)\mathbf{\hat{U}}_{i-1/2}^{+}+\frac{1}{2}\left( \mathbf{A}+S\mathbf{I} \right)\mathbf{\hat{U}}_{i-1/2}^{-} \\ 
 &= -\frac{1}{2}\left( \mathbf{A}+S\mathbf{I} \right)\mathbf{\hat{U}}_{i-1/2}^{+}+\frac{1}{2}\left( \mathbf{A}+S\mathbf{I} \right)\mathbf{\hat{U}}_{i-1/2}^{-}+\mathbf{A\hat{U}}_{i-1/2}^{+} \\ 
 &= -\frac{1}{2}\left( \mathbf{A}+S\mathbf{I} \right)\left( \mathbf{\hat{U}}_{i-1/2}^{+}-\mathbf{\hat{U}}_{i-1/2}^{-} \right)+\mathbf{A\hat{U}}_{i-1/2}^{+} \\ 
 &= -\mathbf{D}_{i-1/2}^{+}+\mathbf{A\hat{U}}_{i-1/2}^{+}
\end{aligned} \label{FDWENO_NONCONS_flux_diff_minus}
\end{equation}
In the above two equations, and for a linear system, we can define the left-going and right-going fluctuations as
$\mathbf{D}_{i+1/2}^{-}$ \ $\equiv $ \ $\left[ \left( \mathbf{A}-S\mathbf{I} \right)/2 \right]$ $\left( \mathbf{\hat{U}}_{i+1/2}^{+}-\mathbf{\hat{U}}_{i+1/2}^{-} \right)$ and $\mathbf{D}_{i-1/2}^{+}\equiv \left[ \left( \mathbf{A}+S\mathbf{I} \right)/2 \right]\left( \mathbf{\hat{U}}_{i-1/2}^{+}-\mathbf{\hat{U}}_{i-1/2}^{-} \right)$ respectively. We will soon point to references where their non-linear extensions can be obtained. 
Putting the above two equations together, we can now show that
\begin{equation}
-\frac{\left( \mathbf{\hat{F}}_{i+1/2}^{+}+\mathbf{\hat{F}}_{i+1/2}^{-} \right)-\left( \mathbf{\hat{F}}_{i-1/2}^{+}+\mathbf{\hat{F}}_{i-1/2}^{-} \right)}{\Delta x}=-\frac{1}{\Delta x}\left( \mathbf{D}_{i+1/2}^{-}+\mathbf{D}_{i-1/2}^{+} \right)-\mathbf{A}\frac{\left( \mathbf{\hat{U}}_{i+1/2}^{-}-\mathbf{\hat{U}}_{i-1/2}^{+} \right)}{\Delta x}.\label{eq:FDWENO_NONCONS_diff}
\end{equation}
While the above derivation was done in a simpler context, it is now easy to identify the full update equation:-
\begin{equation}
{{\partial }_{t}}{{\mathbf{U}}_{i}}=-\frac{1}{\Delta x}\left( \mathbf{D}_{HLLI}^{-}\left( \mathbf{\hat{U}}_{i+1/2}^{-},\mathbf{\hat{U}}_{i+1/2}^{+} \right)+\mathbf{D}_{HLLI}^{+}\left( \mathbf{\hat{U}}_{i-1/2}^{-},\mathbf{\hat{U}}_{i-1/2}^{+} \right) \right)-\mathbf{A}\left( {{\mathbf{U}}_{i}} \right)\frac{\left( \mathbf{\hat{U}}_{i+1/2}^{-}-\mathbf{\hat{U}}_{i-1/2}^{+} \right)}{\Delta x}
\label{eq:FDWENO_NONCONS_dudt}
\end{equation}
The fluctuation $\mathbf{D}_{HLLI}^{-}\left( \mathbf{\hat{U}}_{i+1/2}^{-},\mathbf{\hat{U}}_{i+1/2}^{+} \right)$ captures the contribution from the left-going waves at zone boundary “$i+1/2$”. The fluctuation $\mathbf{D}_{HLLI}^{+}\left( \mathbf{\hat{U}}_{i-1/2}^{-},\mathbf{\hat{U}}_{i-1/2}^{+} \right)$ captures the contribution from the right-going waves at zone boundary “$i-1/2$”. The subscript “\textit{HLLI}” indicates that we are using the \textit{HLLI} Riemann solver from~\cite{dumbser2016new}, which can indeed handle non-linearities in PDE systems of the form shown in eqn.~\eqref{eq:pde_2}. Appendix C of ~\cite{dumbser2016new}  provides further details on evaluating fluctuations for PDEs with non-linearities.
	
In the previous sub-section we have mentioned that it is advantageous to carry out the reconstruction in the characteristic variables. In a \emph{tour de force}, the 54 eigenvectors of FO-Z4 system for the full Einstein--Euler system have been derived in~\cite{DumbserZanottiGaburroPeshkov2023}. However, it would be prohibitive to project the entire system into that eigenspace for carrying out a reconstruction of the solution vector. Fortunately, the eigenspace splits into a 5 component subspace for the hydrodynamic system and an eigenspace for the rest of the system. 
{
We recall that, although no physical shocks can form in the metric, 
since   all the spacetime fields are linearly degenerate, gauge shocks can still form,
as a result of the gauge conditions chosen~\citep{Alcubierre2003,Jimenez2022,Baumgarte2023a,Baumgarte2023b}. 
In this respect, it could be quite beneficial
to treat the reconstruction of the variables in the GR sector using characteristic projection. 
However, due to the complexity of the equations, we have not followed this approach here and we postpone the analysis of such a  specialized
aspect to a future investigation. 
}

It is also useful to observe that eqn.~\eqref{eq:FDWENO_NONCONS_dudt} is not in exact flux conservative form. An exact flux conservative form is essential (because of the Lax-Wendroff theorem) for accurately predicting shock locations when the system is conservative, as is the case for hydrodynamics. The modern trend is to seek out schemes that revert to a conservation form when such is present. However, eqn.~\eqref{eq:FDWENO_NONCONS_dudt} does derive from eqn.~\eqref{eq:fdweno_cons} which is in conservation form, so for most practical cases it does a rather good job in predicting shock locations.  {Besides, since the solutions of the Einstein equations are continuous in the metric terms, except of course when a singularity forms or when gauge shocks are present}, 
the lack of a conservation form is not a major impediment. As a result, eqn.~\eqref{eq:FDWENO_NONCONS_dudt} has an important place in the solution of the Einstein equations. This completes our description of FD-WENO schemes as they are extended to hyperbolic PDEs, like the FO-Z4 system, that are not in conservation form.

\subsection{AFD-WENO Schemes for Conservation Laws} \label{subsec:afdweno_cons}

The previous Sub-section has shown us how to obtain a scheme that is fully in non-conservative form. It is very easy to implement and very suitable for the Einstein equations in vacuum, i.e. when evolving the hydrodynamics is not a priority. However, it is often the case that one has to simultaneously evolve the equations for relativistic hydrodynamics in addition to the Einstein equations. In that case, we will need a scheme that can handle non-conservative products and is, nevertheless, versatile enough to revert to a conservation form when such a conservation form is present in the problem. The scheme in the previous Sub-section does not meet this requirement. We have also seen that a study of conservation laws is a first step towards deriving methods that can handle non-conservative products. Therefore, in this Sub-section we will present an alternative finite difference WENO (AFD-WENO) scheme for conservation laws. This will be an essential first step towards deriving schemes that can seamlessly accommodate both a conservation law as well as non-conservative products -- that will be the task of the next section.

This and the next Sub-section rely on interpolation, which is different from the reconstruction that was used in the previous two Sub-sections. It is, therefore, worthwhile to make a distinction between reconstruction and interpolation. Reconstruction is used in all finite volume astrophysical codes and also in the previous FD-WENO schemes where the fluxes or states are reconstructed. It consists of starting with the zone averages in a given stencil and obtaining therefrom the high degree polynomial whose integration over each of the zones of the stencil matches the original zone averages. Interpolation is used less often in the numerical solution of conservation laws, however it is the approach that will be needed in this and the next Sub-section. It consists of starting with the point values at each of the zone centers of a stencil and obtaining therefrom the high degree polynomial that matches those point values. Therefore, the two words, reconstruction and interpolation, carry different connotations. When applied to the same stencil, reconstruction and interpolation produce polynomials with the same degree. However, the underlying polynomial coefficients that are produced by invoking reconstruction or interpolation on a given stencil can indeed be very different. Standard WENO concepts like linear weights, smoothness indicators, normalized non-linear weights etc. are often the same for reconstruction and interpolation; so there is indeed a beneficial transference of knowledge between them. WENO reconstruction, especially as it relates to FD-WENO schemes, has been amply documented in the literature starting from~\cite{jiang1996efficient},~\cite{balsara2000monotonicity} and continuing through~\cite{balsara2016efficient}, where it was presented in its most polished form. The WENO interpolation, as it relates to AFD-WENO, has also been recently documented in a very polished form in Sections 3 and 4 of~\cite{balsara2024a}.

\begin{figure}[!ht]
	\begin{center}
		\includegraphics[width=0.9\textwidth,clip=]{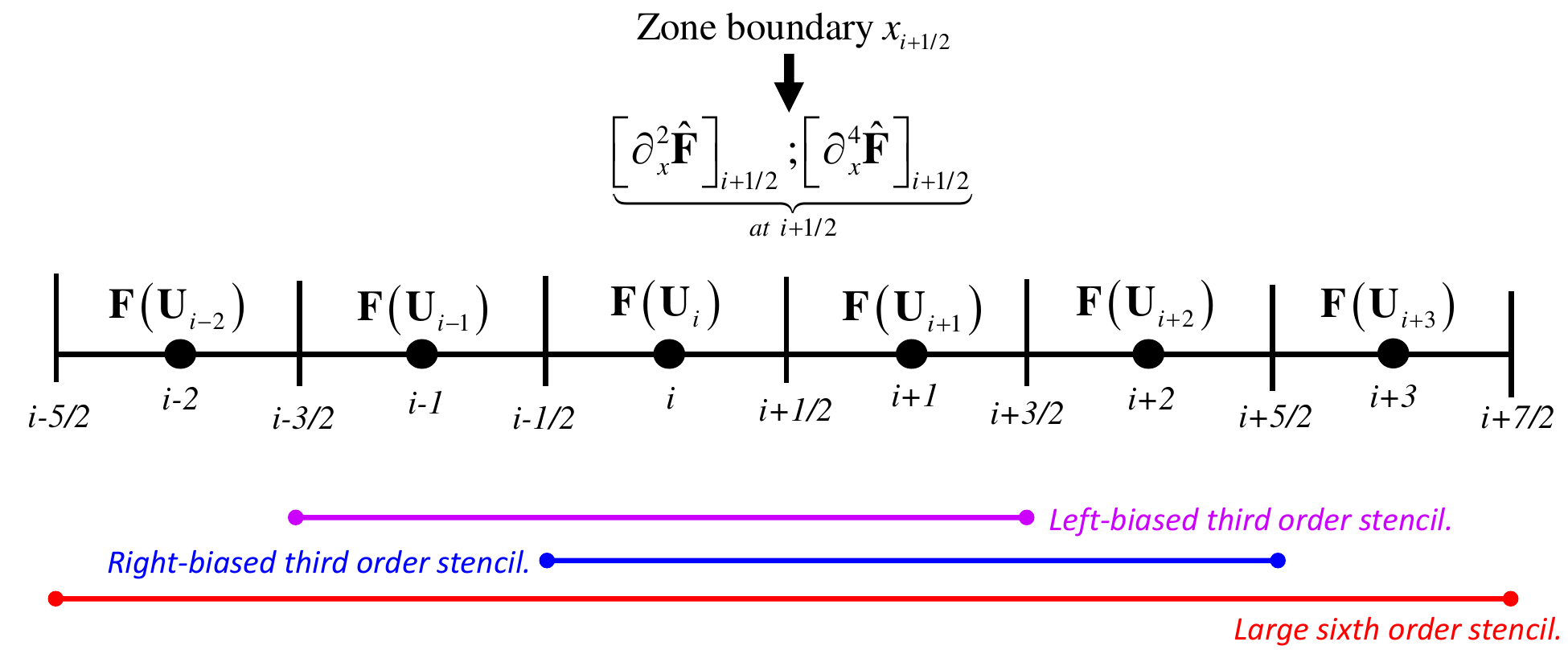}
		\caption{Panel shows part of the mesh around zone boundary “$i+1/2$”. The fluxes are evaluated pointwise at the zone centers, as shown by the thick dots. The zone boundaries are shown by the vertical lines. The figure also shows the stencils associated with the zone boundary “$i+1/2$” for the third and fifth order AFD-WENO schemes. We have two smaller third order stencils and a large sixth order stencil. For a third order AFD-WENO scheme, the two smaller stencils can be non-linearly hybridized. In that case, the second derivatives of the flux can be obtained at the zone boundary when the smoothness in the solution warrants it. For fifth order AFD-WENO, the two smaller stencils can be non-linearly hybridized along with the larger stencil. In that case, the second and fourth derivatives of the flux can be obtained at the zone boundary when the smoothness in the solution warrants it. The process described here can be done for Adaptive Order and Multiresolution WENO interpolation.}
		\label{fig:stencil_3}
	\end{center}
\end{figure}

Let us say that we have a high order pointwise WENO interpolation strategy that is applied to the solution vector of the mesh. At each zone boundary, say “$i+1/2$” we will then have high order interpolants $\mathbf{\hat{U}}_{i+1/2}^{-}$ and $\mathbf{\hat{U}}_{i+1/2}^{+}$. Fig.~\ref{fig:stencil_2} shows us that the same stencils can be used for interpolation and reconstruction. Say also that we invoke the Riemann solver (in pointwise fashion) at each zone boundary to obtain ${{\mathbf{F}}^{*}}\left( \mathbf{\hat{U}}_{i+1/2}^{-},\mathbf{\hat{U}}_{i+1/2}^{+} \right)$. The flux from the previous sentence will indeed be a suitably high order flux at the zone boundary “$i+1/2$”. However, say that we naively assert a discrete in space but continuous in time update in the zone “$i$” of the form
\begin{equation}
    {{\partial }_{t}}{{\mathbf{U}}_{i}}=-\frac{1}{\Delta x}\left( {{\mathbf{F}}^{*}}\left( \mathbf{\hat{U}}_{i+1/2}^{-},\mathbf{\hat{U}}_{i+1/2}^{+} \right)-{{\mathbf{F}}^{*}}\left( \mathbf{\hat{U}}_{i-1/2}^{-},\mathbf{\hat{U}}_{i-1/2}^{+} \right) \right).
    \label{eq:afdcons_1stdudt}
\end{equation}
We would find, to our chagrin, that eqn.~\eqref{eq:afdcons_1stdudt} only has second order spatial accuracy. Even if the solution had been very smooth, and even if the interpolation had been carried out with very high order accuracy, the above equation would only result in a spatially second order scheme. Understanding why this is so will indeed show us the way out of this dilemma. We illustrate this for the simplest case where we assume that we are trying to obtain a third order accurate scheme. Because we have assumed a very smooth solution and a very smooth flux, we can make the Taylor series expansion:-
\begin{equation}
    \mathbf{F}\left( x \right)={{f}_{0}}+x{{\left( {{\partial }_{x}}f \right)}_{0}}+\frac{{{x}^{2}}}{2}{{\left( \partial _{x}^{2}f \right)}_{0}}+\frac{{{x}^{3}}}{6}{{\left( \partial _{x}^{3}f \right)}_{0}}+...
    \label{eq:afdcons_taylor}
\end{equation}
All the terms of the Taylor series, ${{f}_{0}}$, ${{\left( {{\partial }_{x}}f \right)}_{0}}$, ${{\left( \partial _{x}^{2}f \right)}_{0}}$, ${{\left( \partial _{x}^{3}f \right)}_{0}}$are all evaluated at $x=0$. We can evaluate eqn.~\eqref{eq:afdcons_taylor} and its higher derivatives at $x=\pm \Delta x/2$ to get
\begin{equation}
\begin{aligned}
  & {{\left. \mathbf{F}\left( x \right) \right|}_{x=\pm \Delta x/2}}={{f}_{0}}\pm \frac{\Delta x}{2}{{\left( {{\partial }_{x}}f \right)}_{0}}+\frac{\Delta {{x}^{2}}}{8}{{\left( \partial _{x}^{2}f \right)}_{0}}\pm \frac{\Delta {{x}^{3}}}{48}{{\left( \partial _{x}^{3}f \right)}_{0}}+...\quad ; \quad \\ 
 & {{\left. {{\partial }_{x}}\mathbf{F}\left( x \right) \right|}_{x=\pm \Delta x/2}}={{\left( {{\partial }_{x}}f \right)}_{0}}\pm \frac{\Delta x}{2}{{\left( \partial _{x}^{2}f \right)}_{0}}+\frac{\Delta {{x}^{2}}}{8}{{\left( \partial _{x}^{3}f \right)}_{0}}\pm ...\quad ; \quad  \\ 
 & {{\left. \partial _{x}^{2}\mathbf{F}\left( x \right) \right|}_{x=\pm \Delta x/2}}={{\left( \partial _{x}^{2}f \right)}_{0}}\pm \frac{\Delta x}{2}{{\left( \partial _{x}^{3}f \right)}_{0}}+...
\end{aligned}. \label{eq:afdcons_simple_deri}
\end{equation}
Finite differencing the point values of the fluxes at $x=\pm \Delta x/2$ gives:-
\begin{equation}
\frac{1}{\Delta x}\left[ {{\left. \mathbf{F}\left( x \right) \right|}_{x=\Delta x/2}}-{{\left. \mathbf{F}\left( x \right) \right|}_{x=-\Delta x/2}} \right]={{\left( {{\partial }_{x}}f \right)}_{0}}+\frac{\Delta {{x}^{2}}}{24}{{\left( \partial _{x}^{3}f \right)}_{0}}
\label{eq:afdcons_2nddiff}
\end{equation}
The error term, which is proportional to $\Delta {{x}^{2}}$ in eqn.~\eqref{eq:afdcons_2nddiff}, tells us that the scheme is only second order accurate regardless of the accuracy of the interpolation. Now realize that if this were a good $k^{th}$ order scheme, it should have returned ${{\left( {{\partial }_{x}}f \right)}_{0}}$ at $x=0$ with an error term that is proportional to $\Delta {{x}^{k}}$. That is indeed the true meaning of finite differencing the right hand side of eqn.~\eqref{eq:pde}. Instead, the presence of ${\Delta {{x}^{2}}{{\left( \partial _{x}^{3}f \right)}_{0}}}/{24}\;$ prevents eqn.~\eqref{eq:afdcons_2nddiff} from even being a third order accurate expression. However, realize that a third derivative at the zone center is equivalent to finite differencing two second derivatives that are evaluated at the zone boundaries. Thus we can take our numerical fluxes at $x=\pm \Delta x/2$ to be
\begin{equation}
\mathbf{F}_{x=\pm \Delta x/2}^{num}=\left[ {{\left. \mathbf{F}\left( x \right) \right|}_{x=\pm \Delta x/2}} \right]-\frac{\Delta {{x}^{2}}}{24}\left[ {{\left. \partial _{x}^{2}\mathbf{F}\left( x \right) \right|}_{x=\pm \Delta x/2}} \right]
\label{eq:afdcons_2ndfnum}
\end{equation}
By finite differencing the numerical fluxes from eqn.~\eqref{eq:afdcons_2ndfnum} we can easily see that
\begin{equation}
\frac{1}{\Delta x}\left[ \mathbf{F}_{x=\Delta x/2}^{num}-\mathbf{F}_{x=-\Delta x/2}^{num} \right]={{\left( {{\partial }_{x}}f \right)}_{0}}+O\left( \Delta {{x}^{4}} \right)
\label{eq:afdcons_2ndfnumdiff}
\end{equation}
In other words, the correction term in eqn.~\eqref{eq:afdcons_2ndfnum} with the higher order flux derivative was essential for restoring accuracy.

Appendix A of~\cite{balsara2024a} shows us how to design an AFD-WENO scheme for conservation laws up to any desired accuracy. For up to ninth order of accuracy, the resulting AFD-WENO scheme is explicitly given by
\begin{equation}
\begin{aligned}
    & {{\partial }_{t}}{{\mathbf{U}}_{i}}
    =
    -\frac{1}{\Delta x}\left\{ {{\mathbf{F}}^{*}}\left( \mathbf{\hat{U}}_{i+1/2}^{-},\mathbf{\hat{U}}_{i+1/2}^{+} \right)-{{\mathbf{F}}^{*}}\left( \mathbf{\hat{U}}_{i-1/2}^{-},\mathbf{\hat{U}}_{i-1/2}^{+} \right) \right\}
    \\
    & - \frac{1}{\Delta x}
    \Bigg\{
         \left[ -\frac{1}{24}{{\left( \Delta x \right)}^{2}}{{\left[ \partial _{x}^{2}\mathbf{F} \right]}_{i+1/2}}
         +
         {\color{red}
         \frac{7}{5760}{{\left( \Delta x \right)}^{4}}{{\left[ \partial _{x}^{4}\mathbf{F} \right]}_{i+1/2}}}
         -
         {\color{blue}
         \frac{31}{967680}{{\left( \Delta x \right)}^{6}}{{\left[ \partial _{x}^{6}\mathbf{F} \right]}_{i+1/2}}}
         +
         {\color{magenta}
         \frac{127}{154828800}{{\left( \Delta x \right)}^{8}}{{\left[ \partial _{x}^{8}\mathbf{F} \right]}_{i+1/2}}}
         \right]
         \\ &  \quad \quad
         -\left[ 
         -
         \frac{1}{24}{{\left( \Delta x \right)}^{2}}{{\left[ \partial _{x}^{2}\mathbf{F} \right]}_{i-1/2}}
         +
         {\color{red}
         \frac{7}{5760}{{\left( \Delta x \right)}^{4}}{{\left[ \partial _{x}^{4}\mathbf{F} \right]}_{i-1/2}}}
         -
         {\color{blue}
         \frac{31}{967680}{{\left( \Delta x \right)}^{6}}{{\left[ \partial _{x}^{6}\mathbf{F} \right]}_{i-1/2}}}
         +
         {\color{magenta}
         \frac{127}{154828800}{{\left( \Delta x \right)}^{8}}{{\left[ \partial _{x}^{8}\mathbf{F} \right]}_{i-1/2}}}
         \right]
    \Bigg\}
\end{aligned} \label{eq:afdcons_scheme}
\end{equation}

Notice that eqn.~\eqref{eq:afdcons_scheme} is still in conservation form, and therefore, it should be able to capture shock locations accurately. The black terms in the above equation yield a third order scheme. In that case, the second derivatives of $\mathbf{F}$ have to be evaluated at the zone boundaries with a WENO interpolation scheme that is at least third order accurate. If the red terms are also included, in addition to the black terms, the scheme becomes fifth order accurate. In that case, all the derivatives of $\mathbf{F}$ have to be evaluated at the zone boundaries with a WENO interpolation scheme that is at least fifth order accurate. If the blue terms are also included, in addition to the black and red terms, we get a seventh order scheme. In that case, all the derivatives of $\mathbf{F}$ have to be evaluated at the zone boundaries with a WENO interpolation scheme that is at least seventh order accurate. If the magenta terms are included, in addition to the black, red and blue terms, we get a ninth order scheme. In that case, all the derivatives of $\mathbf{F}$ have to be evaluated at the zone boundaries with a WENO interpolation scheme that is at least ninth order accurate. The WENO-based process by which these higher order derivatives can be obtained without engendering any Gibbs phenomenon is described in Section 4 of~\cite{balsara2024a}. The stencils that are used within a fifth order scheme to obtain the higher order flux derivatives at the zone boundaries are also shown schematically in Fig.~\ref{fig:stencil_3} of this paper. This completes our discussion of AFD-WENO schemes for conservation laws.

\subsection{AFD-WENO Schemes for Hyperbolic Systems with Non-Conservative Products} \label{subsec:afdweno_noncons}
The key purpose of the previous section was to lead us to the AFD-WENO schemes for hyperbolic systems with some flux terms and some non-conservative products. Since first order treatments of GR  produce hyperbolic systems that are of this type, such AFD-WENO schemes are most useful for GR. Such schemes have been obtained very recently by~\cite{balsara2024b} and we briefly show the reader why they work in this Sub-section. In one dimension, such systems can be formally written as:-
\begin{align}
    {{\partial }_{t}}\mathbf{U}+{{\partial }_{x}}\mathbf{F}\left( \mathbf{U} \right)+\mathbf{C}\left( \mathbf{U} \right){{\partial }_{x}}\mathbf{U}=0.
    \label{eq:afdnc_pde}
\end{align}
Here $\mathbf{C}\left( \mathbf{U} \right)$ is a solution-dependent matrix of non-conservative products. 
 
The trick for the derivation is to start with a conservation law of the same form as eqn.~\eqref{eq:pde} but to rewrite it differently as
\begin{align}
    {{\partial }_{t}}\mathbf{U}+{{\partial }_{x}}\mathbf{F}\left( \mathbf{U} \right)=0
    \qquad
    \text{ with the flux splitting }
    \qquad
    \mathbf{F}\left( \mathbf{U} \right)={{\mathbf{F}}_{C}}\left( \mathbf{U} \right)+{{\mathbf{F}}_{NC}}\left( \mathbf{U} \right).
    \label{eq:afdnc_pdesimple}
\end{align}
In other words, we will be effecting a \textit{trompe l’oeil} where we will treat the flux ${{\mathbf{F}}_{C}}\left( \mathbf{U} \right)$ in conservation form but we will write ${{\mathbf{F}}_{NC}}\left( \mathbf{U} \right)$ as if it can only be written as a non-conservative product. The Jacobians of the fluxes can now be written as:-
\begin{align}
    \mathbf{A}\left( \mathbf{U} \right)=\frac{\partial \left( {{\mathbf{F}}_{C}}\left( \mathbf{U} \right)+{{\mathbf{F}}_{NC}}\left( \mathbf{U} \right) \right)}{\partial \mathbf{U}}=\mathbf{B}\left( \mathbf{U} \right)+\mathbf{C}\left( \mathbf{U} \right)
    \quad \text{with} \quad
    \mathbf{B}\left( \mathbf{U} \right)\equiv \frac{\partial {{\mathbf{F}}_{C}}\left( \mathbf{U} \right)}{\partial \mathbf{U}}
    \quad \text{and} \quad
    \mathbf{C}\left( \mathbf{U} \right)\equiv \frac{\partial {{\mathbf{F}}_{NC}}\left( \mathbf{U} \right)}{\partial \mathbf{U}}. 
    \label{eq:afdnc_AU}
\end{align}
The upshot is that eqn.~\eqref{eq:afdnc_pdesimple} can now be written in the following equivalent forms:-
\begin{align}
    {{\partial }_{t}}\mathbf{U}+{{\partial }_{x}}\mathbf{F}\left( \mathbf{U} \right)=0
    \quad \Leftrightarrow \quad
    {{\partial }_{t}}\mathbf{U}+\mathbf{A}\left( \mathbf{U} \right){{\partial }_{x}}\mathbf{U}=0
    \quad \Leftrightarrow \quad
    {{\partial }_{t}}\mathbf{U}+{{\partial }_{x}}{{\mathbf{F}}_{C}}\left( \mathbf{U} \right)+\mathbf{C}\left( \mathbf{U} \right){{\partial }_{x}}\mathbf{U}=0.
    \label{eq:afdnc_pde_simplenew}
\end{align}
It is now easy to see that if we initially obtain a scheme for the rightmost equation in eqn.~\eqref{eq:afdnc_pde_simplenew}, then we can eventually obtain a scheme for eqn.~\eqref{eq:afdnc_pde} just by dropping the subscript “$C$” in ${{\mathbf{F}}_{C}}\left( \mathbf{U} \right)$.

We start our derivation by writing the resolved fluxes ${{\mathbf{F}}^{*}}\left( \mathbf{\hat{U}}_{i+1/2}^{-},\mathbf{\hat{U}}_{i+1/2}^{+} \right)$ and ${{\mathbf{F}}^{*}}\left( \mathbf{\hat{U}}_{i-1/2}^{-},\mathbf{\hat{U}}_{i-1/2}^{+} \right)$ in the form of fluctuations by using the following definitions for the left-going and right-going fluctuations:-
\begin{equation}
\begin{aligned}
  {{\mathbf{F}}^{*}}\left( \mathbf{\hat{U}}_{i+1/2}^{-},\mathbf{\hat{U}}_{i+1/2}^{+} \right)
  &=
  \mathbf{D}_{{}}^{*-}\left( \mathbf{\hat{U}}_{i+1/2}^{-},\mathbf{\hat{U}}_{i+1/2}^{+} \right)+{{\mathbf{F}}_{C}}\left( \mathbf{\hat{U}}_{i+1/2}^{-} \right)+{{\mathbf{F}}_{NC}}\left( \mathbf{\hat{U}}_{i+1/2}^{-} \right)
  \quad \text{  ;} 
  \\ 
 {{\mathbf{F}}^{*}}\left( \mathbf{\hat{U}}_{i-1/2}^{-},\mathbf{\hat{U}}_{i-1/2}^{+} \right)
 &=
 -\mathbf{D}_{{}}^{*+}\left( \mathbf{\hat{U}}_{i-1/2}^{-},\mathbf{\hat{U}}_{i-1/2}^{+} \right)+{{\mathbf{F}}_{C}}\left( \mathbf{\hat{U}}_{i-1/2}^{+} \right)+{{\mathbf{F}}_{NC}}\left( \mathbf{\hat{U}}_{i-1/2}^{+} \right) 
\end{aligned}. \label{eq:afdnc_flucta}
\end{equation}
Putting eqn.~\eqref{eq:afdnc_flucta} in eqn.~\eqref{eq:afdcons_scheme} allows us to write the intermediate equation:-

\begin{equation}
\begin{aligned}
   & {{\partial }_{t}}{{\mathbf{U}}_{i}}
   =
   -\frac{1}{\Delta x}\left\{ \mathbf{D}_{{}}^{*-}\left( \mathbf{\hat{U}}_{i+1/2}^{-},\mathbf{\hat{U}}_{i+1/2}^{+} \right)+\mathbf{D}_{{}}^{*+}\left( \mathbf{\hat{U}}_{i-1/2}^{-},\mathbf{\hat{U}}_{i-1/2}^{+} \right) \right\} \\ 
   & \quad -\frac{1}{\Delta x}\left\{ {{\mathbf{F}}_{C}}\left( \mathbf{\hat{U}}_{i+1/2}^{-} \right)-{{\mathbf{F}}_{C}}\left( \mathbf{\hat{U}}_{i-1/2}^{+} \right) \right\}-\frac{1}{\Delta x}\left\{ {{\mathbf{F}}_{NC}}\left( \mathbf{\hat{U}}_{i+1/2}^{-} \right)-{{\mathbf{F}}_{NC}}\left( \mathbf{\hat{U}}_{i-1/2}^{+} \right) \right\} \\ 
 & \quad -\frac{1}{\Delta x}\Bigg\{ 
   \left[ -
   {\color{black}
   \frac{1}{24}{{\left( \Delta x \right)}^{2}}{{\left[ \partial _{x}^{2}\mathbf{F} \right]}_{i+1/2}}}
   +
   {\color{red}
   \frac{7}{5760}{{\left( \Delta x \right)}^{4}}{{\left[ \partial _{x}^{4}\mathbf{F} \right]}_{i+1/2}}}
   -
   {\color{blue}
   \frac{31}{967680}{{\left( \Delta x \right)}^{6}}{{\left[ \partial _{x}^{6}\mathbf{F} \right]}_{i+1/2}}}
   +
   {\color{magenta}
   \frac{127}{154828800}{{\left( \Delta x \right)}^{8}}{{\left[ \partial _{x}^{8}\mathbf{F} \right]}_{i+1/2}}}
   \right]
   \\ 
   & \quad \qquad \
   -\left[ 
   -
   \frac{1}{24}{{\left( \Delta x \right)}^{2}}{{\left[ \partial _{x}^{2}\mathbf{F} \right]}_{i-1/2}}
   +
   {\color{red}
   \frac{7}{5760}{{\left( \Delta x \right)}^{4}}{{\left[ \partial _{x}^{4}\mathbf{F} \right]}_{i-1/2}}}
   -
   {\color{blue}
   \frac{31}{967680}{{\left( \Delta x \right)}^{6}}{{\left[ \partial _{x}^{6}\mathbf{F} \right]}_{i-1/2}}}
   +
   {\color{magenta}
   \frac{127}{154828800}{{\left( \Delta x \right)}^{8}}{{\left[ \partial _{x}^{8}\mathbf{F} \right]}_{i-1/2}}}
   \right] 
 \Bigg\} 
\end{aligned} \label{eq:afdnc_scheme_1stage}
\end{equation}
Now, we will write the term $\left\{ {{\mathbf{F}}_{NC}}\left( \mathbf{\hat{U}}_{i+1/2}^{-} \right)-{{\mathbf{F}}_{NC}}\left( \mathbf{\hat{U}}_{i-1/2}^{+} \right) \right\}$ as a term that looks like $\mathbf{C}\left( {{\mathbf{U}}_{i}} \right){{\left( {{\partial }_{x}}\mathbf{\hat{U}} \right)}_{i}}$ along with some higher order terms which will eventually cancel off. We write
\begin{equation}
\begin{aligned}
   - & \frac{1}{\Delta x}\left\{ {{\mathbf{F}}_{NC}}\left( \mathbf{\hat{U}}_{i+1/2}^{-} \right)-{{\mathbf{F}}_{NC}}\left( \mathbf{\hat{U}}_{i-1/2}^{+} \right) \right\} 
   \\
   & \cong
   -{{\left( {{\partial }_{x}}{{\mathbf{F}}_{NC}} \right)}_{i}}-
   \left\{
   +
   \frac{\Delta {{x}^{2}}}{24}{{\left( \partial _{x}^{3}{{\mathbf{F}}_{NC}} \right)}_{i}}
   +
   {\color{red}
   \frac{\Delta {{x}^{4}}}{1920}{{\left( \partial _{x}^{5}{{\mathbf{F}}_{NC}} \right)}_{i}}
   }
   +
   {\color{blue}
   \frac{\Delta {{x}^{6}}}{322560}{{\left( \partial _{x}^{7}{{\mathbf{F}}_{NC}} \right)}_{i}}
   }
   +
   {\color{magenta}
   \frac{\Delta {{x}^{8}}}{92897280}{{\left( \partial _{x}^{9}{{\mathbf{F}}_{NC}} \right)}_{i}} 
   }
   \right\} 
   \\
   & 
   \cong -\mathbf{C}\left( {{\mathbf{U}}_{i}} \right){{\left( {{\partial }_{x}}\mathbf{\hat{U}} \right)}_{i}}-\frac{1}{\Delta x}
   \Bigg\{
       \left[ 
           \frac{1}{24}{{\left( \Delta x \right)}^{2}}{{\left[ \partial _{x}^{2}{{\mathbf{F}}_{NC}} \right]}_{i+1/2}}
           -
           {\color{red}
           \frac{7}{5760}{{\left( \Delta x \right)}^{4}}{{\left[ \partial _{x}^{4}{{\mathbf{F}}_{NC}} \right]}_{i+1/2}}  
           } \right.
           \\
           & 
           \left. \qquad \qquad \qquad \qquad \qquad \quad
           +
           {\color{blue}
           \frac{31}{967680}{{\left( \Delta x \right)}^{6}}{{\left[ \partial _{x}^{6}{{\mathbf{F}}_{NC}} \right]}_{i+1/2}}
           }
           -
           {\color{magenta}
           \frac{127}{154828800}{{\left( \Delta x \right)}^{8}}{{\left[ \partial _{x}^{8}{{\mathbf{F}}_{NC}} \right]}_{i+1/2}} 
           }
       \right] \\
       & \qquad \qquad \qquad \qquad \qquad  -
       \left[ 
           \frac{1}{24}{{\left( \Delta x \right)}^{2}}{{\left[ \partial _{x}^{2}{{\mathbf{F}}_{NC}} \right]}_{i-1/2}}
           -
           {\color{red}
           \frac{7}{5760}{{\left( \Delta x \right)}^{4}}{{\left[ \partial _{x}^{4}{{\mathbf{F}}_{NC}} \right]}_{i-1/2}}  
           } \right.
           \\
           & 
           \left.
           \qquad \qquad \qquad \qquad \qquad \quad
           +
           {\color{blue}
           \frac{31}{967680}{{\left( \Delta x \right)}^{6}}{{\left[ \partial _{x}^{6}{{\mathbf{F}}_{NC}} \right]}_{i-1/2}}
           }
           -
           {\color{magenta}
           \frac{127}{154828800}{{\left( \Delta x \right)}^{8}}{{\left[ \partial _{x}^{8}{{\mathbf{F}}_{NC}} \right]}_{i-1/2}} 
           }
       \right] 
   \Bigg\}
\end{aligned} \label{eq:afdnc_f_diff_simp}
\end{equation}
We insert eqn.~\eqref{eq:afdnc_f_diff_simp} in eqn.~\eqref{eq:afdnc_scheme_1stage} and use the fact that ${{\mathbf{F}}_{C}}\left( \mathbf{U} \right)=\mathbf{F}\left( \mathbf{U} \right)-{{\mathbf{F}}_{NC}}\left( \mathbf{U} \right)$ to get the penultimate form of our update equation:-
\begin{equation}
\begin{aligned}
   & {{\partial }_{t}}{{\mathbf{U}}_{i}}
   =
   -\frac{1}{\Delta x}\left\{ \mathbf{D}_{{}}^{*-}\left( \mathbf{\hat{U}}_{i+1/2}^{-},\mathbf{\hat{U}}_{i+1/2}^{+} \right)+\mathbf{D}_{{}}^{*+}\left( \mathbf{\hat{U}}_{i-1/2}^{-},\mathbf{\hat{U}}_{i-1/2}^{+} \right) \right\}-\frac{1}{\Delta x}\left\{ {{\mathbf{F}}_{C}}\left( \mathbf{\hat{U}}_{i+1/2}^{-} \right)-{{\mathbf{F}}_{C}}\left( \mathbf{\hat{U}}_{i-1/2}^{+} \right) \right\}-\mathbf{C}\left( {{\mathbf{U}}_{i}} \right){{\left( {{\partial }_{x}}\mathbf{\hat{U}} \right)}_{i}} 
   \\
   &   -\frac{1}{\Delta x}
   \Bigg\{
       \left[ 
           -
           \frac{1}{24}{{\left( \Delta x \right)}^{2}}{{\left[ \partial _{x}^{2}{{\mathbf{F}_{C}}} \right]}_{i+1/2}}
           +
           {\color{red}
           \frac{7}{5760}{{\left( \Delta x \right)}^{4}}{{\left[ \partial _{x}^{4}{{\mathbf{F}_{C}}} \right]}_{i+1/2}}  
           }
           -
           {\color{blue}
           \frac{31}{967680}{{\left( \Delta x \right)}^{6}}{{\left[ \partial _{x}^{6}{{\mathbf{F}}_{C}} \right]}_{i+1/2}}
           }
           +
           {\color{magenta}
           \frac{127}{154828800}{{\left( \Delta x \right)}^{8}}{{\left[ \partial _{x}^{8}{{\mathbf{F}}_{C}} \right]}_{i+1/2}} 
           }
       \right] \\
       &  \quad \ \ -
        \left[ 
           -
           \frac{1}{24}{{\left( \Delta x \right)}^{2}}{{\left[ \partial _{x}^{2}{{\mathbf{F}}_{C} } \right]}_{i-1/2}}
           +
           {\color{red}
           \frac{7}{5760}{{\left( \Delta x \right)}^{4}}{{\left[ \partial _{x}^{4}{{\mathbf{F}}_{C} } \right]}_{i-1/2}}  
           }
           -
           {\color{blue}
           \frac{31}{967680}{{\left( \Delta x \right)}^{6}}{{\left[ \partial _{x}^{6}{{\mathbf{F}}_{C} } \right]}_{i-1/2}}
           }
           +
           {\color{magenta}
           \frac{127}{154828800}{{\left( \Delta x \right)}^{8}}{{\left[ \partial _{x}^{8}{{\mathbf{F}}_{C} } \right]}_{i-1/2}} 
           }
       \right]
   \Bigg\}
\end{aligned} \label{eq:afdnc_dudt_stage1}
\end{equation}
Next, we will derive the final forms of the equations that we need.

The above equation would give us the update strategy for the last equation in eqn.~\eqref{eq:afdnc_pde_simplenew}. But we want to transition from the last equation in eqn.~\eqref{eq:afdnc_pde_simplenew} to eqn.~\eqref{eq:afdnc_pdesimple}. This is easily accomplished by dropping the subscript “$C$” in ${{\mathbf{F}}_{C}}\left( \mathbf{U} \right)$ in eqn.~\eqref{eq:afdnc_dudt_stage1} to get one of our final equations as:-
\begin{equation}
\begin{aligned}
   & {{\partial }_{t}}{{\mathbf{U}}_{i}}
   =
   -\frac{1}{\Delta x}\left\{ \mathbf{D}_{{}}^{*-}\left( \mathbf{\hat{U}}_{i+1/2}^{-},\mathbf{\hat{U}}_{i+1/2}^{+} \right)+\mathbf{D}_{{}}^{*+}\left( \mathbf{\hat{U}}_{i-1/2}^{-},\mathbf{\hat{U}}_{i-1/2}^{+} \right) \right\}-\frac{1}{\Delta x}\left\{ {{\mathbf{F}}}\left( \mathbf{\hat{U}}_{i+1/2}^{-} \right)-{{\mathbf{F}}}\left( \mathbf{\hat{U}}_{i-1/2}^{+} \right) \right\}-\mathbf{C}\left( {{\mathbf{U}}_{i}} \right){{\left( {{\partial }_{x}}\mathbf{\hat{U}} \right)}_{i}} 
   \\
   & \quad -\frac{1}{\Delta x}
   \Bigg\{
       \left[ 
           -
           \frac{1}{24}{{\left( \Delta x \right)}^{2}}{{\left[ \partial _{x}^{2}{{\mathbf{F}}} \right]}_{i+1/2}}
           +
           {\color{red}
           \frac{7}{5760}{{\left( \Delta x \right)}^{4}}{{\left[ \partial _{x}^{4}{{\mathbf{F}}} \right]}_{i+1/2}}  
           }
           -
           {\color{blue}
           \frac{31}{967680}{{\left( \Delta x \right)}^{6}}{{\left[ \partial _{x}^{6}{{\mathbf{F}}} \right]}_{i+1/2}}
           }
           +
           {\color{magenta}
           \frac{127}{154828800}{{\left( \Delta x \right)}^{8}}{{\left[ \partial _{x}^{8}{{\mathbf{F}}} \right]}_{i+1/2}} 
           }
       \right] \\
       & \qquad \quad  -
        \left[ 
           -
           \frac{1}{24}{{\left( \Delta x \right)}^{2}}{{\left[ \partial _{x}^{2}{{\mathbf{F}} } \right]}_{i-1/2}}
           +
           {\color{red}
           \frac{7}{5760}{{\left( \Delta x \right)}^{4}}{{\left[ \partial _{x}^{4}{{\mathbf{F}} } \right]}_{i-1/2}}  
           }
           -
           {\color{blue}
           \frac{31}{967680}{{\left( \Delta x \right)}^{6}}{{\left[ \partial _{x}^{6}{{\mathbf{F}} } \right]}_{i-1/2}}
           }
           +
           {\color{magenta}
           \frac{127}{154828800}{{\left( \Delta x \right)}^{8}}{{\left[ \partial _{x}^{8}{{\mathbf{F}} } \right]}_{i-1/2}} 
           }
       \right]
   \Bigg\}
\end{aligned} \label{eq:afdnc_dudt_stage2}
\end{equation}
The above equation is very illustrative. Notice that when the fluxes are absent, i.e. when the entire system is represented by the non-conservative products, the above equation reduces to
\begin{align}
    {{\partial }_{t}}{{\mathbf{U}}_{i}}=-\frac{1}{\Delta x}\left\{ \mathbf{D}_{{}}^{*-}\left( \mathbf{\hat{U}}_{i+1/2}^{-},\mathbf{\hat{U}}_{i+1/2}^{+} \right)+\mathbf{D}_{{}}^{*+}\left( \mathbf{\hat{U}}_{i-1/2}^{-},\mathbf{\hat{U}}_{i-1/2}^{+} \right) \right\}-\mathbf{C}\left( {{\mathbf{U}}_{i}} \right){{\left( {{\partial }_{x}}\mathbf{\hat{U}} \right)}_{i}}
    \label{eq:afdnc_dudt_onlync}
\end{align}
which is almost identical to eqn.~\eqref{eq:FDWENO_NONCONS_dudt}. (Recall that $\mathbf{C}\left( {{\mathbf{U}}_{i}} \right)\to \mathbf{A}\left( {{\mathbf{U}}_{i}} \right)$ in that limit.)  In other words, when solving just the Einstein equations, without additional hydrodynamic equations, AFD-WENO and FD-WENO become practically identical to one another. Of course, eqn.~\eqref{eq:FDWENO_NONCONS_dudt} relies on reconstruction whereas eqn.~\eqref{eq:afdnc_dudt_onlync} relies on interpolation. A form that is completely equivalent to eqn.~\eqref{eq:afdnc_dudt_stage2} at the analytical level, but has slightly better stability properties, can be written as:-
\begin{equation}
\begin{aligned}
   & {{\partial }_{t}}{{\mathbf{U}}_{i}}
   =
   -\frac{1}{\Delta x}\left\{ \mathbf{D}_{i+1/2}^{*-}\left( \mathbf{\hat{U}}_{i+1/2}^{-},\mathbf{\hat{U}}_{i+1/2}^{+} \right)+\mathbf{D}_{i-1/2}^{*+}\left( \mathbf{\hat{U}}_{i-1/2}^{-},\mathbf{\hat{U}}_{i-1/2}^{+} \right) \right\}-\frac{1}{\Delta x}\left\{ \mathbf{F}\left( \mathbf{\hat{U}}_{i+1/2}^{-} \right)-\mathbf{F}\left( \mathbf{\hat{U}}_{i-1/2}^{+} \right) \right\}
   \\
   & \quad
    -\frac{1}{\Delta x}\mathbf{C}\left( {{\mathbf{U}}_{i}} \right)\left( \mathbf{\hat{U}}_{i+1/2}^{-}-\mathbf{\hat{U}}_{i-1/2}^{+} \right)-\frac{1}{\Delta x}\mathbf{C}\left( {{\mathbf{U}}_{i}} \right)
    \Bigg\{
        \left[ 
           -
           \frac{1}{24}{{\left( \Delta x \right)}^{2}}{{\left[ \partial _{x}^{2}{\mathbf{\hat{U}}} \right]}_{i+1/2}}
           +
           {\color{red}
           \frac{7}{5760}{{\left( \Delta x \right)}^{4}}{{\left[ \partial _{x}^{4}\mathbf{\hat{U}} \right]}_{i+1/2}}  
           } \right.
           \\
           & 
           \left. \qquad \qquad \qquad  \qquad  \qquad \qquad \qquad \qquad \qquad \quad
           -
           {\color{blue}
           \frac{31}{967680}{{\left( \Delta x \right)}^{6}}{{\left[ \partial _{x}^{6}\mathbf{\hat{U}} \right]}_{i+1/2}}
           }
           +
           {\color{magenta}
           \frac{127}{154828800}{{\left( \Delta x \right)}^{8}}{{\left[ \partial _{x}^{8}\mathbf{\hat{U}} \right]}_{i+1/2}} 
           }
       \right] \\
       & \qquad \qquad   \qquad  \qquad \qquad \qquad \qquad \qquad \qquad  -
       \left[ 
           -
           \frac{1}{24}{{\left( \Delta x \right)}^{2}}{{\left[ \partial _{x}^{2}{\mathbf{\hat{U}}} \right]}_{i-1/2}}
           +
           {\color{red}
           \frac{7}{5760}{{\left( \Delta x \right)}^{4}}{{\left[ \partial _{x}^{4}\mathbf{\hat{U}} \right]}_{i-1/2}}  
           } \right.
           \\
           & 
           \left. \qquad \qquad \qquad  \qquad  \qquad \qquad \qquad \qquad \qquad \quad
           -
           {\color{blue}
           \frac{31}{967680}{{\left( \Delta x \right)}^{6}}{{\left[ \partial _{x}^{6}\mathbf{\hat{U}} \right]}_{i-1/2}}
           }
           +
           {\color{magenta}
           \frac{127}{154828800}{{\left( \Delta x \right)}^{8}}{{\left[ \partial _{x}^{8}\mathbf{\hat{U}} \right]}_{i-1/2}} 
           }
       \right]
    \Bigg\}
   \\
   & \quad -\frac{1}{\Delta x}
   \Bigg\{
       \left[ 
           -
           \frac{1}{24}{{\left( \Delta x \right)}^{2}}{{\left[ \partial _{x}^{2}{{\mathbf{F}}} \right]}_{i+1/2}}
           +
           {\color{red}
           \frac{7}{5760}{{\left( \Delta x \right)}^{4}}{{\left[ \partial _{x}^{4}{{\mathbf{F}}} \right]}_{i+1/2}}  
           }
           -
           {\color{blue}
           \frac{31}{967680}{{\left( \Delta x \right)}^{6}}{{\left[ \partial _{x}^{6}{{\mathbf{F}}} \right]}_{i+1/2}}
           }
           +
           {\color{magenta}
           \frac{127}{154828800}{{\left( \Delta x \right)}^{8}}{{\left[ \partial _{x}^{8}{{\mathbf{F}}} \right]}_{i+1/2}} 
           }
       \right] \\
       & \qquad \quad  -
        \left[ 
           -
           \frac{1}{24}{{\left( \Delta x \right)}^{2}}{{\left[ \partial _{x}^{2}{{\mathbf{F}} } \right]}_{i-1/2}}
           +
           {\color{red}
           \frac{7}{5760}{{\left( \Delta x \right)}^{4}}{{\left[ \partial _{x}^{4}{{\mathbf{F}} } \right]}_{i-1/2}}  
           }
           -
           {\color{blue}
           \frac{31}{967680}{{\left( \Delta x \right)}^{6}}{{\left[ \partial _{x}^{6}{{\mathbf{F}} } \right]}_{i-1/2}}
           }
           +
           {\color{magenta}
           \frac{127}{154828800}{{\left( \Delta x \right)}^{8}}{{\left[ \partial _{x}^{8}{{\mathbf{F}} } \right]}_{i-1/2}} 
           }
       \right]
   \Bigg\}
\end{aligned} \label{eq:afdnc_dudt_final}
\end{equation}
This completes our derivation of AFD-WENO schemes for systems with non-conservative products as they arise in GR. Along with the derivations, we have striven to provide insights on the strength of each scheme and its suitability for NR. Notice that the evaluation of the flux derivatives in eqns.~\eqref{eq:afdnc_dudt_stage2} or~\eqref{eq:afdnc_dudt_final} is the same as was sketched in Fig.~\ref{fig:stencil_3}. The evaluation of the higher derivatives of the states at the zone boundaries in eqn.~\eqref{eq:afdnc_dudt_final} also follows a process that is practically identical to Fig.~\ref{fig:stencil_3} and is described in detail in~\cite{balsara2024b}.


Some overarching comments that connect the multiple finite difference WENO formulations are in order.
\begin{enumerate}
    \item Realize that the scheme in Sub-section~\ref{subsec:fdweno} relies on a flux splitting into an LLF flux and can, therefore, be somewhat dissipative. The schemes described in Sub-sections~\ref{subsec:fdweno_noncons},~\ref{subsec:afdweno_cons} and~\ref{subsec:afdweno_noncons} can use improved Riemann solvers. Any good Riemann solver can be used in the schemes described in those three Sub-scections. The \textit{HLLI} Riemann solver from~\cite{dumbser2016new} might be very beneficial because it can handle all different types of intermediate waves including the contact discontinuities that are needed in well-balancing of hydrodynamic problems.
    %
    %
    \item The AFD-WENO scheme in this sub-section has an advantage over the FD-WENO scheme in Sub-section~\ref{subsec:fdweno_noncons} in that it recovers the conservation form when such a form is present. This can be very useful if the hydrodynamical equations are to be evolved in conjunction with the Einstein equations.
    %
    %
    \item The term ${{\left( {{\partial }_{x}}\mathbf{\hat{U}} \right)}_{i}}$ in eqn.~\eqref{eq:afdnc_dudt_stage2} has to be evaluated using the same WENO interpolation that was used for the state vector; as a result, eqn.~\eqref{eq:afdnc_dudt_stage2} usually shows one lower order of accuracy compared to eqn.~\eqref{eq:afdnc_dudt_final}. Eqn.~\eqref{eq:afdnc_dudt_stage2} may have some advantages for GR nevertheless because when the solution is smooth, it gives very good quality solutions at a lower computational cost.
    %
    %
    \item When strong shocks are present, eqn.~\eqref{eq:afdnc_dudt_final} has slightly better stability properties compared to eqn.~\eqref{eq:afdnc_dudt_stage2} because the term $\mathbf{C}\left( {{\mathbf{U}}_{i}} \right)\left( \mathbf{\hat{U}}_{i+1/2}^{-}-\mathbf{\hat{U}}_{i-1/2}^{+} \right)$ is itself treated in a finite difference form. However, since strong shocks are only present in the hydrodynamic sector in a NR calculation, the treatment of flux terms is identical in both equations.
    %
    %
    \item We have only addressed spatial accuracy in this Section. Like all finite difference WENO schemes, the schemes presented here are to be coupled with strong stability preserving Runge-Kutta (SSP-RK) timestepping. Temporally high-order SSP-RK timestepping has been described in~\cite{shu1988efficient} or~\cite{spiteri2002new,spiteri2003non}.
    %
    %
    \item In general, the treatment of the source terms in GR is such that the source terms are always non-stiff. As a result, we shouldn’t need variants of SSP-RK timestepping that include stiff source terms.
    %
    %
    \item It would be useful to keep an eye on the ecosystem of ideas that are emerging around AFD-WENO methods. Such ideas include well-balancing, physical realizability, and preservation of divergence and curl constraints. These enhancements retain the low cost of the original AFD-WENO scheme.
\end{enumerate}
This completes our discussion of FD-WENO and AFD-WENO schemes for systems with non-conservative products as they arise in GR.

\subsection{Well balancing of FD-WENO and AFD-WENO schemes}
\label{subsec:WB}
Preserving the stationarity of an equilibrium solution $\mathbf{U}_e$ is far from trivial, and,
unless suitable steps are undertaken, a general numerical scheme will not be able to 
satisfy the condition $\partial_t\mathbf{U}_e=0$ exactly over arbitrarily long timescales. This 
is particularly frustrating, for instance, if one is interested in studying numerically the oscillation modes of astrophysical objects in equilibrium. In these circumstances, 
if the order of magnitude of the discretization  errors introduced by the numerical scheme are close
to those of the physical initial perturbation under investigation, 
it will be  impossible to distinguish among 
a physical effect and a numerical artifact. In the worse cases, the accumulation of numerical errors
can even spoil completely the equilibrium solution over sufficiently long timescales.
Hence, a whole field of research exists to ensure the so-called \emph{well--balancing} of a numerical scheme. Initially proposed in the context of the shallow water equations, well--balanced numerical schemes have been proposed in a variety of different techniques, sometimes tailored to specific
PDE systems \citep{Bermudez1994,leveque1998balancing,gosse2001well,audusse2004fast,BottaKlein,Castro2017Book,castro2020well,Gaburro2021WBGR1D,xu2024,bhoriya2024}.

Here we follow the pragmatic approach chosen by \cite{DumbserZanottiGaburroPeshkov2023}, who,
motivated by the complexity of the Einstein equations, 
subtracted the equilibrium solution alltogether from the governing PDE, see also \cite{PareschiRey,berberich2021high}, therefore solving the following augmented system 
\begin{align}
    {{\partial }_{t}}\mathbf{U}+{{\partial }_{x}}\mathbf{F}\left( \mathbf{U} \right)-{{\partial }_{x}}\mathbf{F}\left( \mathbf{U}_e \right)
    +\mathbf{C}\left( \mathbf{U} \right){{\partial }_{x}}\mathbf{U}
    -\mathbf{C}\left( \mathbf{U}_e \right){{\partial }_{x}}\mathbf{U}_e=\mathbf{S}(\mathbf{U})-\mathbf{S}(\mathbf{U}_e); \qquad {{\partial }_{t}}\mathbf{U}_e = 0. 
\end{align}
In this way, the discretization errors introduced by the numerical scheme are removed and
exact equilibrium can be maintained. There is a price to pay, of course, in so far the
vector of evolved quantities is effectively doubled, becoming $\tilde{\mathbf{U}}=[\mathbf{U},\mathbf{U}_e]$. The equilibrium sector of the extended vector $\tilde{\mathbf{U}}$, i.e. $\mathbf{U}_e$, is ultimately
excluded from the time evolution, but it enters all the spatial discretization procedures of the numerical scheme.

We have added this property to our numerical schemes as an extra feature that can be activated, or not, depending on the problem considered. For more details on the numerical approach, see \cite{DumbserZanottiGaburroPeshkov2023,PareschiRey,berberich2021high}. 

\section{Numerical tests}
\label{sec:tests}

In this section, we explore two distinct computational schemes outlined in Section~\ref{sec:scheme}. First, we consider the scheme described by Eqn.~\eqref{eq:FDWENO_NONCONS_dudt}, explained in Section~\ref{subsec:fdweno_noncons}. It is noteworthy that employing a $k^{th}$-order accurate WENO--AO reconstruction in Eqn.~\eqref{eq:FDWENO_NONCONS_dudt} results in a $k^{th}$-order accurate FD--WENO scheme. We label the resultant scheme as FD--WENO-k. Next, we explore the AFD--WENO scheme outlined in Section~\ref{subsec:afdweno_noncons} and defined by Eqn.~\eqref{eq:afdnc_dudt_stage2}. The scheme given by Eqn.~\eqref{eq:afdnc_dudt_stage2} usually shows one  order of accuracy lower compared to eqn.~\eqref{eq:afdnc_dudt_final}; nevertheless, it may have some advantages for GR because, when the solution is smooth, it gives very good quality solutions at a lower computational cost. Implementing a WENO--AO interpolation of order '$k$' in Eqn.~\eqref{eq:afdnc_dudt_stage2} yields a $(k-1)^{th}$ order scheme. We label the resultant scheme as AFD--WENO-(k-1). For both schemes, we explore the scenarios where $k$ is set to 3, 5, 7, and 9, providing a comprehensive examination across different order schemes.

In the following, we show the performances of the new schemes over a representative sample of standard
tests in NR. While doing so, we are also going to monitor 
the energy constraint $H$ and the momentum constraints $M_i$, also known as ADM constraints, which, 
{in a vacuum spacetime,} are defined as
\begin{eqnarray}
\label{eqn.adm1}
H &=& R - K_{ij} K^{ij} + K^2\,, \\
\label{eqn.adm2}
M_i &=& \gamma^{jl} \left( \partial_l
K_{ij} - \partial_i K_{jl} - \Gamma^m_{jl} K_{mi} + \Gamma^m_{ji} K_{ml} 
\right)\,.
\end{eqnarray}

\subsection{Gauge wave}
\label{sec:gaugewave}

We start our validation by considering the propagation of a gauge wave, which
is generated after a suitable change of coordinates in a flat Minkowski spacetime~\citep{Alcubierre2004},
producing the following metric
\begin{equation}
	\label{eqn.gw.metric}
	d s^2 = - H(x,t) \, d t^2 + H(x,t) \, d x^2 + d y^2 + d z^2, \quad \text{where} \quad H(x,t) = 
	1-A\,\sin \left( 2\pi(x-t) \right)\,.
\end{equation}
In practice, it describes a sinusoidal wave of amplitude $A$ that propagates along the $x$-axis, with
the lapse $\alpha = \sqrt{H}$ and 
the extrinsic curvature initialized through its definition $K_{ij} = -\partial_t \gamma_{ij}/ (2\alpha)$.
As reported by several authors, this test cannot be evolved stably and accurately over long timescales by the
the classical BSSNOK system, either in second or first-order formulations~\citep{Babiuc_2008, Brown2012},
while it has been successfully solved within the first--order CCZ4
and Z4 formulations~\citep{Dumbser2017strongly, DumbserZanottiGaburroPeshkov2023}, in both cases
with no need to activate the damping mechanisms. We have also solved this problem by selecting
$\kappa_1=0$, $\kappa_2=0$, 
no \emph{gamma--driver} ($s=0$),
harmonic gauge conditions, i.e. $g(\alpha)=1$ in Eq~\eqref{eqn.slicing}, and $c=1$.
The numerical domain is the rectangle $[-0.5,0.5] \times [-0.05, 0.05]$, that is covered by
a $N_x\times N_y$ uniform grid. Periodic boundary conditions are selected.

\begin{table}[h]
	\centering
	\begin{tabular}{|c||c|c|c|c|c|}
		\hline
		Scheme  & Grid size & $L_1-$errors & Order & $L_\infty-$errors & Order \\
           \hline
           \hline
		\multirow{4}{*}{FD--WENO-3}
          & 32 $\times$ 2	  &   1.32436e-04 &   --   &  2.68393e-04 &  -- 	\\
          & 64 $\times$ 4	  &   1.65963e-05 & 	3.00  &  3.41976e-05 & 	2.97  \\
          & 128 $\times$ 8  &   2.07707e-06 & 	3.00  &  4.29149e-06 & 	2.99  \\
          & 256 $\times$ 16 & 	2.59697e-07 & 	3.00  &  5.36977e-07 & 	3.00  \\
           \hline
		\multirow{4}{*}{FD--WENO-5}
           &  32 $\times$ 2   &  1.41137e-06 &  --   &  4.11999e-06 & --    \\
            &  64 $\times$ 4   &  4.42906e-08 & 4.99  &  1.31668e-07 & 4.97   \\
            &  128 $\times$ 8  &  1.38883e-09 & 5.00  &  4.13672e-09 & 4.99   \\
            &  256 $\times$ 16 &  4.34472e-11 & 5.00  &  1.29485e-10 & 5.00   \\
        \hline
        \multirow{4}{*}{FD--WENO-7}
            &  16 $\times$ 1 &  9.83235e-06 &  --  & 2.75912E-05 &   --   \\ 	
            &  32 $\times$ 2 &  4.15493e-08 & 7.89 & 1.50863E-07 & 7.51  \\
            &  48 $\times$ 3 &  2.43860e-09 & 6.99 & 8.53961E-09 & 7.08  \\
            &  64 $\times$ 4 &  3.29132e-10 & 6.96 & 1.13484E-09 & 7.02  \\
           \hline
        \multirow{4}{*}{FD--WENO-9}
            &  8 $\times$ 1 &  1.70885e-04 &  --   & 3.55896e-04 &   --   \\ 	
            &  16 $\times$ 2 &  8.48349e-07 & 7.65 & 2.66925e-06 & 7.06  \\
            &  24 $\times$ 3 &  2.57895e-08 & 8.62 & 9.82247e-08 & 8.14  \\
            &  32 $\times$ 4 &  2.17566e-09 & 8.60 & 8.25465e-09 & 8.61   \\
           \hline
	\end{tabular}
  \vspace{0.3cm}
	\caption{\nameref{sec:gaugewave}: $L_1$ errors, $L_\infty$ errors, and the corresponding order of convergence for the third, fifth, seventh and ninth order accurate FD--WENO schemes. The lapse $\alpha$ variable has been shown. A final time of $t=1.0$ has been used for the study.}
	\label{table:accuracyFD--WENO}
\end{table}

\begin{table}[h]
	\centering
	\begin{tabular}{|c||c|c|c|c|c|}
		\hline
		Scheme  & Grid size & $L_1-$errors & Order & $L_\infty-$errors & Order \\
           \hline
           \hline
		\multirow{4}{*}{AFD--WENO-2}
          &  32 $\times$ 2   & 3.45649E-04 &  --  & 5.68896E-04 &  --  \\ 
          &  64 $\times$ 4   & 8.25823E-05 & 2.07 & 1.34221E-04 &  2.08   \\
          &  128 $\times$ 8  & 2.03870E-05 & 2.02 & 3.30125E-05 &  2.02   \\
          &  256 $\times$ 16 & 5.07952E-06 & 2.00 & 8.21743E-06 &  2.01   \\
           \hline
		\multirow{4}{*}{AFD--WENO-4}
            &  32 $\times$ 2   & 3.69051E-06 &  --  & 8.72790E-06 &  --    \\
            &  64 $\times$ 4   & 1.98678E-07 & 4.22 & 4.35846E-07 &  4.32    \\
            &  128 $\times$ 8  & 1.22491E-08 & 4.02 & 2.63489E-08 &  4.05    \\
            &  256 $\times$ 16 & 7.63165E-10 & 4.00 & 1.63497E-09 &  4.01    \\
        \hline
		\multirow{4}{*}{AFD--WENO-6}
            &  16 $\times$ 1 &  9.80463e-06 &  --  & 2.80057e-05 &   --   \\ 	
            &  32 $\times$ 2 &  7.37080e-08	& 7.06 & 2.24641e-07 & 6.96  \\
            &  48 $\times$ 3 &  6.23277e-09	& 6.09 & 1.81759e-08 & 6.20  \\
            &  64 $\times$ 4 &  1.13379e-09	& 5.92 & 3.27951e-09 & 5.95  \\
           \hline
       \multirow{4}{*}{AFD--WENO-8}
            &  8 $\times$ 1  &  2.77564e-04 &  --   & 5.40091e-04 &   --  \\ 	
            &  16 $\times$ 2 &  8.74626e-07 & 8.31 & 2.78709e-06 & 7.60  \\
            &  24 $\times$ 3 &  3.62176e-08 & 7.85 & 1.31702e-07 & 7.53  \\
            &  32 $\times$ 4 &  4.60690e-09 & 7.17 & 1.68872e-08 & 7.14  \\
           \hline
	\end{tabular}
 \vspace{0.3cm}
	\caption{\nameref{sec:gaugewave}: $L_1$ errors, $L_\infty$ errors, and the corresponding order of convergence for the second, fourth, sixth, and eighth order accurate AFD--WENO schemes. The lapse $\alpha$ variable has been shown. A final time of $t=1.0$ has been used for the study.}
	\label{table:accuracyAFD--WENO}
\end{table}

This test is also rather useful for validating the order of accuracy of the numerical scheme.
We have therefore considered a battery of tests with a short final time, $t=1.0$, producing
results that are reported in Tab.~\ref{table:accuracyFD--WENO}--\ref{table:accuracyAFD--WENO},
corresponding to our FD--WENO and AFD--WENO schemes, respectively. As it can be appreciated,
the nominal order of convergence is successfully reproduced.

\begin{figure}[h]
	\begin{center}
		\includegraphics[width=0.75\textwidth,clip=]{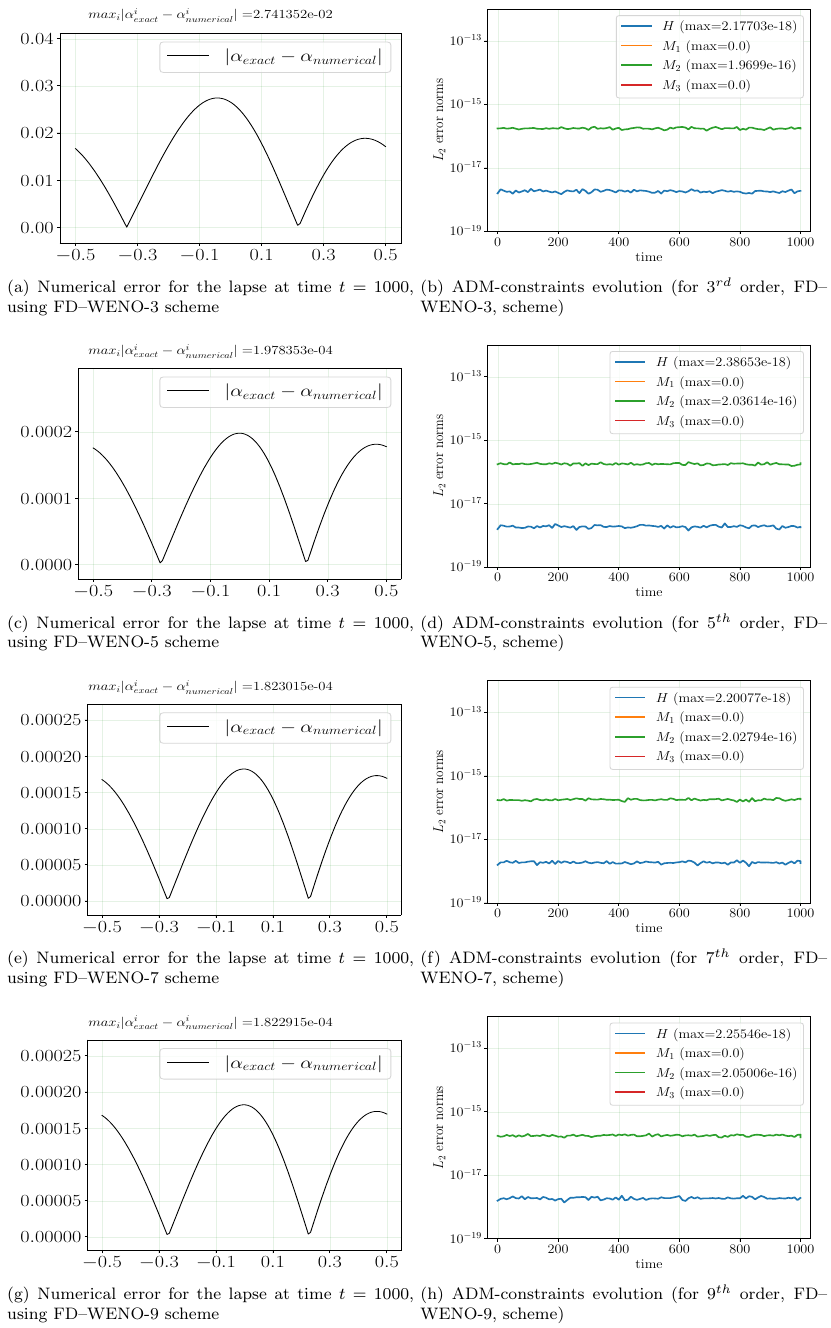}
		\caption{\nameref{sec:gaugewave}: The left panel (Figs.~\ref{fig:FDgauge}a,~\ref{fig:FDgauge}c,~\ref{fig:FDgauge}e,~\ref{fig:FDgauge}g) shows
        {the difference among the numerical solution and the exact one for the variable  $\alpha$ (lapse) at time $t=1000$}  for various order accurate 
        FD--WENO schemes. The right panel (Figs.~\ref{fig:FDgauge}b,~\ref{fig:FDgauge}d,~\ref{fig:FDgauge}f,~\ref{fig:FDgauge}h) shows the $L_2$-error evolution of the ADM constraints for the corresponding order scheme.}
  		\label{fig:FDgauge}
	\end{center}
\end{figure}

\begin{figure}[h]
	\begin{center}
		\includegraphics[width=0.75\textwidth,clip=]{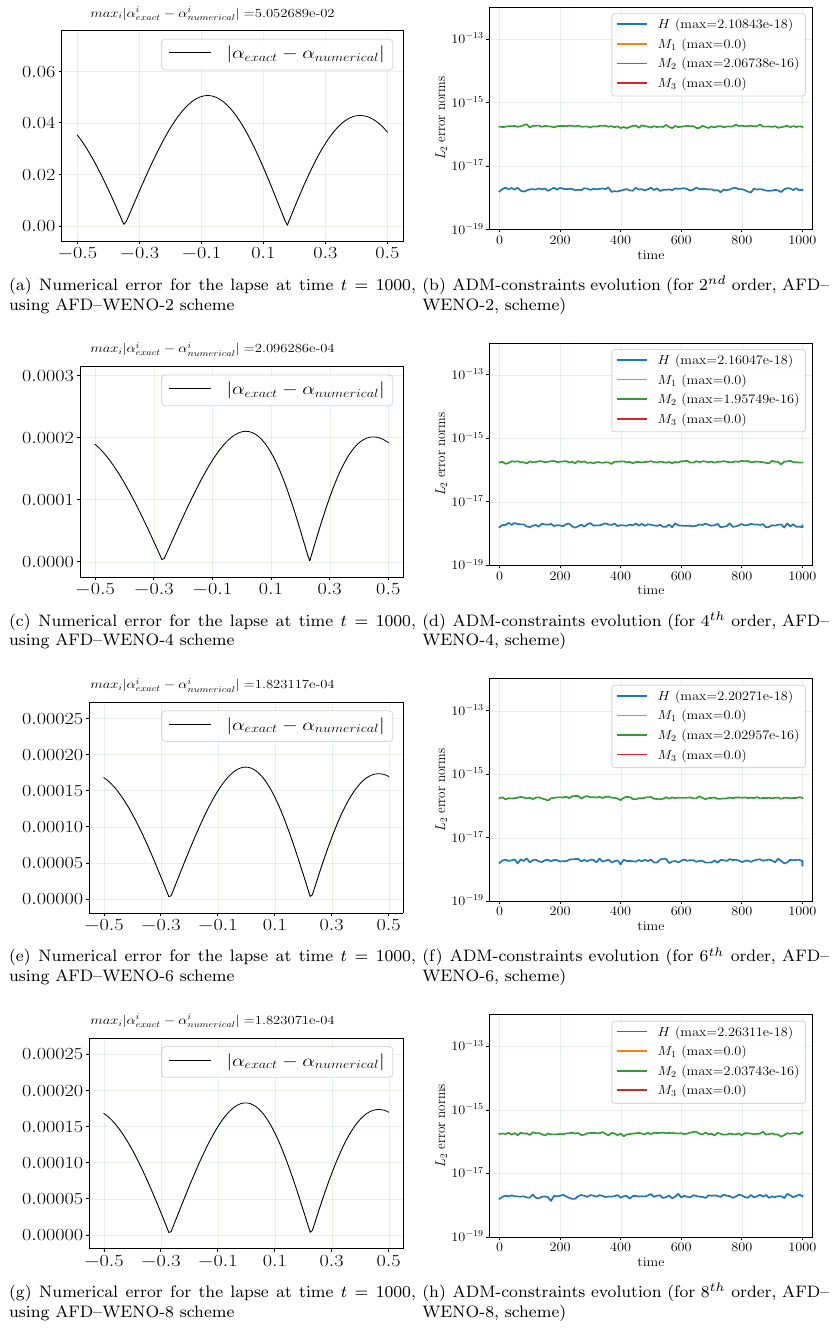}
		\caption{\nameref{sec:gaugewave}: The left panel (Figs.~\ref{fig:AFDgauge}a,~\ref{fig:AFDgauge}c,~\ref{fig:AFDgauge}e,~\ref{fig:AFDgauge}g) shows 
  {the difference among the numerical solution and the exact one for the variable  $\alpha$ (lapse) at time $t=1000$}  for various order accurate 
        FD--WENO schemes. The right panel (Figs.~\ref{fig:AFDgauge}b,~\ref{fig:AFDgauge}d,~\ref{fig:AFDgauge}f,~\ref{fig:AFDgauge}h) shows the $L_2$-error evolution of the ADM constraints for the corresponding order scheme.}
		\label{fig:AFDgauge}
	\end{center}
\end{figure}

Figure~\ref{fig:FDgauge}, on the contrary,
reports the results obtained with FD--WENO schemes,
by showing, in the first column, 
{the numerical errors of the lapse $\alpha$ 
with respect to the exact solution at 
 time $t=1000$}, for various orders, using  a wave amplitude $A=0.1$ {and $128$ zones}.
 In the right column we display the evolution of the ADM constraints,
which remain perfectly under control all along the simulation.
Figure~\ref{fig:AFDgauge} serves the same purpose,
but  this time for the results obtained with AFD--WENO schemes.
{We note that in such a long simulation both FD--WENO-3 and AFD--WENO-2 schemes 
{lead to relatively large errors at the given resolution, 
as it is visible from the numbers} in the top--left panels of 
Fig.~\ref{fig:FDgauge} and Fig.~\ref{fig:AFDgauge}}, with a slightly
better performance of FD--WENO-3 with respect to AFD--WENO-2.
For the higher-order FD--WENO and AFD--WENO schemes ($k=5,7,9$), {the numerical solutions reproduce the analytic ones
with very good accuracy.}

\begin{figure}[h]
	\begin{center}
		\includegraphics[width=0.75\textwidth,clip=]{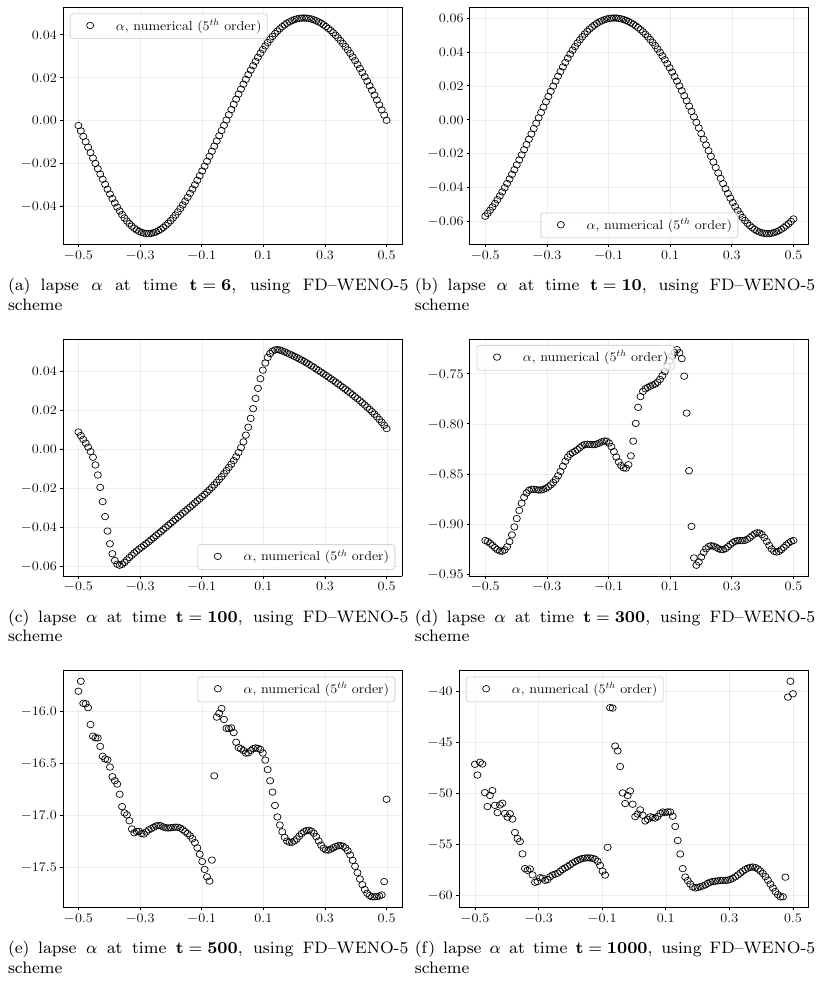}
		\caption{\nameref{sec:gaugewave}: {Time evolution of the lapse profile $\alpha$ at different time levels when \emph{1+log gauge condition} is being used.}}
		\label{fig:FDgauge_shock}
	\end{center}
\end{figure}

{In view of possible studies of critical phenomena 
in gravitational collapse, we have also evolved the same gauge wave but with the \emph{1+log gauge condition}. This gauge condition is known to produce \emph{gauge shocks}~\citep{Alcubierre1997,Baumgarte2022,Baumgarte2023b} which can easily make the code crash. 
The results of our computations are shown in Fig.~\ref{fig:FDgauge_shock}, where strong
gradients are formed in the lapse, but the numerical scheme can still cope with them and the code does not crash, not even after very long times.
Even if at the level of an academic case, 
this test shows the ability of the Z4 formulation, combined with our numerical scheme, to resolve the propagation of gauge shocks. Because the shocks are only of modest strength, interpolation on the primal variables was sufficient; and interpolation in characteristic space was not needed.}

\subsection{Robust stability test}
\label{sec:robuststability}

\begin{figure}[!ht]
   \begin{center}
      \includegraphics[width=0.9\textwidth,clip=]{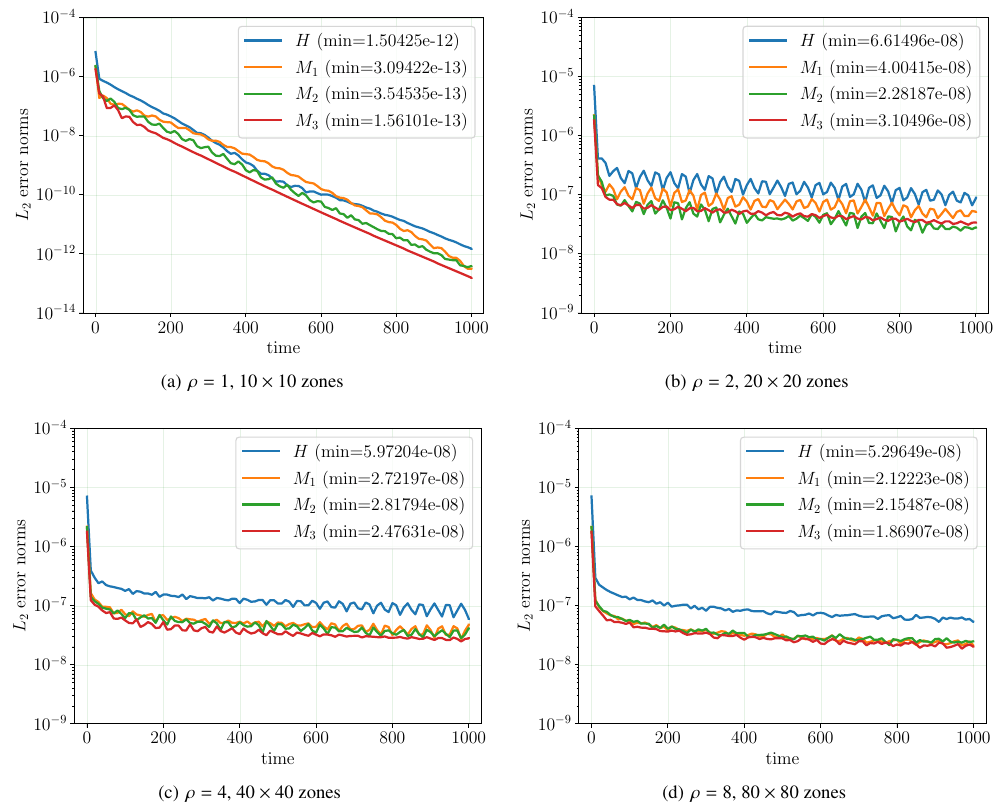}
      \caption{\nameref{sec:robuststability}: Time evolution of $L_2-$norms of Einstein constraints for the Robust Stability Test with a random initial perturbation of amplitude $10^{-7}/\rho^2$ in all quantities on a sequence of successively refined meshes on the square domain in 2D. The final time is $t=1000$, and we used a fifth-order accurate FD-WENO scheme (FD-WENO-5) to obtain the results.}
      \label{fig:FD_RSTAB}
   \end{center}
\end{figure}

\begin{figure}[!ht]
   \begin{center}
      \includegraphics[width=0.9\textwidth,clip=]{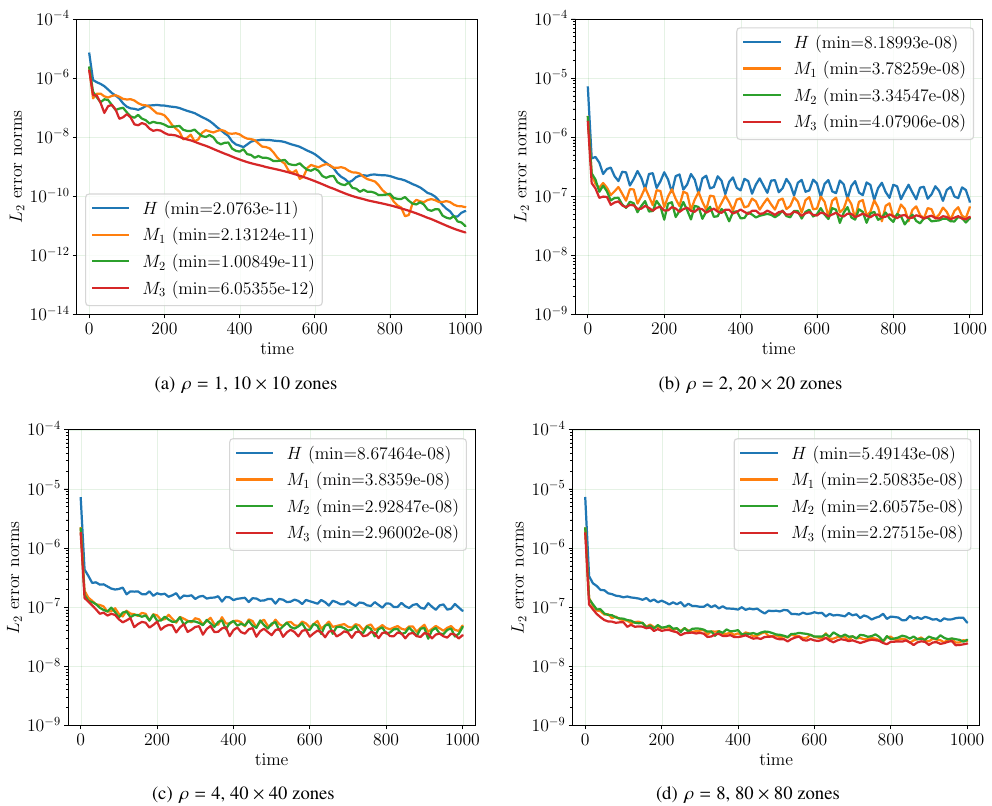}
      \caption{\nameref{sec:robuststability}: Time evolution of $L_2-$norms of Einstein constraints for the Robust Stability Test with a random initial perturbation of amplitude $10^{-7}/\rho^2$ in all quantities on a sequence of successively refined meshes on the square domain in 2D. The final time is $t=1000$, and we used a fourth-order accurate AFD-WENO scheme (AFD-WENO-4) to obtain the results.}
      \label{fig:AFD_RSTAB}
   \end{center}
\end{figure}

The so--called \emph{robust stability test} is used to provide an empiric indication 
of the hyperbolicity of the PDE system, 
and it was introduced when there were very few theoretical analysis of the
second order BSSNOK formulation~\citep{Beyer2004}. In view of the huge
progress witnessed by NR in the last twenty years, and especially because
strong hyperbolicity of the present first--order Z4 formulation has been verified
by \cite{DumbserZanottiGaburroPeshkov2023}, this test has lost part of its
relevance.
Nevertheless, it is still quite useful to exclude possible exponential
unstable modes which could be in principle triggered by a given numerical algorithm~\citep{Calabrese2006}. We should also comment on the fact that, to the best
of our knowledge, none of the first--order formulations of the Einstein equations have shown any problem with this test~\citep{Bona:2004,Brown2012,Dumbser2017strongly,DumbserZanottiGaburroPeshkov2023}.
We have therefore prepared the usual setup, given by a flat  Minkowski spacetime in two space dimensions,
in which all the metric terms are perturbed with 
a random  perturbation of  amplitude $\pm 10^{-7}/\varrho^2$. The integer $\rho$ 
is also used to control the resolution in a set of increasingly refined meshes.
The computational domain is given by the square $[-0.5;0.5]\times[-0.5;0.5]$,
{with periodic boundary conditions along each direction.}
As customary for this test, the \emph{gamma-driver} is activated, with $\eta=0.2$ in Eq.~\eqref{g-driver2}.
The time evolution of the Einstein constraints for the FD-WENO and for the AFD-WENO schemes are shown in Fig.~\ref{fig:FD_RSTAB} and Fig.~\ref{fig:AFD_RSTAB}, respectively,
establishing quite a successful outcome for this test.

\subsection{Gowdy wave}
\label{sec:gowdywave}
There is a relatively simple solution to test the numerical behavior of the PDE system in the strong field regime, and this is given by the so--called Gowdy solution~\citep{Gowdy1971}.
It describes  a polarized gravitational wave (for us along the x-direction) via a spacetime metric given by
\begin{equation}
\label{Gowdymetric}
    ds^2=t^{-1/2}e^{Q/2}(-dt^2+dx^2)+t(e^P dy^2 + e^{-P}dz^2)\,,
\end{equation}
with 
\begin{eqnarray}
    P(x,t)&=&J_0(2\pi t)\cos(2\pi x)\,,\\
    Q(x,t)&=&\pi J_0(2\pi)J_1(2\pi)-2\pi\, t J_0(2\pi t)J_1(2\pi t) \cos^2(2\pi x)+2\pi^2 t^2[ J_0^2(2\pi t) + J_1^2(2\pi t) - J_0^2(2\pi ) - J_1^2(2\pi )]\,,
\end{eqnarray}
where $J_0$ and $J_1$ are the Bessel functions~\citep{Bona:2004,Alcubierre2004}. This test becomes interesting when used to simulate the formation of a singularity. This is better obtained after performing the time change of coordinate $t\rightarrow \tau$, i.e.
\begin{equation}
\label{change-t}
t=t_0 e^{-\tau/\tau_0}\,,
\end{equation}
implying that the singularity is reached in the limit of $\tau\rightarrow +\infty$, $\tau$ now being the new time coordinate of the code.
Since the roots of $J_0(2\pi\,t)$ determines the times at which the lapse is momentarily constant 
(in space), we conform to the standard choice and we select 
$t_0$ as the 20th root of $J_0(2\pi\,t)$, giving $t_0=9.8753205829098$. 
The corresponding value of the lapse resulting from the coordinate transformation \eqref{change-t}
is $\alpha=t^{3/4}\,e^{Q/4}/\tau_0$, and, after imposing $\alpha=1$ at $\tau=0$ ($t=t_0$), we get the value $\tau_0=471.806749033034$.
The time coordinate transformation does not affect the metric terms $\gamma_{ij}$, which follow directly from \eqref{Gowdymetric}, while the extrinsic curvature can be easly computed from
the definition $K_{ij}=-1/(2\alpha)\,\partial_\tau\gamma_{ij}$.
The simulation then starts at $\tau=0$,
reaching the singularity in the limit of $\tau\rightarrow +\infty$ ($t\rightarrow 0)$.

\begin{figure}[!ht]
   \begin{center}
      \includegraphics[width=0.75\textwidth,clip=]{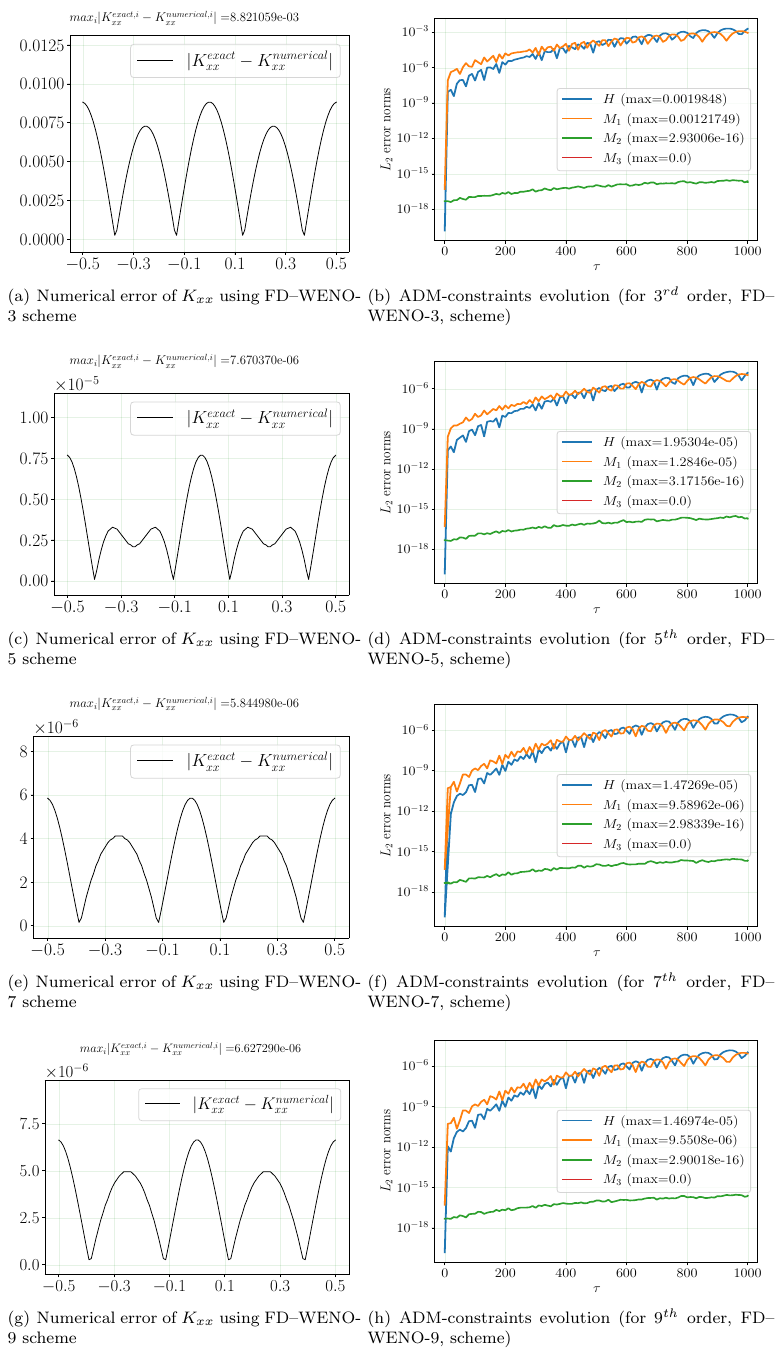}
      \caption{\nameref{sec:gowdywave}: The left panel (Figs.~\ref{fig:FDgowdy}a,\ref{fig:FDgowdy}c,\ref{fig:FDgowdy}e,\ref{fig:FDgowdy}g) shows
      {the difference among the numerical solution and the exact one for the variable  $K_{xx}$ at time $\tau=1000$}
  for various order accurate FD--WENO schemes. The right panel (Figs.~\ref{fig:FDgowdy}b,\ref{fig:FDgowdy}d,\ref{fig:FDgowdy}f,\ref{fig:FDgowdy}h) shows the $L_2$-error evolution of the ADM constraints at the corresponding order.}
      \label{fig:FDgowdy}
   \end{center}
\end{figure}

\begin{figure}[!ht]
   \begin{center}
        \includegraphics[width=0.75\textwidth,clip=]{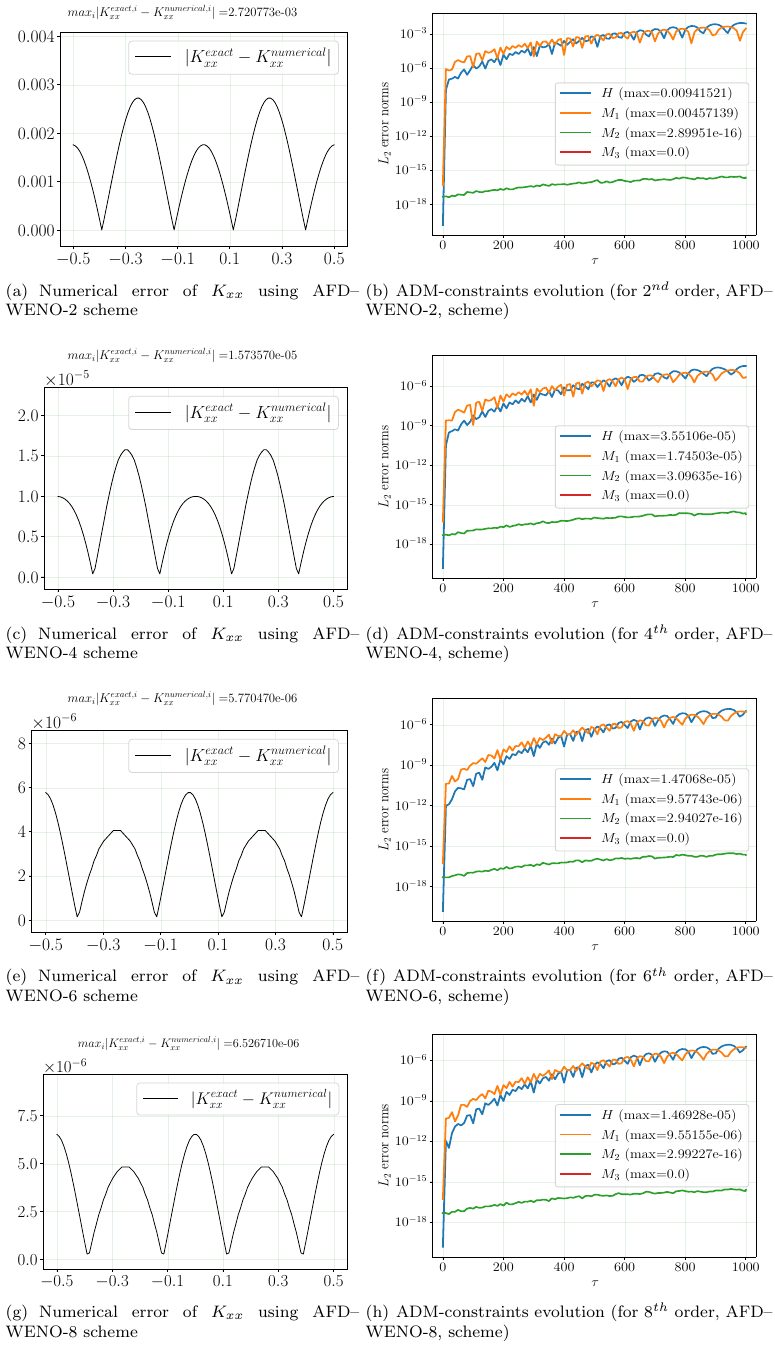}
      \caption{\nameref{sec:gowdywave}: The left panel (Figs.~\ref{fig:AFDgowdy}a,\ref{fig:AFDgowdy}c,\ref{fig:AFDgowdy}e,\ref{fig:AFDgowdy}g) shows
{the difference among the numerical solution and the exact one for the variable  $K_{xx}$ at time $\tau=1000$}      
 for various order accurate AFD--WENO schemes. The right panel (Figs.~\ref{fig:AFDgowdy}b,\ref{fig:AFDgowdy}d,\ref{fig:AFDgowdy}f,\ref{fig:AFDgowdy}h) shows the $L_2$-error evolution of the ADM constraints at the corresponding order.}
      \label{fig:AFDgowdy}
   \end{center}
\end{figure}

We have solved this test as a one-dimensional problem using a mesh composed of 120 zones, up to the final time $\tau=1000$. We impose periodic boundary conditions.
In Fig.~\ref{fig:FDgowdy} and Fig.~\ref{fig:AFDgowdy} we show the results obtained with our FD-WENO and AFD-WENO schemes, respectively. The left panels
report 
{the numerical error of the variable $K_{xx}$ at the final time.} The right panels, on the other hand, show the evolution of the Einstein constraints, which, in spite of a linear increase already documented in the literature~\citep{Bona:2004}, remain under control.

\subsection{Stationary black holes}
\label{sec:BH}
Contrary to naive expectations, keeping  
a simple Schwarzschild or Kerr black hole stationary
in three dimensions via a NR code is far from trivial.
{In this subsection we refer to the stationary BH solutions of Schwarzschild (1916) and Kerr (1963), both of them in horizon penetrating coordinates,
and not to puncture black holes.}. Within the second order BSSNOK formulation,
promising attempts were
performed by \cite{Alcubierre2001}, who, however, performed the calculations on one octant only
(because of spherical symmetry) while they reported the appearance of an unstable mode
when they extended the simulation in the full grid space.
{A substantial progress is reported in \cite{Tichy2009}, with the Hamiltonian constraint that, however, manifests an exponential growth on long timescales.
}
In 2001, after using an alternative first--order formulation of the Einstein equations, henceforth the KST formulation, \cite{Kidder2001} were able to 
evolve a Schwarzschild black hole in three space dimensions, though
the momentum constraints did not seem to be fully under control.
Better results were obtained  by \cite{Bruegmann2013}, who 
kept a Schwarzschild black hole in spherical Kerr--Schild coordinates stationary 
for very long timescales, using spectral methods 
in combination with a first--order formulation of the Einstein equations in the generalized harmonic gauge. 
Again in this formulation, \cite{Tichy2023} obtained rather good results as well for a non-rotating black hole, but using DG methods. 
Finally, after using a well--balanced version of the 
first--order Z4 formulation of the Einstein equations, in combination with an ADER--DG scheme,
\cite{DumbserZanottiGaburroPeshkov2023} were able to preserve
perfect stationary equilibria of even extreme Kerr black holes over arbitrarily long timescales.

\begin{figure}[!ht]
   \begin{center}
      \subfigure[$\alpha$, lapse profile at $t=1000$ (slice plot) and vector field of the shift.]{
         \includegraphics[width=0.55\textwidth,clip=]{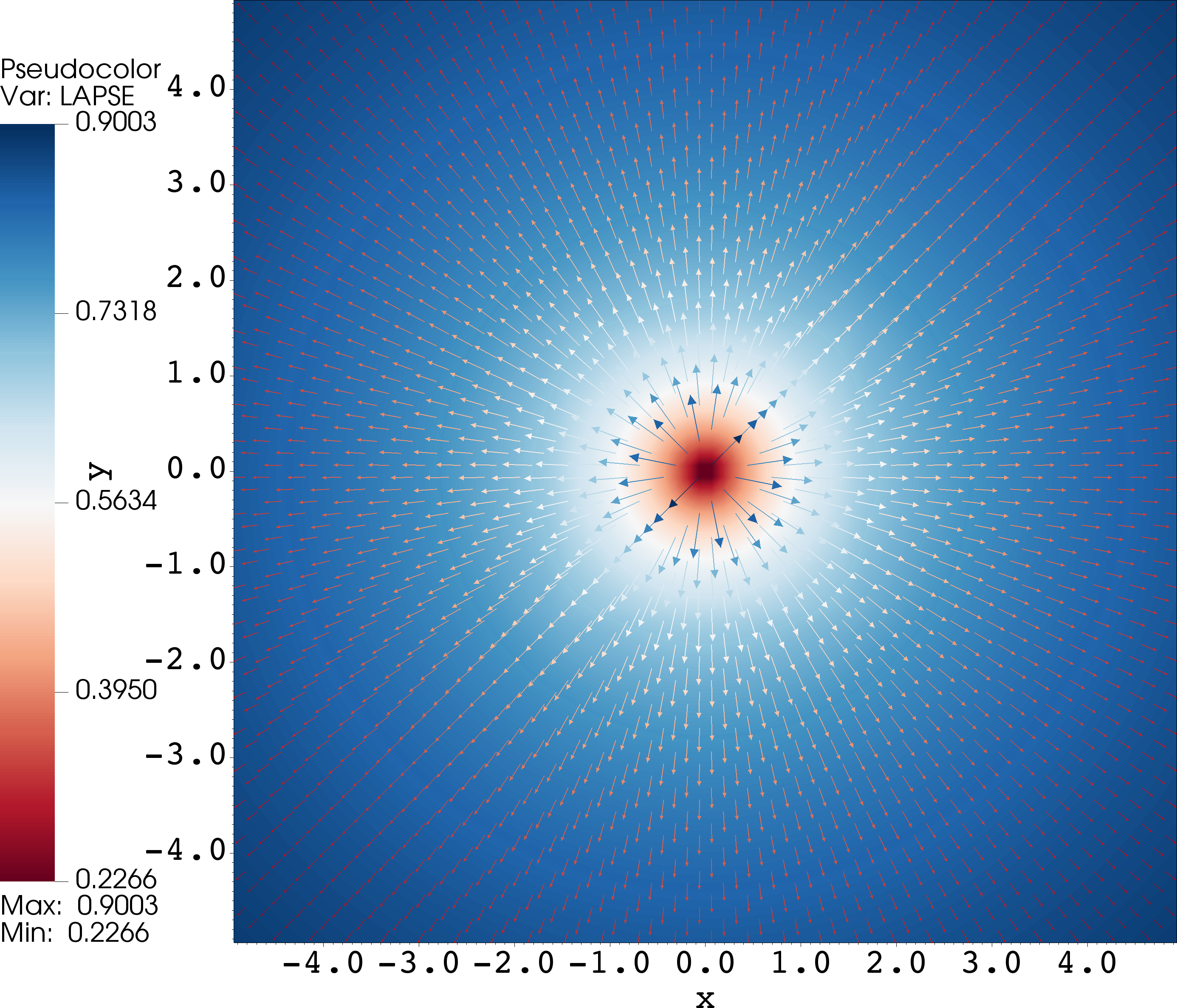}
         \label{subfig:BH3D_FD_0p0_lapse_slice}
            }
      \subfigure[{profile of the lapse error along $x=y=0$.}]{
         \includegraphics[width=0.45\textwidth,clip=]{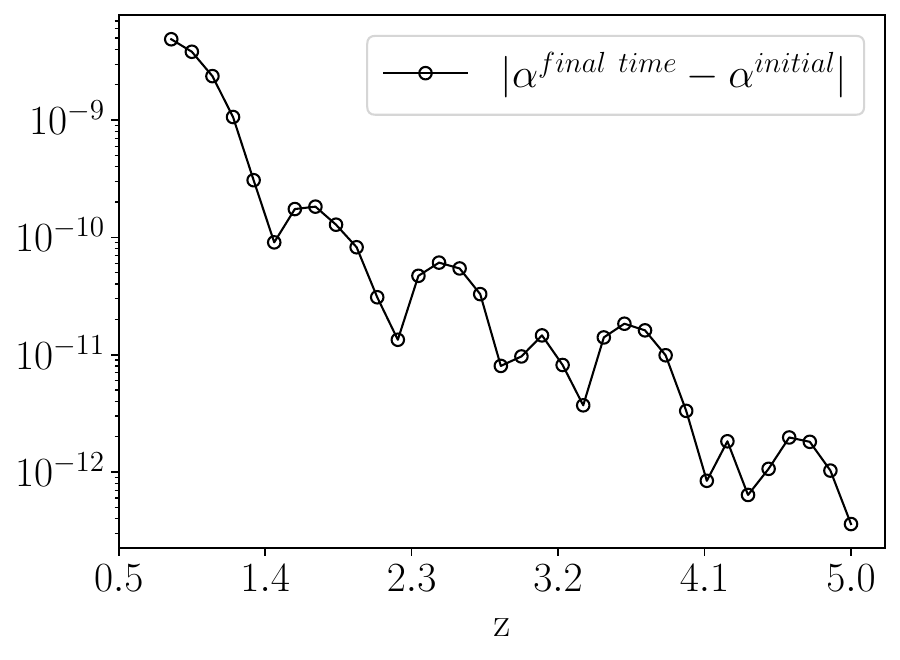}
         \label{subfig:BH3D_FD_0p0_lapse_cut}
            }
      \subfigure[{profile of the $K_{zz}$ error along $x=y=0$.}]{
         \includegraphics[width=0.45\textwidth,clip=]{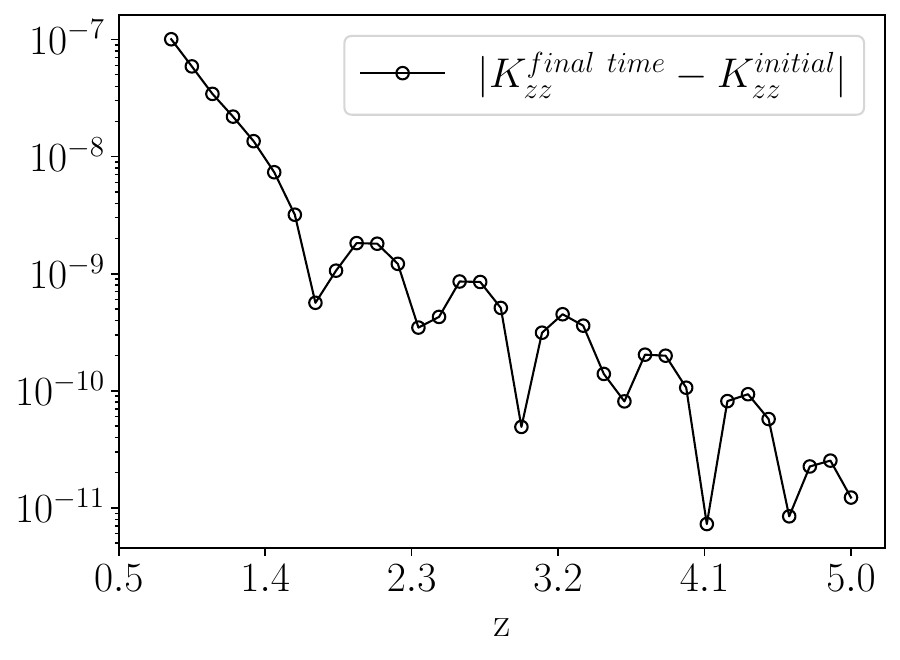}
         \label{subfig:BH3D_FD_0p0_K33_cut}
            }
      \subfigure[{profile of the $\gamma_{zz}$ error along $x=y=0$.}]{
         \includegraphics[width=0.45\textwidth,clip=]{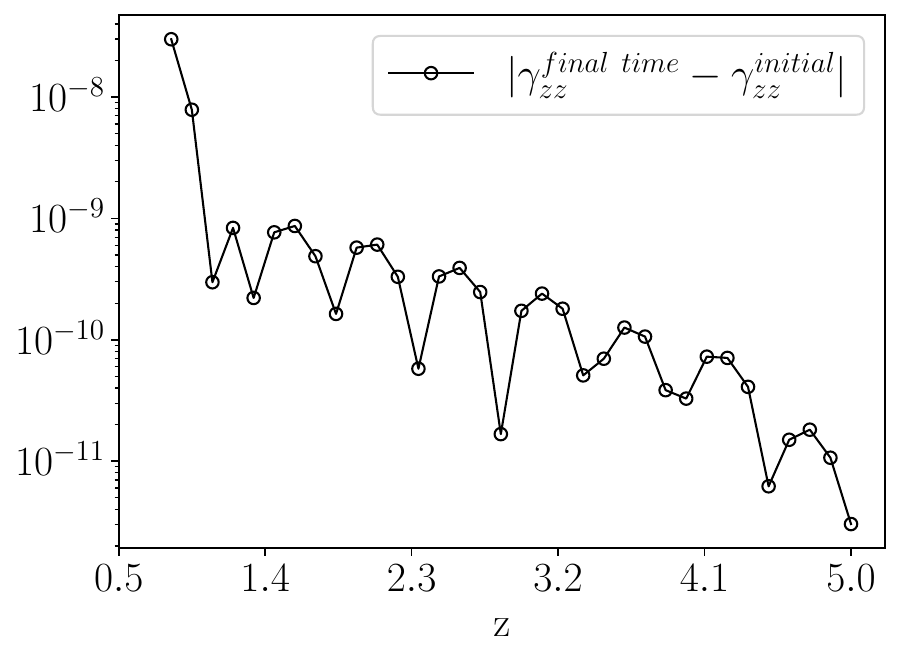}
         \label{subfig:BH3D_FD_0p0_G33_cut}
            }
      \subfigure[Evolution of $L_2$-error norms for the ADM-constraints]{
         \includegraphics[width=0.45\textwidth,clip=]{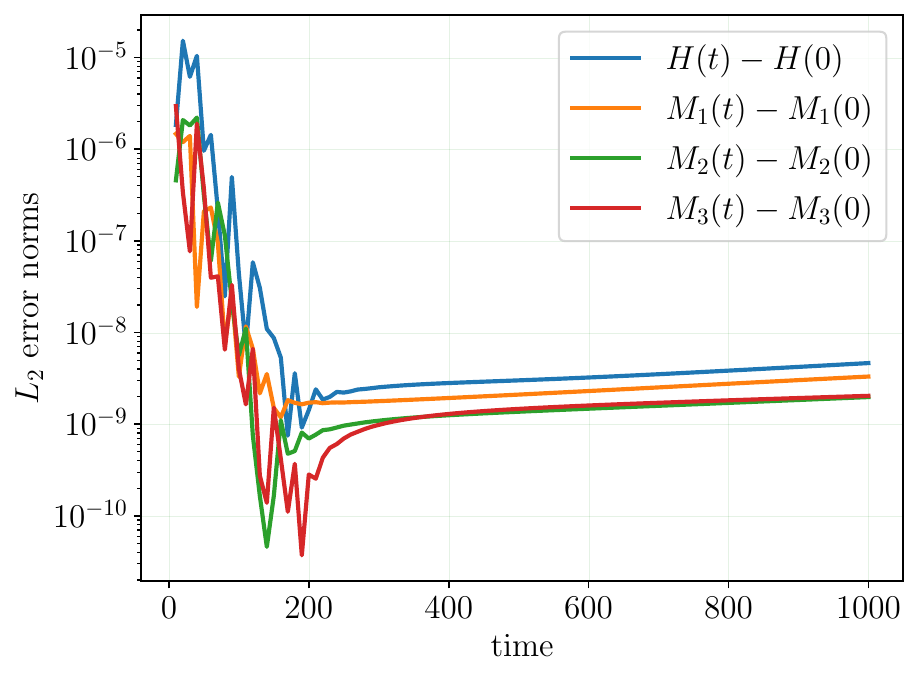}
         \label{subfig:BH3D_FD_0p0_ADM}
            }
      \caption{Schwarzschild black hole: Figs.~\ref{fig:BH3D_FD_aom_0p0}a-\ref{fig:BH3D_FD_aom_0p0}d show {the absolute errors of $\alpha$, $K_{zz}$ and $\gamma_{zz}$ for the Schwarzschild black hole ($a=0$), computed with respect to the exact solution at the final time $t=1000$. Fig.~\ref{fig:BH3D_FD_aom_0p0}e shows the violation of the ADM-constraints.} A seventh-order accurate FD-WENO scheme has been used.}
      \label{fig:BH3D_FD_aom_0p0}
   \end{center}
\end{figure}

\begin{figure}[!ht]
   \begin{center}
      \subfigure[$\alpha$, lapse profile at $t=1000$ (slice plot) and vector field of the shift.]{
         \includegraphics[width=0.55\textwidth,clip=]{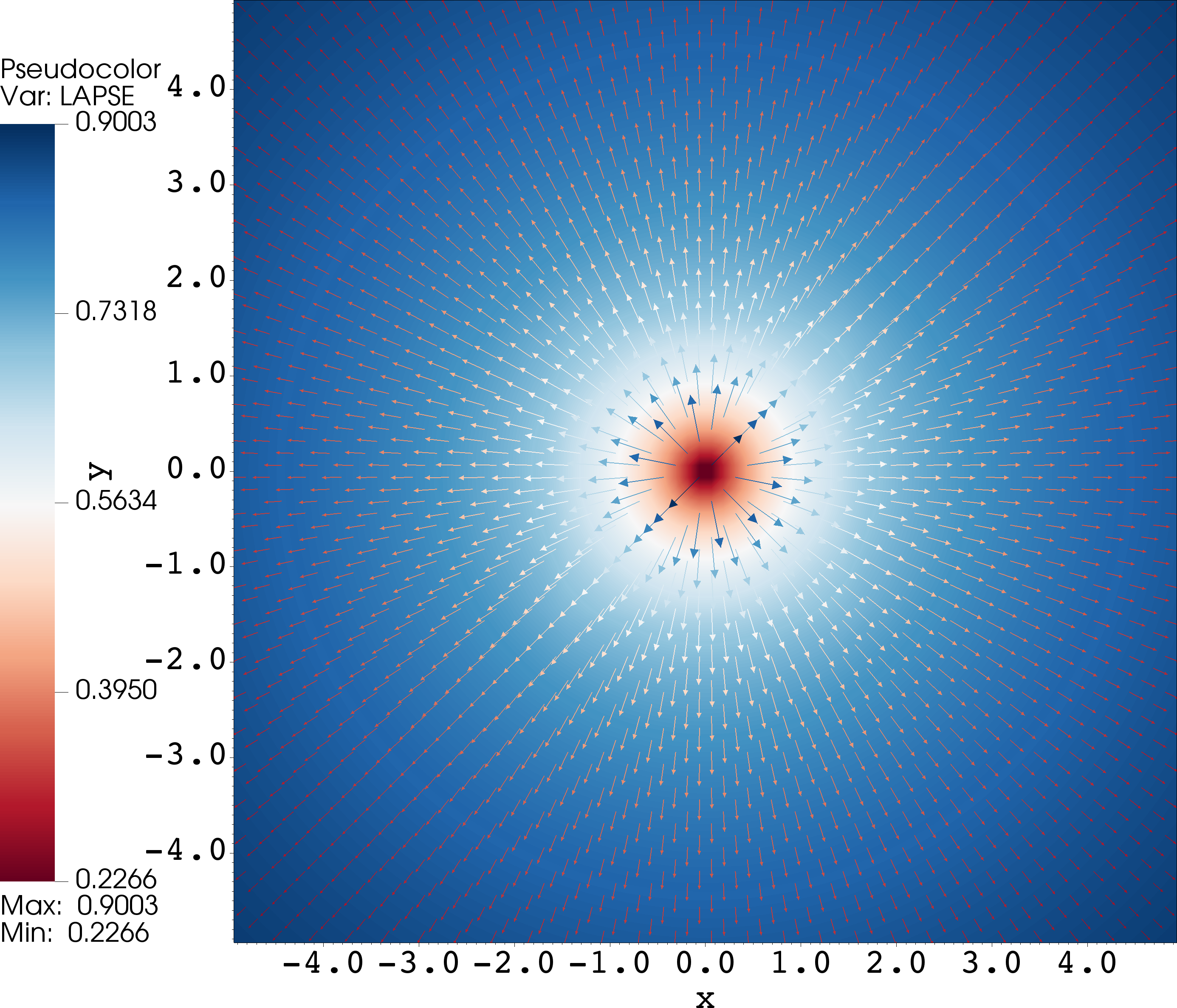}
         \label{subfig:BH3D_AFD_0p0_lapse_slice}
            }
      \subfigure[{profile of the lapse error along $x=y=0$.}]{
         \includegraphics[width=0.45\textwidth,clip=]{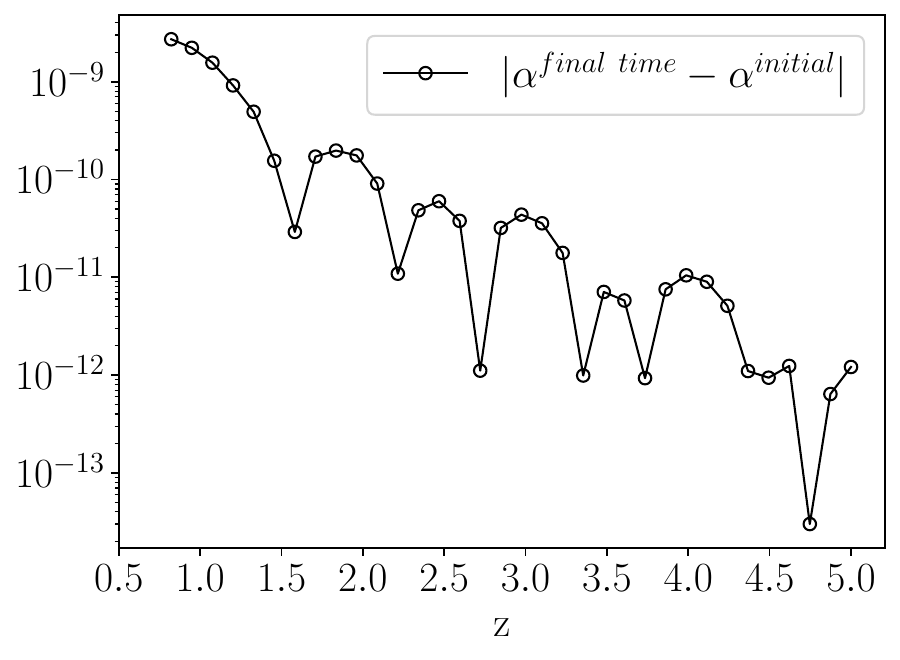}
         \label{subfig:BH3D_AFD_0p0_lapse_cut}
            }
      \subfigure[{profile of the $K_{zz}$ error along $x=y=0$.}]{
         \includegraphics[width=0.45\textwidth,clip=]{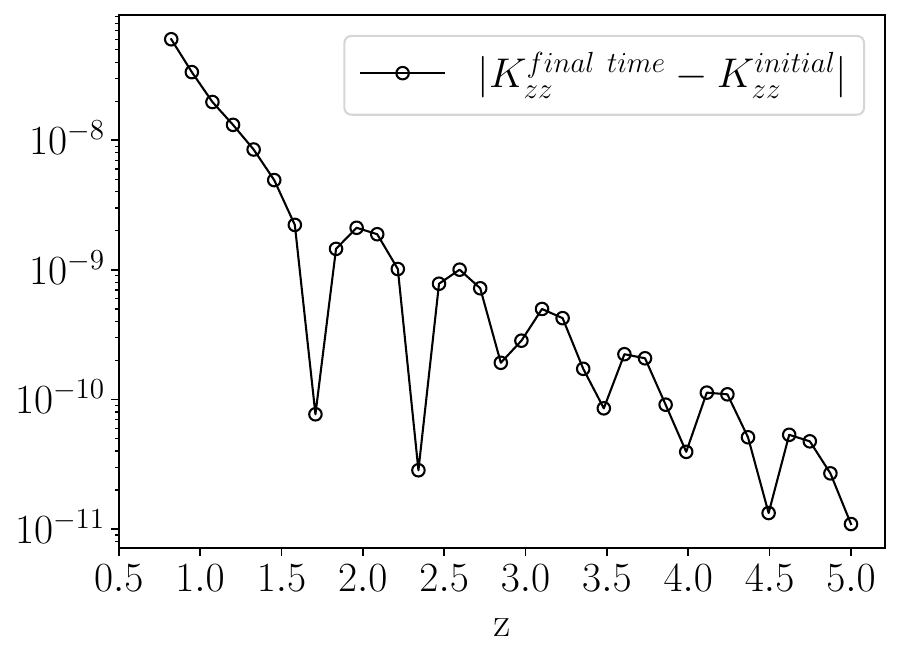}
         \label{subfig:BH3D_AFD_0p0_K33_cut}
            }
      \subfigure[{profile of the $\gamma_{zz}$ error along $x=y=0$.}]{
         \includegraphics[width=0.45\textwidth,clip=]{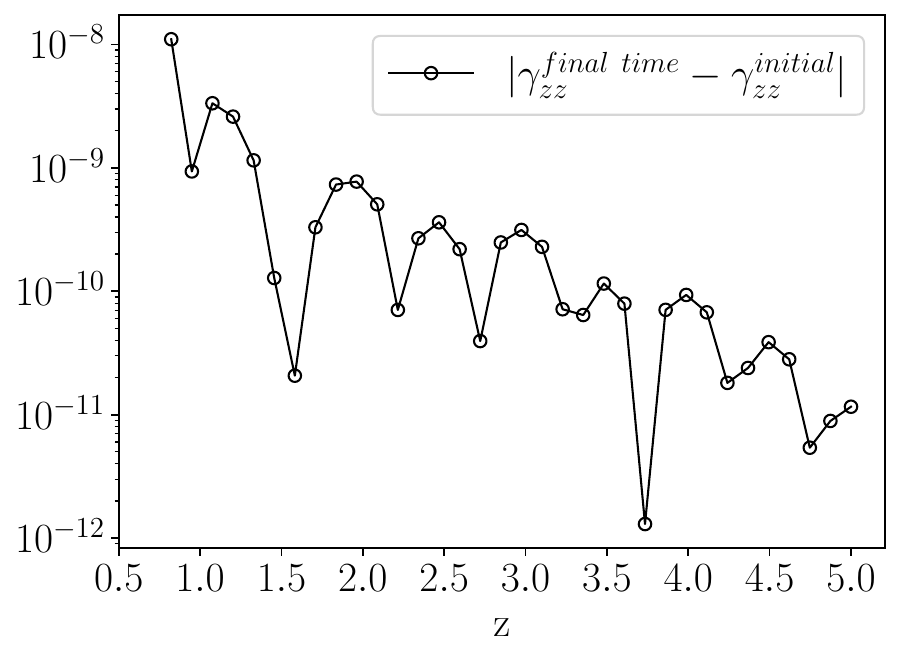}
         \label{subfig:BH3D_AFD_0p0_G33_cut}
            }
      \subfigure[Evolution of $L_2$-error norms for the ADM-constraints]{
         \includegraphics[width=0.45\textwidth,clip=]{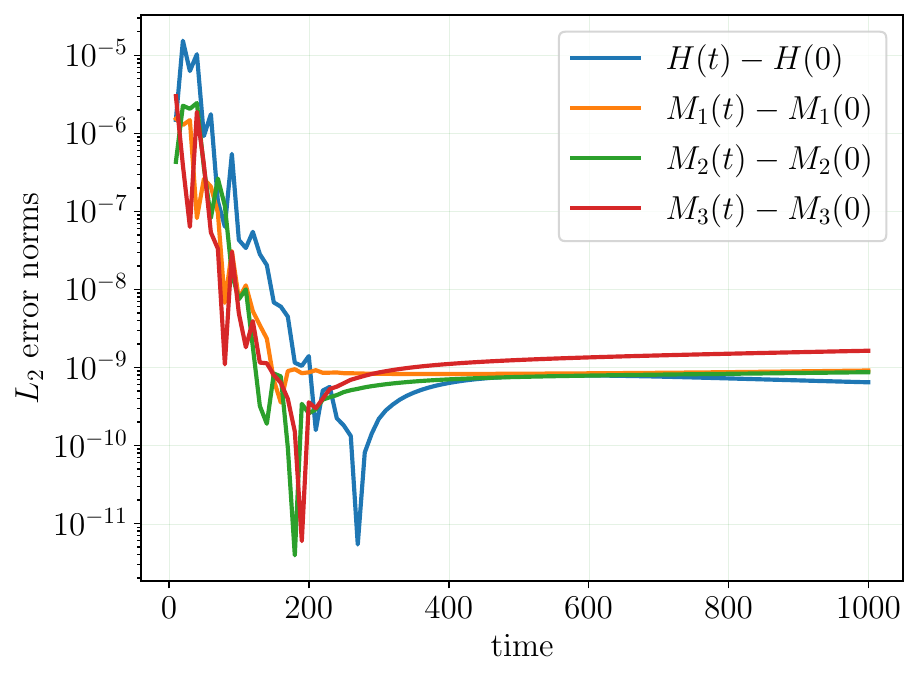}
         \label{subfig:BH3D_AFD_0p0_ADM}
            }
      \caption{Schwarzschild black hole: Figs.~\ref{fig:BH3D_AFD_aom_0p0}a-\ref{fig:BH3D_AFD_aom_0p0}d show {the absolute errors of $\alpha$, $K_{zz}$ and $\gamma_{zz}$ for the Schwarzschild black hole ($a=0$), computed with respect to the exact solution at the final time $t=1000$. Fig.~\ref{fig:BH3D_AFD_aom_0p0}e shows the violation of the ADM-constraints.} A sixth-order accurate AFD-WENO scheme has been used.}
      \label{fig:BH3D_AFD_aom_0p0}
   \end{center}
\end{figure}

\begin{figure}[!ht]
   \begin{center}
      \subfigure[$\alpha$, lapse profile at $t=1000$ (slice plot) and vector field of the shift.]{
         \includegraphics[width=0.55\textwidth,clip=]{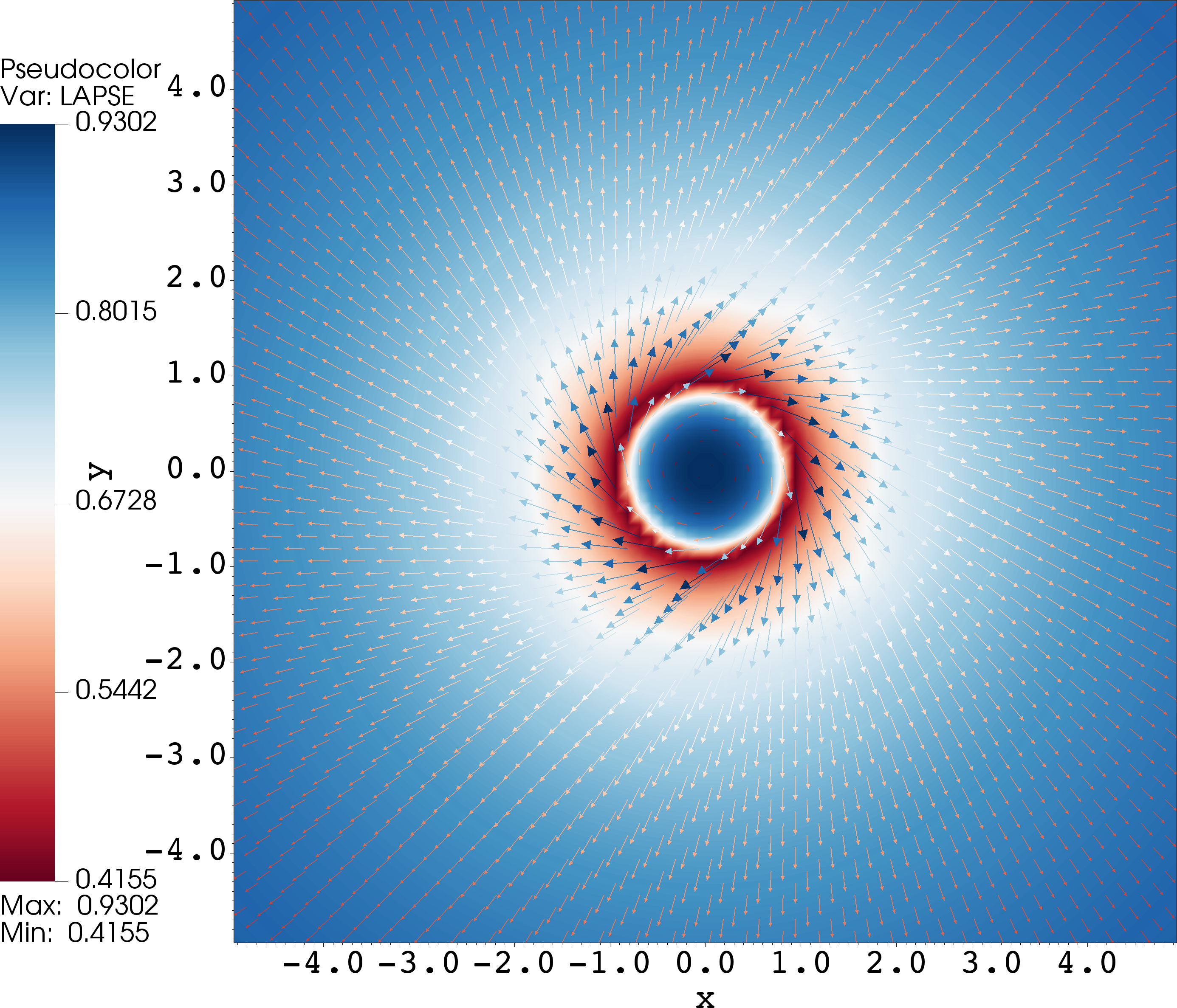}
         \label{subfig:BH3D_FD_0p9_lapse_slice}
            }
      \subfigure[{profile of the lapse error along $x=y=0$.}]{
         \includegraphics[width=0.45\textwidth,clip=]{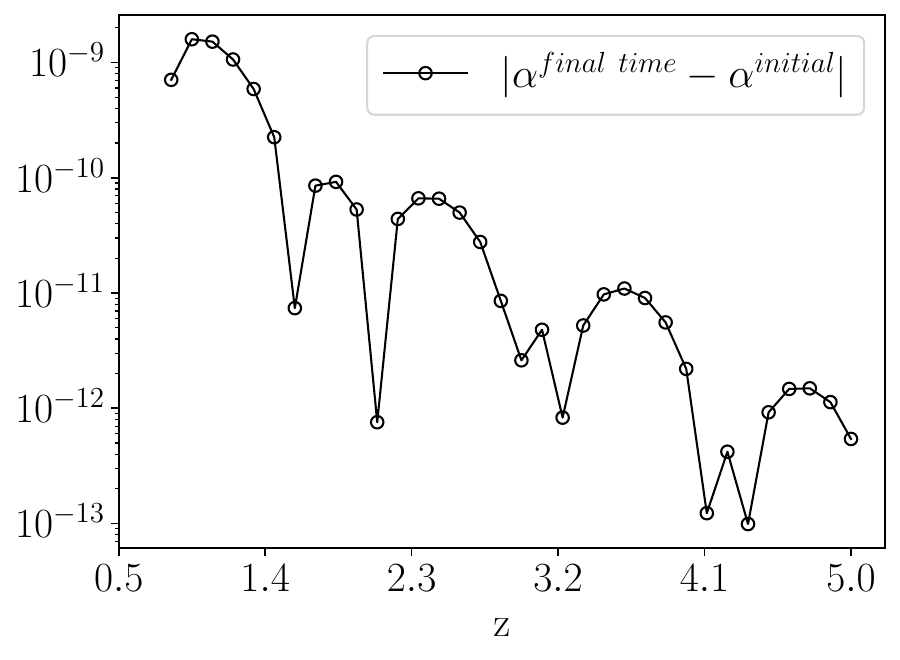}
         \label{subfig:BH3D_FD_0p9_lapse_cut}
            }
      \subfigure[{profile of the $K_{zz}$ error along $x=y=0$.}]{
         \includegraphics[width=0.45\textwidth,clip=]{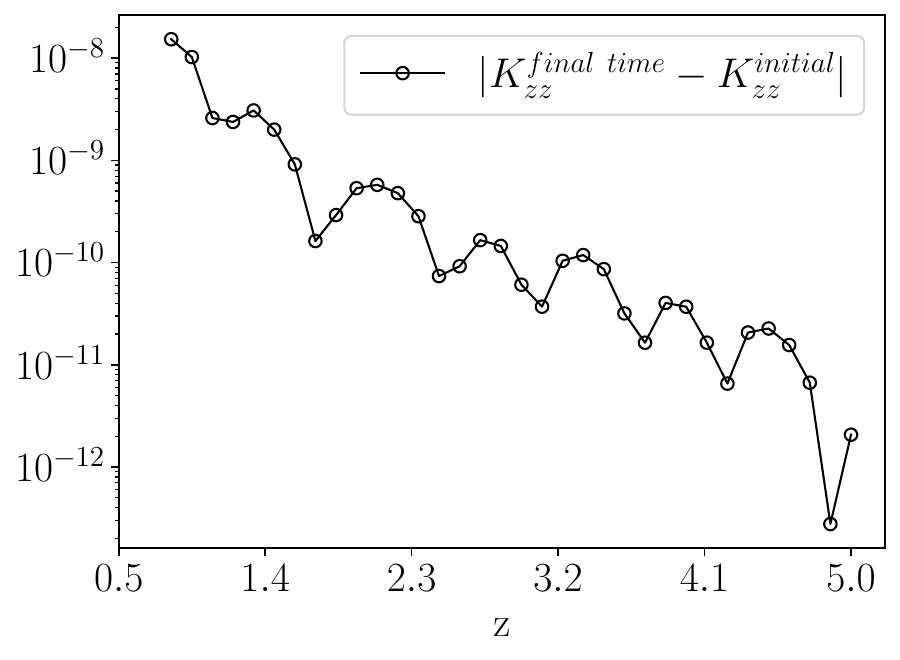}
         \label{subfig:BH3D_FD_0p9_K33_cut}
            }
      \subfigure[{profile of the $\gamma_{zz}$ error along $x=y=0$.}]{
         \includegraphics[width=0.45\textwidth,clip=]{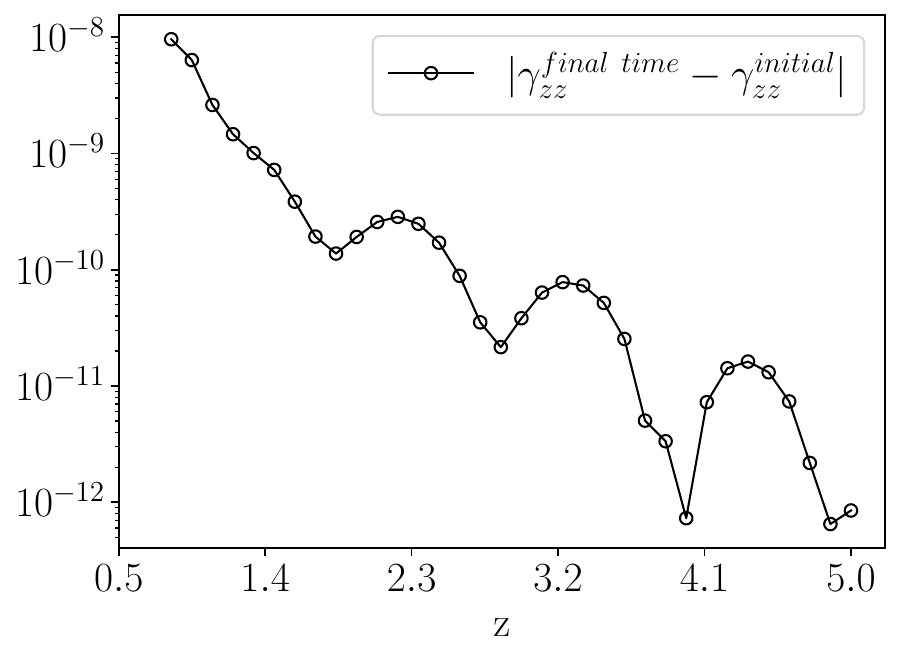}
         \label{subfig:BH3D_FD_0p9_G33_cut}
            }
      \subfigure[Evolution of $L_2$-error norms for the ADM-constraints]{
         \includegraphics[width=0.45\textwidth,clip=]{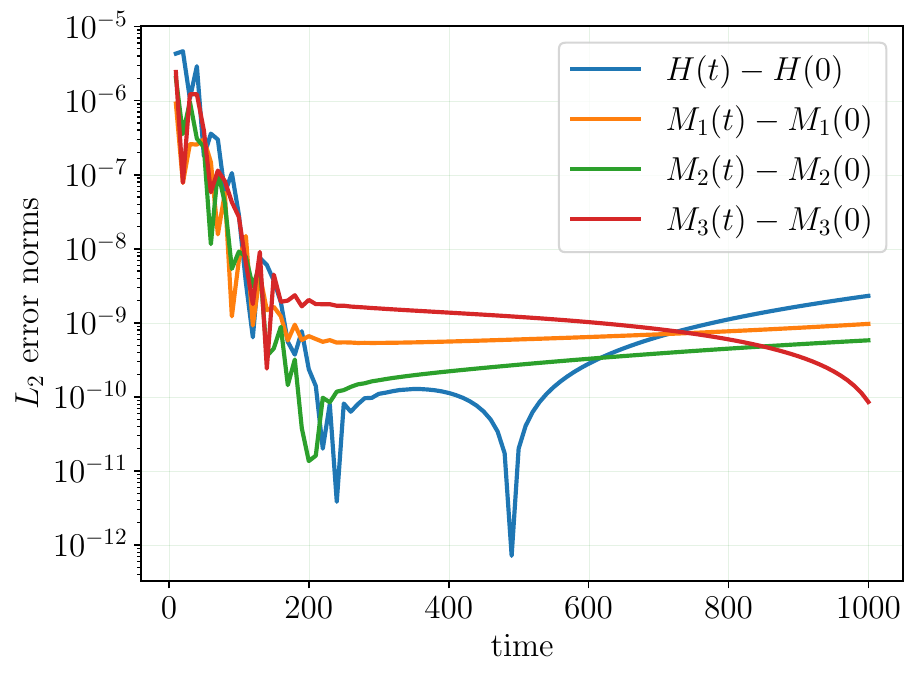}
         \label{subfig:BH3D_FD_0p9_ADM}
            }
      \caption{Kerr black hole: Figs.~\ref{fig:BH3D_FD_aom_0p9}a-\ref{fig:BH3D_FD_aom_0p9}d show {the absolute errors of $\alpha$, $K_{zz}$ and $\gamma_{zz}$ for the Kerr black hole ($a=0.9$), computed with respect to the exact solution at the final time $t=1000$. Fig.~\ref{fig:BH3D_FD_aom_0p9}e shows the violation of the ADM-constraints.} A seventh-order accurate FD-WENO scheme has been used.}
      \label{fig:BH3D_FD_aom_0p9}
   \end{center}
\end{figure}

\begin{figure}[!ht]
   \begin{center}
      \subfigure[$\alpha$, lapse profile at $t=1000$ (slice plot) and vector field of the shift.]{
         \includegraphics[width=0.55\textwidth,clip=]{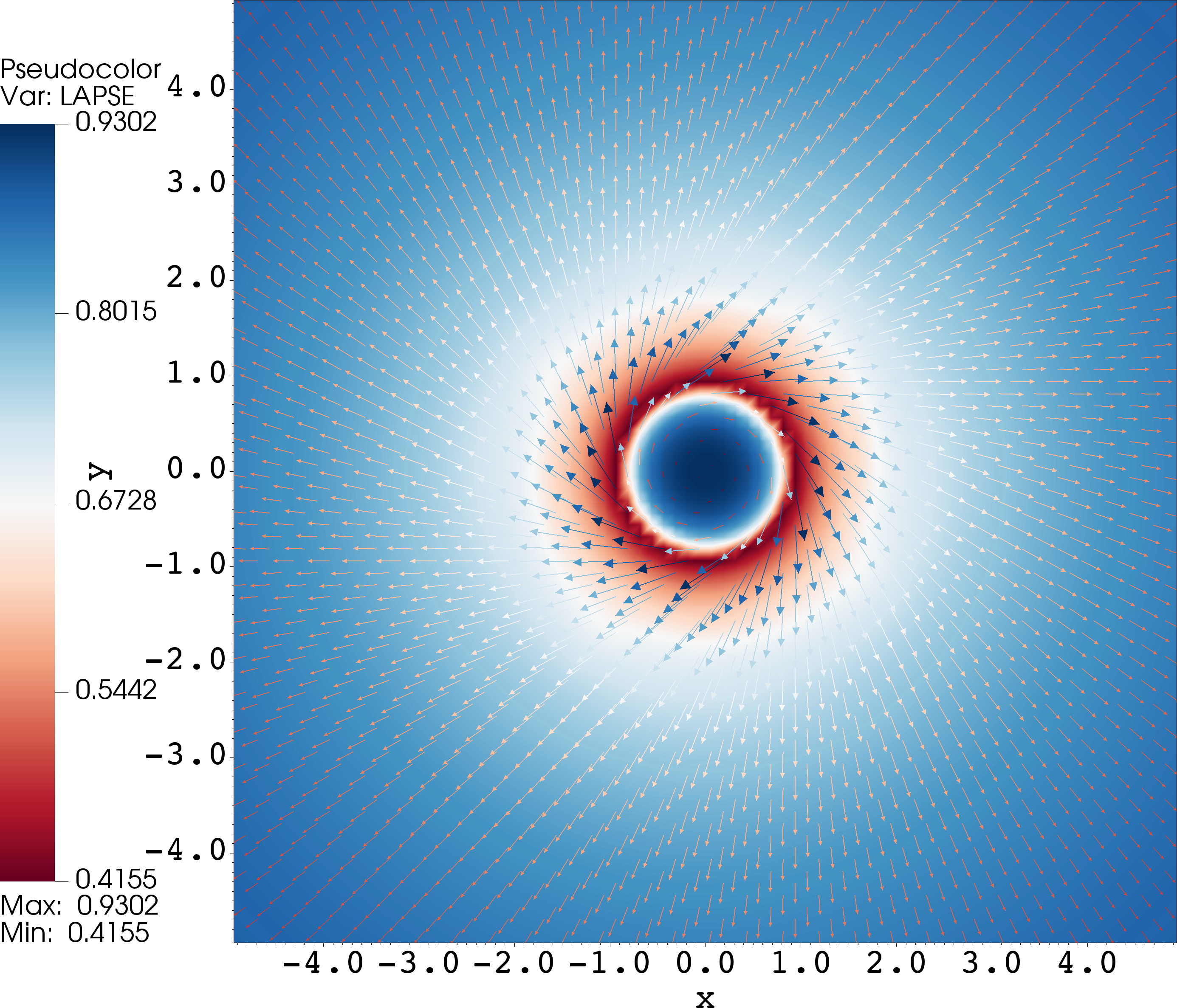}
         \label{subfig:BH3D_AFD_0p9_lapse_slice}
            }
      \subfigure[{profile of the lapse error along $x=y=0$.}]{
         \includegraphics[width=0.45\textwidth,clip=]{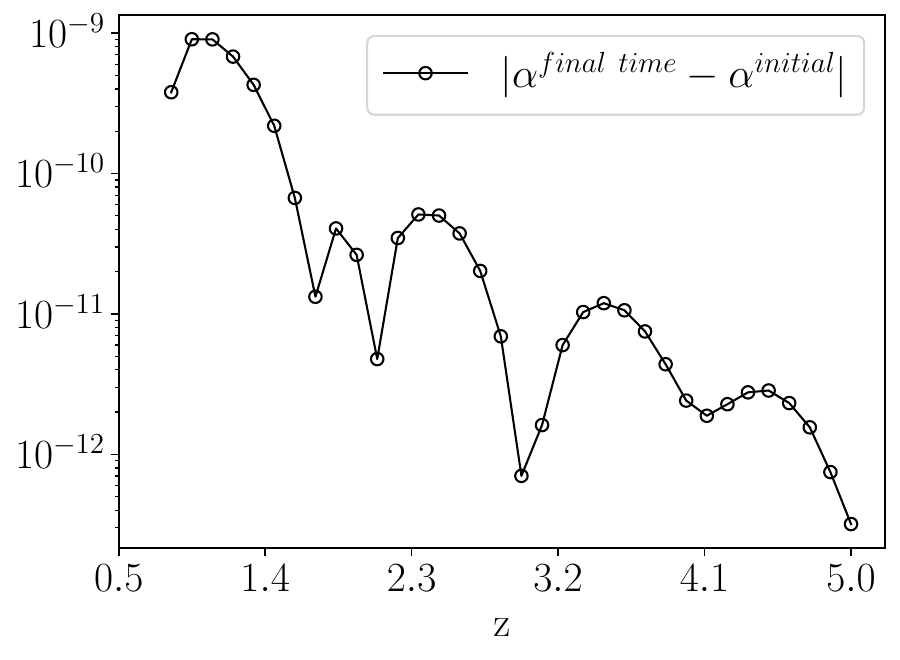}
         \label{subfig:BH3D_AFD_0p9_lapse_cut}
            }
      \subfigure[{profile of the $K_{zz}$ error along $x=y=0$.}]{
         \includegraphics[width=0.45\textwidth,clip=]{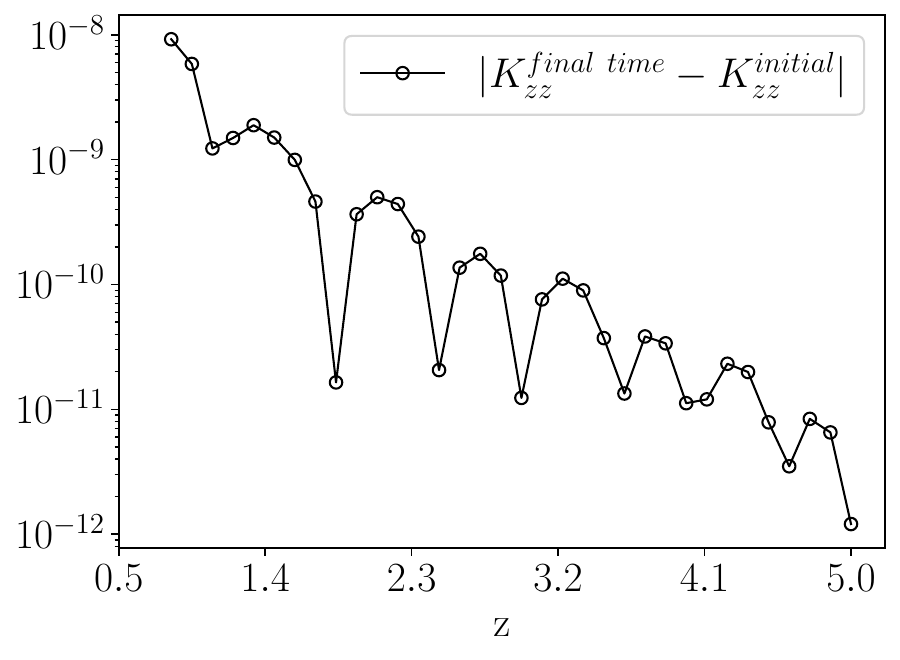}
         \label{subfig:BH3D_AFD_0p9_K33_cut}
            }
      \subfigure[{profile of the $\gamma_{zz}$ error along $x=y=0$.}]{
         \includegraphics[width=0.45\textwidth,clip=]{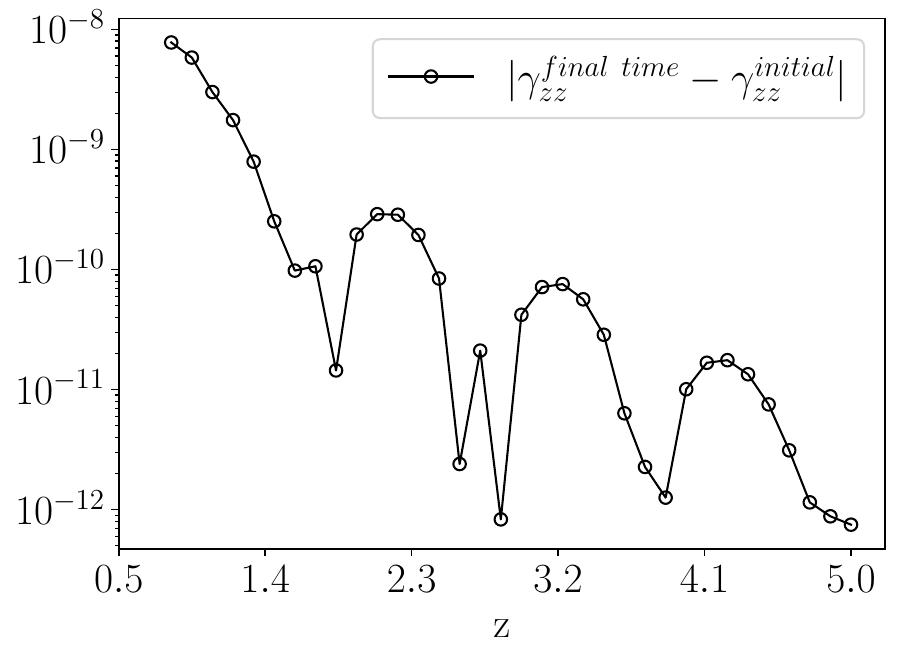}
         \label{subfig:BH3D_AFD_0p9_G33_cut}
            }
      \subfigure[Evolution of $L_2$-error norms for the ADM-constraints]{
         \includegraphics[width=0.45\textwidth,clip=]{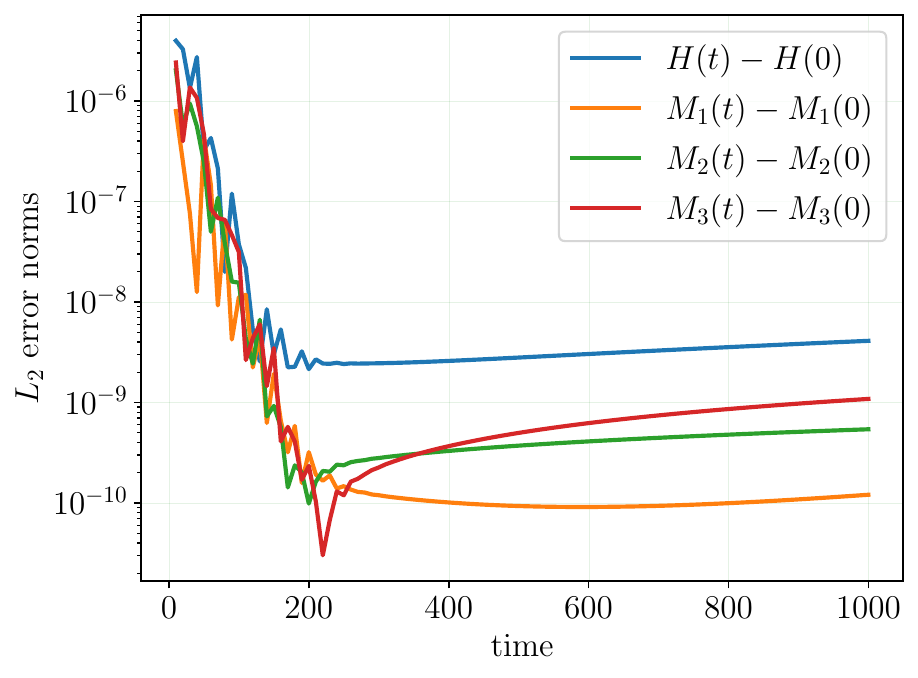}
         \label{subfig:BH3D_AFD_0p9_ADM}
            }
      \caption{Kerr black hole: Figs.~\ref{fig:BH3D_AFD_aom_0p9}a-\ref{fig:BH3D_AFD_aom_0p9}d show {the absolute errors of $\alpha$, $K_{zz}$ and $\gamma_{zz}$ for the Kerr black hole ($a=0.9$), computed with respect to the exact solution at the final time $t=1000$. Fig.~\ref{fig:BH3D_AFD_aom_0p9}e shows the violation of the ADM-constraints.} A sixth-order accurate AFD-WENO scheme has been used.}
      \label{fig:BH3D_AFD_aom_0p9}
   \end{center}
\end{figure}

We have therefore considered the stationary solution represented by the Kerr spacetime in pseudo-Cartesian Kerr--Schild coordinates, as they can be found, for instance, in \cite{deFelice90} (Sect. 11.4) or in \cite{Visser2009}.
Here, we reconsider that problem by adopting a seventh-order FD-WENO scheme and a sixth-order AFD-WENO scheme, applied in three-dimensional (spatial) simulations.
The three-dimensional computational domain is given by $\Omega = [-5;5]^3$ covered by a $80\times 80 \times 80$ uniform grid, while, as boundary conditions, we impose the stationary equilibrium solution at the outer border. 
{
The gauge condition adopted is the \emph{1+log gauge condition}.}
We also recall that, in the coordinates adopted,
the physical singularity is mapped into the the so--called ring singularity in
z = 0 plane, with a radius given by the spin of the black hole. For this reason we have
excised the ring singularity from the numerical domain, which remains inside an appropriate excision box.
{
Moreover, we insert an initial perturbation in the cleaning variable $\Theta$, which  is chosen as 
\begin{equation}
	\Theta(t=0) = A_0 \exp \left( -\halb \frac{(x-2.5)^2 + (y-0)^2 + (z-0)^2}{\sigma^2} \right)\,,
\end{equation}   
with  $A_0 = 10^{-3}$, $\sigma=0.2$. 
}
We have found crucial for a successful evolution of these equilibrium tests
over arbitrarily long times
to activate the \emph{well balancing} property described in Sect.~\ref{subsec:WB}. Without 
well balancing, small but ever growing deviations from equilibrium where detected on times $t\geq 100$, even at the sixth order of convergence of our numerical schemes.
Under these settings, we have considered the following two representative cases:
\begin{itemize}
\item {\it{A Schwarzschild black hole}} ($a=0$). Fig.~\ref{fig:BH3D_FD_aom_0p0} and Fig.~\ref{fig:BH3D_AFD_aom_0p0} show the results of 
a long run with $t_f=1000\,M$, obtained with the  FD-WENO scheme and with the AFD-WENO scheme, respectively. 
The central top panel reports a contour-color representation of the lapse $\alpha$ through the $z=0$ plane, while the arrows visualize the vector field of the shift $\beta^i$, which, in spite of the
black hole being non/rotating, is non zero due to the choice of the coordinates.
{
The remaining panels, with 1D profiles along the $z$ axis, show the numerical error of three representative quantities at the final time $t=1000$ with respect to the initial equilibrium solution.}
\item {\it{A Kerr black hole}} ($a=0.9$).
In a rather similar way,
Fig.~\ref{fig:BH3D_FD_aom_0p9} and Fig.~\ref{fig:BH3D_AFD_aom_0p9} show the results of  a long run with $t_f=1000\,M$, obtained with the  FD-WENO scheme and with the AFD-WENO scheme, respectively.
The vector field of the shift, represented in the central top panels of those figures, manifests
the clockwise rotation of the black hole.
\end {itemize}
Finally, the bottom right panels of all figures from Fig.~\ref{fig:BH3D_FD_aom_0p0} to  Fig.~\ref{fig:BH3D_AFD_aom_0p9}
monitor the evolution of the 
{
constraint violations $H(t) - H(0)$
and $M_i(t) - M_i(0)$ for the Hamiltonian and the momentum constraints. 
As it can be seen,
the effects of the introduced perturbation decay 
exponentially and the solution converges  back to the exact equilibrium. These results confirm
what also found by \cite{DumbserZanottiGaburroPeshkov2023}.
}
%

\subsection{Head on collision of two black holes}
\label{sec:HeadOnCollision}

The correct simulation of binary mergers (either black holes or neutron stars) is of course
the final goal of numerical relativity, which is proving to be an invaluable scientific 
tool in the preset gravitational wave era~\citep{Abbott2019}. 
In this work, still at the level of an exploratory
calculation, before a dedicated analysis is performed, we have 
considered the head-on collision of two nonrotating black holes,  which are modelled as two moving punctures. We compute
the initial conditions through the
\texttt{TwoPunctures} initial data code of~\cite{Ansorg:2004ds}, selecting two black holes with 
equals masses, $M_1=M_2=1$, initial positions given by $\boldsymbol{x}^- =
(-1,0,0)$ and $\boldsymbol{x}^+ = (+1,0,0)$, zero individual spins and zero linear momenta\footnote{{As an alternative way to obtain the initial conditions, one could also adopt the analytic prescription introduced by \cite{Brill1963}}. See also \cite{Alcubierre2003b}.}.

\begin{figure}[!ht]
   \begin{center}
      \includegraphics[width=0.9\textwidth,clip=]{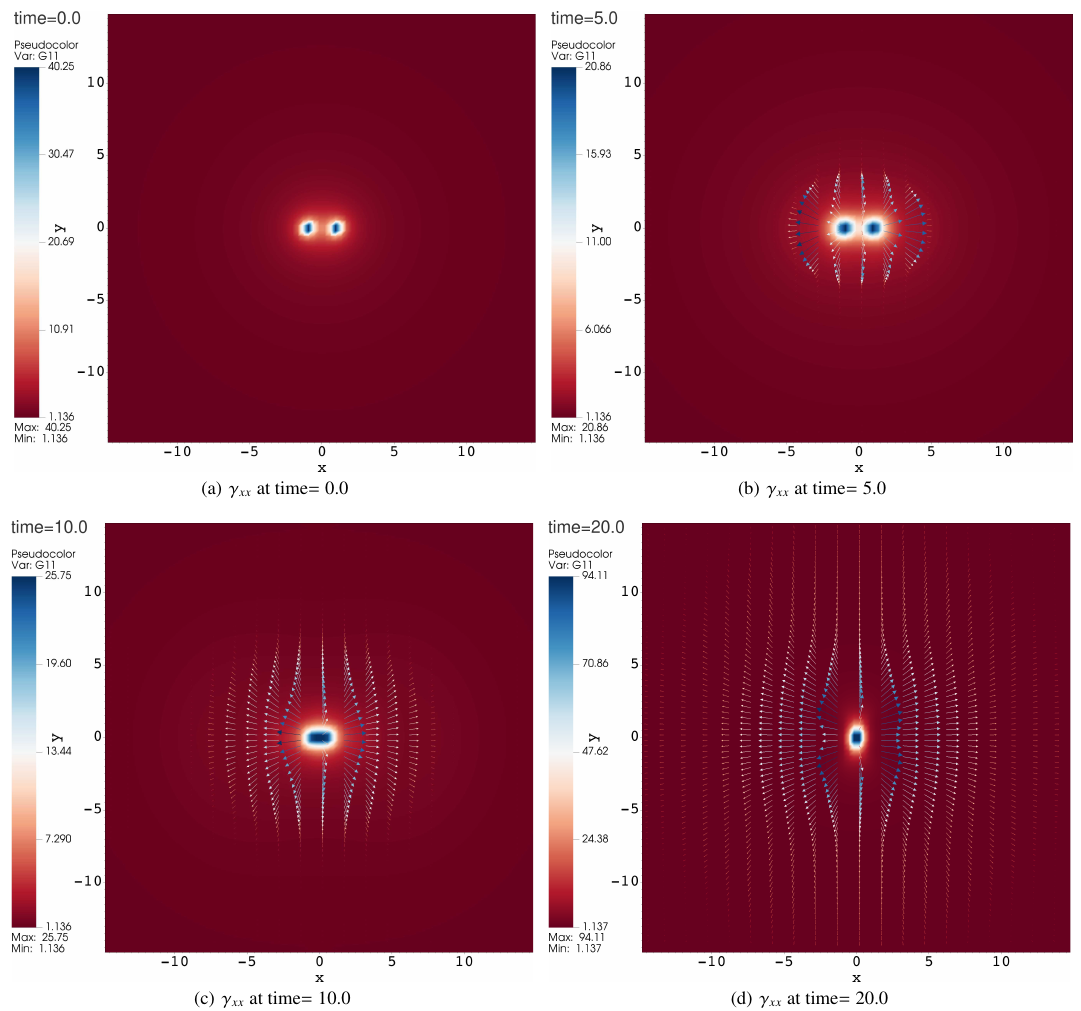}
      \caption{\nameref{sec:HeadOnCollision}: Profiles for the metric term $\gamma_{xx}$ shown at various time levels $(t=0,5,10,20)$. The computations were performed using a ninth--order accurate FD--WENO scheme.}
      \label{fig:fluct_tp_BH}
   \end{center}
\end{figure}

\begin{figure}[!ht]
   \begin{center}
      \includegraphics[width=0.9\textwidth,clip=]{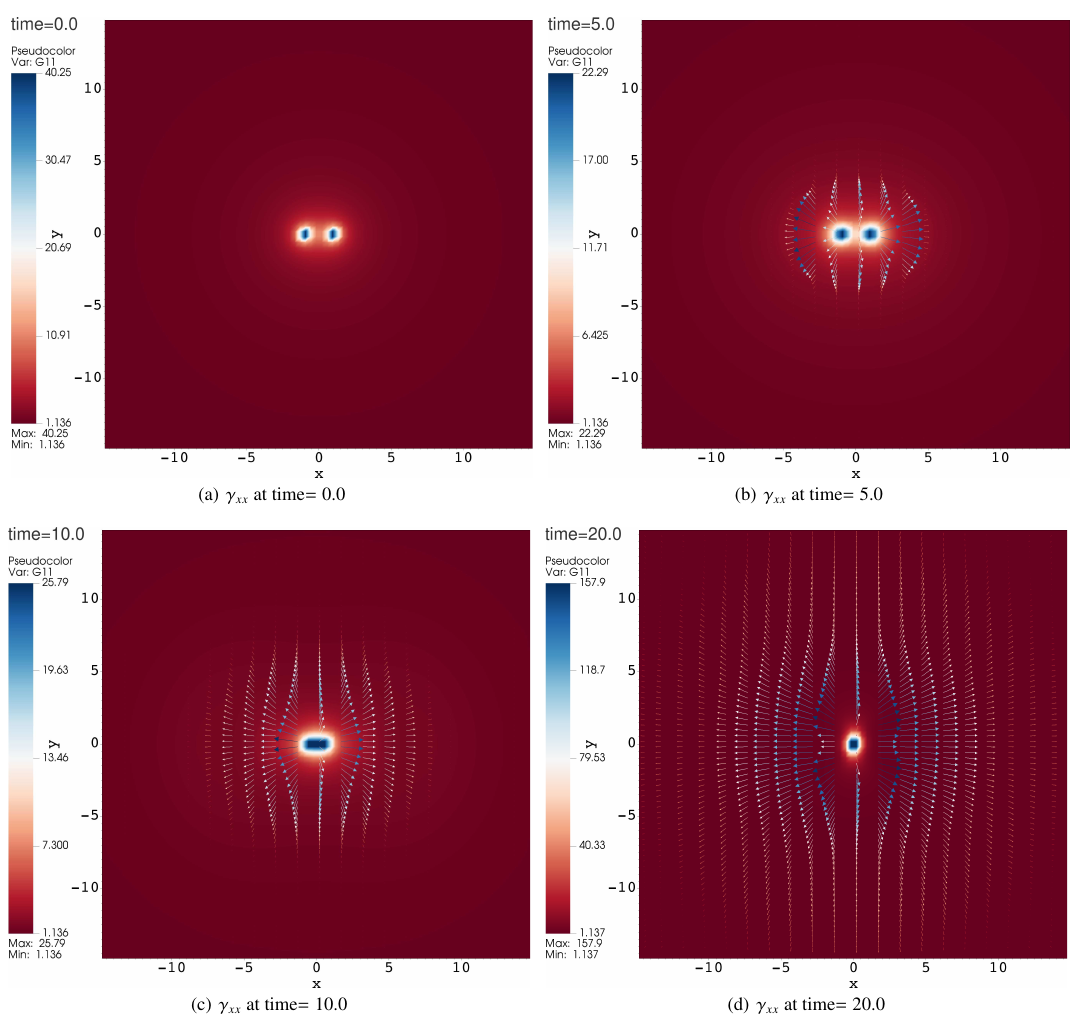}
      \caption{\nameref{sec:HeadOnCollision}: Profiles for the metric term $\gamma_{xx}$ shown at various time levels $(t=0,5,10,20)$. The computations were performed using an eighth--order accurate AFD--WENO scheme.}
      \label{fig:AFD_tp_BH}
   \end{center}
\end{figure}

The three-dimensional computational domain is given by $\Omega = [-15;15]^3$ covered by a $80\times 80 \times 80$ uniform grid, while outgoing boundary conditions are imposed at the border of the computational domain. As shown by \cite{DumbserZanottiGaburroPeshkov2023}, {to avoid division by zero} at time $t=0$
we find it useful to filter the lapse as
\begin{equation}
\alpha=\frac{\alpha r^6 + \epsilon \alpha_{min}}{r^6 + \epsilon}\,,
\end{equation}
where $\alpha_{min}=0.01$, $\epsilon=10^{-4}$.
Moreover, during the time evolution, all the metric terms are {naturally smoothed out by the \textit{numerical dissipation} present in our Riemann-solver-based WENO finite difference schemes. Note that the use of approximate Riemann solvers means upwinding, which is dissipative, and that WENO is a nonlinear shock capturing feature for high order schemes that has been deliberately designed to deal with discontinuities and singularities  within the discrete solution.} This avoids metric spikes, reaching a rather smooth maximum value at $\gamma_{max}\sim 25$. 
With these caveats, we have solved this test using $\kappa_1=0.2$, $\kappa_2=0.2$, $c=0$, $\mu=0.1$
and with the {\emph {gamma--driver}} activated. The results of our calculations are reported in Fig.~\ref{fig:fluct_tp_BH} and Fig.~\ref{fig:AFD_tp_BH}, respectively, where the contour plots of $\gamma_{xx}$, as well as the shift
vector field, are shown at four representative times. 
Fig.~\ref{fig:fluct_tp_BH} refers to the calculations performed with the FD-WENO scheme,
while  Fig.~\ref{fig:AFD_tp_BH} refers to those obtained with the AFD-WENO scheme.

By the time $t\sim 10$ {(with units of time being $M_1=M_2$)}, the merger of the two black holes is almost complete, and a single  black hole is hence produced,
{remaining stationary on longer times.}
The vector field of the shift $\beta^i$
manifests the expected behaviour, being oriented in opposite direction with respect to the merger. 

\section{Conclusions}
\label{sec:conclusions}
Advances in the sensitivities of gravitational wave interferometers are such that they will make it possible to make very accurate observations. As a result, we will soon need very high order schemes for numerical relativity that can match those accurate observations. 
{This motivated} us to look at suitable high order accurate discretizations of first order formulations of the Einstein field equations. Because first order formulations have a hyperbolic PDE structure that is common to other very well-explored PDE systems, higher order solution methods have been very well-developed for this class of problems. Therefore, it is very attractive to look at high accuracy methods for solving first order hyperbolic PDE systems. The well-developed higher-order methodologies include DG schemes, finite volume WENO schemes, and finite difference WENO schemes. All three of these classes of schemes present natural pathways to high order of accuracy in space. However, the first order formulations for numerical GR have the additional complication that they result in extremely large non-conservative hyperbolic PDE systems. These systems are indeed so large that the memory requirements for DG, and perhaps even finite volume WENO schemes, are quite {demanding}. This forces us to take a hard look at recent advances in finite difference WENO schemes. 

The Z4 formulation of the Einstein equations, written as a first--order system, naturally gives rise to a hyperbolic PDE system, that we have indicated with the acronym FO-Z4 and that is briefly described in Section 2. While the original Z4 system
was proposed in first--order \emph{conservative} form~\cite{Bona:2004,Bona-and-Palenzuela-Luque-2005:numrel-book}, 
a close analysis performed by \cite{DumbserZanottiGaburroPeshkov2023} has shown that
its hyperbolicity is better preserved when it is 
cast as a hyperbolic PDE system in \emph{non-conservation form}. 
This is of course  problematic from the numerical viewpoint,
because finite difference WENO methods for systems that are in non-conservation form are not well known to the astrophysical community and general-purpose solution methods for PDEs with non-conservative products have only been recently discovered. It is, therefore, of great value to the astrophysical community that these methods should all be concatenated in one place and shown to work for the FO-Z4 system of general relativity. While we focus on the FO-Z4 system, the methods that we present here are much more broadly useful for any first--order formulation of GR. Indeed, the Introduction documents other first--order formulations for GR that also result in hyperbolic PDEs with non-conservative products.

Section 3 shows us the different finite difference WENO technologies that are at our disposal. These bifurcate broadly into FD-WENO schemes that are based on higher order reconstruction and AFD-WENO schemes that are based on higher order interpolation. FD-WENO schemes are described in Sub-sections 3.1 and 3.2. The FD-WENO scheme in Sub-section 3.2 is particularly easy to implement, very low cost, and especially useful when the relativistic hydrodynamical equations don't need to be solved in conjunction with the Einstein field equations. But we also realize that the relativistic hydrodynamics equations are naturally written in conservation-law form. As a result, we will quite frequently need the AFD-WENO formulation that is described in Sub-sections 3.3 and 3.4. While AFD-WENO schemes cost marginally more than their FD-WENO cousins, they have the great advantage that they can retrieve a conservation form when conservation needs to be respected in the physical problem. We also present discussions on reconstruction, interpolation, non-linear hybridization and upwinding which are all the algorithmic niceties that a robust high order scheme must have if it is to operate stably. Visual descriptions of the algorithms have also been provided in Figs. 1, 2 and 3 in order to make the methods accessible to the greater community. 

It is also worth pointing out that these days one desires not just a baseline scheme but also an ecosystem of supporting ideas around the baseline scheme. Examples of this ecosystem of ideas include well-balancing, physical constraint preservation, and divergence- and curl-free evolution of vector fields. Wherever possible, we have documented the additional availability of this ecosystem of supporting capabilities for the FD-WENO and AFD-WENO schemes described here.

We have therefore implemented the FD-WENO and AFD-WENO schemes 
in a three dimensional code that uses Cartesian coordinates within the standard $3+1$ foliation of spacetime. We have successfully
reproduced some of the most relevant standard tests of NR in vacuum, as outlined in \cite{Alcubierre2004}, which include a gauge wave, the robust stability test, the Gowdy wave, stationary isolated black holes (with excision) and a simple head-on collision of two puncture black holes~\citep{Ansorg:2004ds}. Concerning the gauge wave, we could evolve it on a long time scale with {very small} 
deviations from the analytic solution, provided the order of the scheme is at least 4. 
This points once more to the conclusion that high order of accuracy might become absolutely crucial in NR. 
Concerning single stationary black holes
{(not of puncture type but rather the exact solution of \cite{Kerr63})}, we could evolve them stably over long timescales, thus confirming the positive results obtained by \cite{DumbserZanottiGaburroPeshkov2023} who also adopted the same FO-Z4 formulation, but with an entirely different numerical scheme. In this test case, just like in 
\cite{DumbserZanottiGaburroPeshkov2023}, we had to augment our numerical scheme with a suitable \emph{well-balancing} feature, in such a way to preserve perfect stationarity. We recall that, on the contrary, the standard BSSNOK formulation has known problems in performing such a test. Academic as it may seem at first sight, this test is nevertheless very important since it opens the door to a rather accurate numerical study of the normal modes of oscillations of black holes~\citep{Berti2009,Mamani2022}.
Finally, concerning binary black holes mergers, we confirm the ability of the FO-Z4 formulation in treating simple set-ups with two moving punctures. However, the lack of the conformal factor, that is typical of this formulation, must be compensated by appropriate filtering of the metric spikes that might be produced during the merger. Further investigations are required to make this approach more efficient while at the same time physically neutral. 


\section{Acknowledgments}
This work was financially supported by the Italian Ministry of Education, University 
and Research (MIUR) in the framework of the PRIN 2022 project \textit{High order structure-preserving semi-implicit schemes for hyperbolic equations} and via the  Departments of Excellence  Initiative 2018--2027 attributed to DICAM of the University of Trento (grant L. 232/2016). DSB acknowledges support via NSF grant NSF-AST-2009776, NASA grant NASA-2020-1241 and NASA grant 80NSSC22K0628.

\noindent
We acknowledge the CINECA award under the ISCRA initiative, for the availability of high-performance computing resources and support.

\bibliography{ms}{}
\bibliographystyle{aasjournal}

\end{document}